\documentclass[aps,prb,twocolumn,longbibliography,notitlepage,nofootinbib,superscriptaddress]{revtex4-2}

\usepackage{comment}
\usepackage{multirow}
\usepackage{ifthen}
\usepackage{physics}

\usepackage{amsmath}\allowdisplaybreaks
\usepackage{upgreek}
\usepackage{amsfonts}
\usepackage{amssymb}
\usepackage{dsfont}
\usepackage{braket}
\usepackage{graphicx}
\usepackage{tikz-cd}

\def\ua{\uparrow}
\def\da{\downarrow}

\usepackage{afterpage}
\graphicspath{ {./figs/} }

\usepackage{epsfig}

\usepackage[scr=boondox]{mathalpha}

\usepackage{pst-all}

\usepackage{mathtools}
\usepackage{tikz}
\usetikzlibrary{intersections,calc,3d,arrows,decorations.markings}
\usepackage{color}
\usepackage{hyperref}
\hypersetup{colorlinks=true,citecolor=blue,linkcolor=blue, urlcolor=blue}
\usepackage{diagbox}

\def\H{\mathcal{H}}

\def\Z{\mathbb{Z}}
\def\R{\mathbb{R}}

\def\TT{\mathsf{T}}

\def\U{\text{U}}
\def\AI{\mathsf{AI}}
\def\Gspace{G_{\text{space}}}
\newcommand{\NN}{{\mathbb N}}
\newcommand{\ZZ}{{\mathbb Z}}
\newcommand{\RR}{{\mathbb R}}
\newcommand{\CC}{{\mathbb C}}

\newcommand{\EE}{{\mathbb E}}

\newcommand{\om}{{\omega_2}}
\def\Id{\mathrm{Id}}
\newcommand{\hh}{\mathrm{h}}
\newcommand{\zz}{\mathrm{z}}
\renewcommand{\aa}{\mathrm{a}}

\renewcommand{\tt}{\mathtt{t}}

\newcommand{\SO}{\mathscr{S}_{\text{o}}}
\newcommand{\PO}{\vec{\mathscr{P}}_{\text{o}}}
\newcommand{\OO}{\text{o}}
\def\Res{\mathrm{Res}}

\newcommand{\eps}{\epsilon}

\newcommand{\Cmop}{\tilde{C}_{M_{\text{o}}}^+}
\newcommand{\Cmom}{\tilde{C}_{M_{\text{o}}}^-}
\newcommand{\Cmopm}{\tilde{C}_{M_{\text{o}}}^{\pm}}
\newcommand{\MO}{M_\text{o}}

\def\rbs{\boldsymbol{r}}
\def\tbs{\boldsymbol{t}}
\def\ee{\mathrm{e}}
\def\ii{\mathrm{i}}
\def\Rbs{\boldsymbol{R}}

\usepackage{environ}
\NewEnviron{eqs}{%
\begin{equation}\begin{split}
    \BODY
\end{split}\end{equation}
}

\def\Mmc{\mathcal{M}}

\def\WP{\mathsf{WP}}

\def\pt{\mathrm{pt}}

\def\RU{\mathrm{RU}}
\def\NN{\mathbb{N}}

\def\Nmc{\mathcal{N}}

\def\Gwp{G_{\text{space}}}
\def\Gpt{G_{\text{pt}}}
\def\hbf{\mathbf{h}}
\def\Gs{G_{\text{s}}}
\def\pt{\text{pt}}
\def\Qmc{\mathcal{Q}}

\def\zero{\boldsymbol{0}}
\def\abss{\boldsymbol{a}}
\def\half{\tfrac{1}{2}}

\def\Pmc{\mathcal{P}}

\def\Rmc{\mathcal{R}}
\def\BZ{\mathrm{BZ}}
\def\kbs{\boldsymbol{k}}
\def\HSM{\mathsf{HSM}}
\def\HSMsp{{\mathsf{HSM}^{\star}}'}

\def\TT{\mathbb{T}}
\def\Rep{\mathsf{Rep}}
\def\Ch{C}

\def\Rbs{\boldsymbol{R}}
\def\Haldane{\underline{\mathrm{H}}}

\renewcommand{\selectlanguage}[1]{}

\begin{document}

\author{Naren Manjunath}
\affiliation{Department of Physics, Condensed Matter Theory Center, and Joint Quantum Institute, University of Maryland, College Park, Maryland 20742, USA}
\affiliation{Perimeter Institute for Theoretical Physics, Waterloo, Ontario N2L 2Y5, Canada}

\author{Vladimir Calvera}
\affiliation{Department of Physics, Stanford University, Stanford, California 94305, USA}

\author{Maissam Barkeshli}
\affiliation{Department of Physics, Condensed Matter Theory Center, and Joint Quantum Institute, University of Maryland, College Park, Maryland 20742, USA}

\title{Characterization and classification of interacting (2+1)D topological crystalline \\ insulators with orientation-preserving wallpaper groups}

\begin{abstract}
    While free fermion topological crystalline insulators have been largely classified, the analogous problem in the strongly interacting case has been only partially solved. In this paper, we develop a characterization and classification of interacting, invertible fermionic topological phases in (2+1) dimensions with charge conservation, discrete magnetic translation and $M$-fold point group rotation symmetries, which form the group $G_f = \U(1)^f \times_{\phi} [\Z^2\rtimes \Z_M]$ for $M=1,2,3,4,6$. $\phi$ is the magnetic flux per unit cell. We derive a topological response theory in terms of background crystalline gauge fields, which 
    gives a complete classification of different phases and a physical characterization in terms of quantized response to symmetry defects. We then derive the same classification in terms of a set of real space invariants $\{\Theta_{\text{o}}^\pm\}$ that can be obtained from ground state expectation values of suitable partial rotation operators. We explicitly relate these real space invariants to the quantized coefficients in the topological response theory, and find the dependence of the invariants on the chiral central charge $c_-$ of the invertible phase. Finally, when $\phi = 0$ we derive an explicit map between the free and interacting classifications. 
\end{abstract}
\maketitle
\tableofcontents

\section{Introduction}

The classification and characterization of topological phases of matter with crystalline symmetries is a major direction in condensed matter physics. (For a partial list of references, see e.g. \cite{hasan2010,fu2011topological,Benalcazar2014,ando2015topological,Chiu2016review,watanabe2015filling,watanabe2016filling,schindler2018higher,khalaf2018higher,Benalcazar2019HOTI,Kruthoff2017TIBandComb,Bradlyn2017tqc,Po2017symmind,watanabe2018structure,khalaf2018symmetry,tang2019comprehensive,Cano_2021,Elcoro2021tqc,herzogarbeitman2022interacting,Essin2014spect,Essin2013SF,YangPRL2015,hermele2016,zaletel2017,song2017,huang2017,Thorngren2018,Song2020,Miert2018dislocationCharge,Li2020disc,Liu2019ShiftIns,manjunath2021cgt,Manjunath2020fqh,zhang2022fractional,zhang2022pol,zhang2023complete,manjunath2022mzm} ).
By definition, two gapped ground states that preserve a certain symmetry are said to belong to the same topological phase if they can be adiabatically connected without breaking that symmetry, and otherwise are said to be in distinct topological phases. For crystalline systems in particular, the problems of classifying the topological invariants associated to these phases and understanding how to extract them from microscopic models is still not fully solved, despite enormous recent progress.

\begin{table*}[t]
    \centering
    \begin{tabular}{l|l|l|p{0.2\textwidth}|p{0.21\textwidth}|l}
        $M$& Free classification &  Interacting classification & Quantization of response coefficients & Relation to real-space invariants  & Free vs interacting map   \\ \hline
         1 & $\Z^2$ &   $\Z^3$ & Eq.~\eqref{eq:FTresult_C1}&--  & \\
         2 &$\Z^6$ & $\Z^3\times\Z_4^3\times\Z_2$ & Eq.~\eqref{eq:FTresult_C2} & Eq.~\eqref{eq:RSdef_C2}  & Eq.~\eqref{eq:FImap-C2} \\
         3 & $\Z^8$ & $\Z^3\times\Z_3^5$ & Eq.~\eqref{eq:FTresult_C3}  & Eq.~\eqref{eq:RSdef_C3} & Eq.~\eqref{eq:FImap-C3}\\
         4 & $\Z^9$ & $\Z^3\times\Z_8\times\Z_4^2\times\Z_2$ & Eq.~\eqref{eq:FTresult_C4}  & Eq.~\eqref{eq:RSdef_C4} & Eq.~\eqref{eq:FImap-C4}\\
         6 & $\Z^{10}$ & $\Z^3\times\Z_{12}\times\Z_6\times\Z_3$ & Eq.~\eqref{eq:FTresult_C6}  & Eq.~\eqref{eq:RSdef_C6} &Eq.~\eqref{eq:FImap-C6} \\
    \end{tabular}
    \caption{Summary of the free and interacting classification of invertible fermionic states with symmetry $G_f = \U(1)^f \times_{\phi}[\Z^2\rtimes \Z_M]$ for $M=1,2,3,4,6$. In the first and last columns, we assume a flux per unit cell $\phi = 0$; the other results hold for arbitrary $\phi$.}
    \label{tab:Classif_Summary}
\end{table*}

In this paper we focus on \textit{invertible} fermionic topological states in (2+1) space-time dimensions \cite{barkeshli2021invertible,aasen2021characterization}. A many-body state $\ket{\Psi}$ is said to be invertible if there exists an `inverse' state $\ket{\Psi^{-1}}$, such that $\ket{\Psi} \otimes \ket{\Psi^{-1}}$ can be adiabatically connected to a product state. Special cases of invertible phases include all topological phases that can be realized in free fermion models, and symmetry-protected topological (SPT) states \cite{hasan2010,qiRMP2011,bernevig2013topological,Kitaev2009periodic,ryu2010,Chiu2016review,Chen2013,kapustin2014SPTbeyond,Gu2014Supercoh,kapustin2015fSPT,senthil2015,Wang2020fSPT}. Important examples of invertible states in (2+1)D (D is the space-time dimension) include the integer quantum Hall states, Chern insulators, the quantum spin Hall insulator, and topological insulators and superconductors; note that this definition excludes topologically ordered states with anyonic excitations, which are non-invertible states.  

The classification of (2+1)D invertible states with crystalline symmetries is now almost complete for free fermion systems with $\U(1)$ charge conservation symmetry and additional wallpaper groups, which are the symmetries of interest in this paper \cite{bernevig2013topological,Chiu2016review,Kruthoff2017TIBandComb,Bradlyn2017tqc,Po2017symmind,watanabe2018structure,khalaf2018symmetry,tang2019comprehensive,Cano_2021,Elcoro2021tqc}. Here we understand the group structure of the classification, how to obtain a complete set of momentum-space band invariants, and we have a partial understanding of how to obtain real-space invariants. Prior works have studied systems with zero chiral central charge (denoted $c_-$), in which there is an equal number of left- and right-moving gapless edge states, as well as the case with $c_- \ne 0$. These free fermion classifications are based on methods from Wannier representation theory and $K$-theory, which classify the distinct topological band structures allowed by symmetry. An important situation not covered by these classifications is when the system has a nonzero flux $\phi$ per unit cell, which changes the symmetry group, replacing translations with magnetic translations. 

In the case of interacting fermionic systems, our understanding is less complete. For systems with internal symmetries, several classification approaches have been employed previously, including fixed-point wavefunction constructions (these are applicable only when $c_-=0$) \cite{Gu2014Supercoh,Wang2020fSPT}, cobordism theory \cite{kapustin2015fSPT} and invertible topological quantum field theory (TQFT) \cite{freed2016}. Another general framework, which is the one used in this paper, is based on $G$-crossed braided tensor category theory \cite{barkeshli2019,Barkeshli2020Anomaly,barkeshli2021invertible,aasen2021characterization}, which characterizes the algebraic properties of symmetry defects in terms of a concrete set of data, consistency equations, and equivalence relations. The G-crossed BTC approach is more physically transparent and computationally tractable than the cobordism and invertible TQFT approaches, which assume at the outset the existence of a topologically invariant path integral on any space-time manifold. Nevertheless, the different approaches, when their domain of validity overlaps, are expected to be mathematically equivalent. The G-crossed BTC, cobordism, and invertible TQFT approaches will collectively be referred to here as the `TQFT' approach. 

If we consider interacting systems with crystalline symmetries, there are two broad classification approaches which are physically distinct but are nonetheless mathematically in close correspondence. The first is based on real-space constructions of ideal ground state wavefunctions \cite{song2017,huang2017,zhang2020realspace}. These have been applied systematically to different two-dimensional crystalline symmetry groups, but this approach only captures states with $c_- = 0$. The second general approach is to use TQFT methods after replacing the desired crystalline symmetry group by an effective internal symmetry group. This approach relies on the so-called `fermionic crystalline equivalence principle' (fCEP), discussed previously in Refs.~\cite{Thorngren2018,Else2019,debray2021invertible,zhang2020realspace,manjunath2022mzm}. The fCEP states that the classification of invertible states with spatial symmetry $G_f$ is in one-to-one correspondence with that of invertible states with an internal symmetry $G_{f}^{\text{eff}}$ that can be fully deduced from $G_f$. The TQFT approach is applied to crystalline systems after assuming the fCEP, which has not been proven in full generality. However, its usefulness is that it can also be applied to states with $c_- \ne 0$, where the above real-space constructions no longer hold.

The TQFT-based approach has been applied to both classify and characterize invertible states with orientation-preserving wallpaper group symmetries \cite{Manjunath2020fqh,manjunath2021cgt,zhang2022fractional,zhang2022pol,zhang2023complete}. Orientation-reversing symmetries such as reflections and glides have not yet been analyzed in this framework. Also note that many prior works have studied how to extract crystalline topological invariants which are equivalent to those that appear in TQFT, from microscopic models. These works primarily focus on $c_- = 0$ \cite{Fang2012PGS,Jadaun2013PGS,Biswas2016,Miert2018dislocationCharge,Liu2019ShiftIns,Song2019_2pi,Song2021polarization,herzogarbeitman2022interacting}, but in some cases study $c_- \ne 0$ as well \cite{coh2009,shiozaki2017invt,Li2020disc}. 

There are however important gaps in the above literature. First, although specific crystalline topological invariants and response properties predicted by TQFT have been studied in detail \cite{zhang2022fractional,zhang2022pol,zhang2023complete}, a thorough derivation of the entire TQFT classification has not appeared previously, even in the orientation-preserving case. Secondly, although the TQFT and real-space classifications are closely related (especially when $c_- = 0$), their precise relationship has not been spelled out mathematically. Thirdly, we know that there is a map between the free and interacting classification of invertible states with a given symmetry~\cite{Senthil2015SPT}, but for (2+1)D crystalline topological states this map has not been systematically studied, despite the huge progress in understanding the free and interacting classifications on their own. Knowing this map is important because two distinct free fermion invertible states may be equivalent within the interacting classification if they can be adiabatically connected by turning on suitable interactions \cite{fidkowski2010effects,Gu2014interaction,Morimoto2015}; moreover certain invertible states are \textit{intrinsically} interacting in that they cannot be realized by free fermions \cite{ning2021enforced,manjunath2022mzm}. Therefore this map explains, in non-perturbative fashion, how free fermion topological invariants get modified once interactions are considered.

The goal of this paper is to address these three shortcomings. We consider the symmetries $G_f = \U(1)^f \times_{\phi} G_{\text{space}}$, where $G_{\text{space}}$ is an orientation preserving wallpaper group in two space dimensions. This notation is explained below. We classify invertible topological states in (2+1)D using three completely different approaches: a TQFT, a real-space construction, and a `band structure combinatorics' approach. The TQFT and real-space classifications apply to interacting systems while the `band structure combinatorics' classification applies to free fermion systems. We discuss in detail the relationship between the TQFT and real-space classifications, as well as the map between the free (band structure combinatorics) and interacting (real-space) classifications. These results are summarized in Table~\ref{tab:Classif_Summary}.

Mathematically we define  $G_{\text{space}} = \Z^2 \rtimes \Z_M$, where $\Z^2$ refers to discrete translations in two space dimensions and $\Z_M$ for $M = 1,2,3,4,6$ refers to elementary $M$-fold point group rotations. $G_{\text{space}}$ is also referred to as the wallpaper group pM. The symbol $\times_{\phi}$ implies that elementary translations in the $x$ and $y$ directions do not commute, due to a magnetic flux $\phi \mod 2\pi$ in each unit cell.

We first derive a topological `response' action from TQFT that classifies invertible $G_f$ symmetric topological states in (2+1) dimensions with interactions; the result is Eq.~\eqref{eq:mainresponse}. The topological action directly expresses the desired crystalline topological invariants in terms of the system's response to inserting symmetry defects such as magnetic fluxes, or lattice disclinations and dislocations. While some individual terms in the effective action have been studied previously \cite{zhang2022fractional,zhang2022pol,zhang2023complete}, this work gives the first complete derivation in the case of fermionic systems. Prior work gave a complete derivation in the case of bosonic topological phases \cite{manjunath2021cgt,Manjunath2020fqh}. While the set of topological terms is the same as in the bosonic case, the quantization of the coefficients is different and is a main result of this paper. This classification allows for $c_- \ne 0$ and is also valid for $\phi \ne 0$.

Next, we discuss a physically distinct classification based on real-space invariants, which is again applicable to interacting systems. For each high symmetry point $\OO$ of the real-space unit cell, we define a pair of invariants $\Theta^{\pm}_{\OO}$. Physically, these invariants can be obtained from the expectation value of the many-body ground state under suitable partial rotation operators \cite{zhang2023complete}. We should emphasize that although the previous real-space constructions in Refs.~\cite{huang2017,zhang2020realspace,Song2020} only apply when $c_- = 0$, the invariants $\Theta^{\pm}_{\OO}$ can be defined even when $c_- \ne 0$, and thus generalize the above works (see Ref.~\cite{herzogarbeitman2022interacting} for a related study of many-body real space invariants). If $\Theta^{\pm}_{\OO}$ are known for each $\OO$, we can recover all the invariants in the classification that are classified by cyclic groups. If we additionally know the Chern number $C$, the chiral central charge $c_-$ and the filling per unit cell $\nu$, which are fixed by integer-valued invariants, we can completely determine each response coefficient in the TQFT. We show that the above real-space framework gives all the invariants predicted by TQFT except for $C,c_-$ and $\nu$, and we relate the topological response coefficients to $\Theta^{\pm}_{\OO}$ through explicit formulas. 

The response coefficients in Eq.~\eqref{eq:mainresponse}, as well as the real-space invariants $\Theta^{\pm}_{\OO}$, generally have a dependence on $c_-$. An important result of this paper is to clearly derive this $c_-$-dependence. 
Previous real-space classifications of invertible fermionic states with crystalline symmetries \cite{zhang2020realspace} assumed $c_-=0$ and therefore do not study such $c_-$-dependence. Also note that the above classification schemes apply to arbitrary rational values of $\phi/2\pi$, in contrast to several previous works that all assumed $\phi = 0$. 

Finally, we classify free fermion invertible states with $G_f$ as above but specializing to $\phi = 0$, using methods based on band structure combinatorics \cite{Kruthoff2017TIBandComb,Li2020disc}. Although the classification itself is well known in this case \cite{Kruthoff2017TIBandComb}, the free-to-interacting map is poorly understood. We explicitly derive the relation between $\Theta^{\pm}_{\OO}$ and free fermion band invariants, for each $\OO$; note that this is equivalent to specifying a free-to-interacting map. See Table~\ref{tab:Classif_Summary} for the relevant equations. In our derivation we assume $\phi = 0$, but include $c_- \ne 0$. To our knowledge, this is the first instance where a complete free-to-interacting map has been derived for a crystalline symmetry, accounting for $c_- \ne 0$.

This work addresses several of the major remaining gaps in the classification and characterization problem for (2+1)D invertible fermionic states with orientation-preserving wallpaper group symmetries. Our results may be of relevance to recent experiments simulating Chern insulators in various platform \cite{Dean2013,Hunt2013hb,Ponomarenko2013,Spanton2018FCI,Saito2021,Jaksch2003hofstadter,aidelsburger2013,miyake2013,kennedy2015,Aidelsburger2015,Tai2017hofstadter,hafezi2013,ozawa2019,Owens2018hofstadter,Zondiner2020-jn,Nuckolls2020-jw,Serlin2020-fc,Wu2021-ef}. Apart from the Chern number, our theory predicts that these states have many additional nontrivial topological invariants which depend on their underlying crystalline symmetries, and which can potentially be used as a basis to characterize them.

\subsection*{Organization of paper}
The rest of this paper is organized as follows. Sec.~\ref{sec:Defs} sets up the notation we will use in the paper, along with some background material. In Section~\ref{sec:FT} we state the results of the topological field theory classification. In Section~\ref{sec:RSI} we state the definition and properties of the real-space invariants, both when $c_- = 0$ and $c_- \ne 0$, and establish the relationship between the TQFT and real-space classifications. In Sec.~\ref{sec:FImap}, we discuss the corresponding free fermion invariants and state the map between the free and interacting classifications of invertible states with the given symmetry $G_f$. Note that the technical derivations relevant to Secs.~\ref{sec:FT},~\ref{sec:RSI},~\ref{sec:FImap} can be found in Apps.~\ref{app:DerTopAct},~\ref{app:RSI} and~\ref{app:BandInsulators}-\ref{app:RelationC6-RSI} respectively. In Sec.~\ref{sec:Disc} we conclude and discuss future directions.

\section{Definitions}\label{sec:Defs}

\subsection{Unit cells}
\begin{figure}
    \centering
    \includegraphics[width=0.47\textwidth]{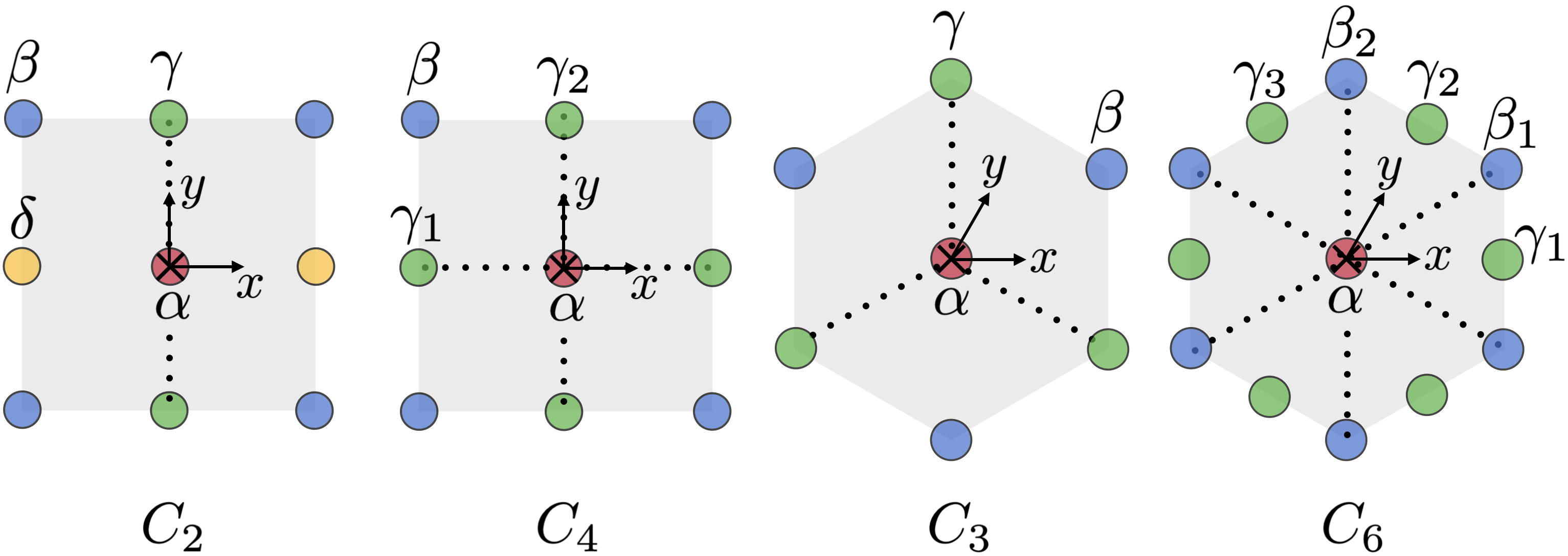}
    \caption{Maximal Wyckoff positions for the orientation preserving wallpaper groups.}
    \label{fig:wyckoff}
\end{figure}

In this work we consider the rotation-symmetric lattices p2, p3, p4 and p6. Representative unit cells for these lattices are shown in Fig.~\ref{fig:wyckoff}. The different high symmetry points are denoted by Greek letters $\alpha, \beta, \gamma, \dots$. The center of the unit cell is denoted $\alpha$ and is chosen to have $M$-fold rotational symmetry. If two points in the same unit cell are related by rotations about some origin (that is, they are in the same maximal Wyckoff position) but are inequivalent under lattice translations, they are distinguished by subscripts (for example we have $\gamma_1, \gamma_2$ on the square lattice).

\subsection{Symmetry operators on the lattice}

Consider any high symmetry point $\OO$ of the unit cell. Let $\MO$ be the order of the rotational symmetry group which preserves $\OO$. For the space group pM, the largest possible value of $\MO$ equals $M$, and in general $M$ is always divisible by $\MO$. 

The symmetry-protected topological invariants for a given topological phase are defined with respect to a choice of symmetry operators. For the case of point-group rotation symmetry, we must specify an operator $\tilde{C}_{M_{\OO}}^+$, corresponding to $M_{\OO}$-fold rotations about $\OO$, and which satisfies $(\tilde{C}_{\MO}^+)^{\MO} = 1$. We use the tilde superscript to signify that $\Cmop$ in general is a `magnetic rotation,' meaning that it comes with a $\U(1)$ gauge transformation which depends on the chosen vector potential. 

There is an ambiguity in the above definition, since given any operator $\Cmop$ satisfying the above conditions we could equally consider the operators $\Cmop \times e^{i \frac{2\pi j}{\MO} \hat{N}}$ for $j = 0, 1, \dots , \MO-1$. Here $\hat{N}$ is the number operator. However, for all the cases studied in Refs.~\cite{zhang2022fractional,zhang2022pol,zhang2023complete}, there is a canonical choice of $\Cmop$. This is identified by using a cut-and-glue procedure to create a disclination of angle $2\pi/\MO$ centered at $\OO$, starting from the infinite plane. As explained in Ref.~\cite{zhang2022fractional}, there is a unique operator $\Cmop$ which attaches zero excess flux at such a disclination; the other possible rotation operators $\Cmop \times e^{i \frac{2\pi j}{\MO} \hat{N}}$ and will insert an additional flux $\frac{2\pi j}{\MO}$ at the disclination, relative to $\Cmop$. Now in a generic lattice model with a magnetic field, properly defining `zero excess flux' around a disclination can be subtle, but Ref.~\cite{zhang2022pol} gives a concrete method to do so.

We also define a set of operators 
$\tilde{C}_{M_\OO}^- := e^{i \frac{\pi}{M_{\OO}} \hat{N}} \tilde{C}_{M_{\OO}}^+$,
and $\tilde{C}_{M_{\OO}, \chi}^{\pm} := e^{i\chi \frac{2\pi}{\MO} \hat{N}} \tilde{C}_{M_{\OO}}^\pm$. Note that
$\tilde{C}_{M_{\OO}}^- = \tilde{C}_{M_{\OO}, 1/2}^+$ and $(\Cmom)^{\MO} = (-1)^F$ where $(-1)^F$ is the fermion parity operation, which is order two and generates a group denoted as $\Z_2^f$.  

\subsection{Background gauge fields}

Next we briefly review the background gauge fields used to derive the topological response theory in Sec.~\ref{sec:FT}. For a more thorough discussion, see App.~\ref{app:DerTopAct}. 

Consider $G_f = \U(1)^f \times_{\phi} [\Z^2 \rtimes \Z_M]$. Using the fCEP (see App.~\ref{app:fcep}), we find that the equivalent internal symmetry group corresponding to $G_f$ is actually isomorphic to $G_f$. Therefore it will be sufficient to construct a background gauge field on a closed 3-manifold $\mathcal{M}^3$ for an \textit{internal} symmetry group isomorphic to $G_f$. Although the gauge fields are for an effective internal symmetry, we will interpret the fluxes of this gauge field as if they were fluxes of the original spatial symmetry. The validity of this interpretation ultimately depends on numerically checking the predictions of the topological response theory in lattice models. A number of such checks have been performed in Refs~\cite{zhang2022fractional,zhang2022pol,zhang2023complete}, giving strong support to this interpretation. More discussion of the fCEP can be found in Sec.~\ref{sec:Disc}. 

There are two equivalent ways to formulate the gauge fields on $\mathcal{M}^3$. First there is a simplicial formulation, where the gauge field is defined on a triangulation of $\mathcal{M}^3$. This is the formulation we use for the actual derivation of the response theory in App.~\ref{app:DerTopAct}, and in the rest of the paper. There is also an equivalent continuum formulation, where we define the gauge fields as real-valued differential 1-forms on $\mathcal{M}^3$ with quantized holonomies. We have only presented the main result (Eq.~\eqref{eq:mainresponse}) in the continuum language because this is the more standard notation in the condensed matter literature. To write Eq.~\eqref{eq:mainresponse} in the simplicial formulation, we replace the wedge products with cup products and impose quantization conditions on the values of the gauge fields. 

In the simplicial formulation, our gauge field of interest can be written as $B = (\delta A,\vec{R},\omega)$ where $\delta A$ is a real valued $\U(1)^f$ gauge field with $\delta A \sim \delta A + 2\pi$, $\vec{R} = (X,Y)$ is a $\Z^2$ gauge field with $\frac{1}{2\pi}\vec{R} \in \Z^2$, and $\omega$ is a $\Z_M$ gauge field with $\omega \in \frac{2\pi}{M}\Z$ and $\omega \sim \omega + 2\pi$. (Note that $\delta$ should not be confused with a coboundary operator, which we denote below as $d$.) We use the notation $\delta A$ because $A$ is identified with the usual vector potential and $\delta A$ represents the deviation of $A$ from some uniform background value that assigns flux $\phi$ per unit cell.

It is important to note that each component of $B$ is defined with respect to a fixed $M$-fold rotationally symmetric origin $\OO$ in the unit cell. Therefore the flux of $B$ carries an origin dependence; this explains why the response coefficients to be introduced below (Eq.~\eqref{eq:mainresponse}) also carry an origin dependence. However, for ease of notation we will leave the dependence of the gauge fields on $\OO$ implicit. 

The flux of $B$ can be written as $(d \delta A + \frac{\phi}{2\pi} A_{XY}, \vec{T}, d\omega)$. $A_{XY}$ and $\vec{T}$ can be written in terms of $\vec{R}$ and $\omega$.\footnote{When $\omega = 0$, $A_{XY} = \frac{1}{2\pi}X \cup Y $ and $\vec{T} = d\vec{R}$. In general, if $B$ is flat, then on a 2-simplex [012], $A_{XY}[012]$ is coboundary-equivalent to the quantity $\frac{1}{4\pi} \vec{R}_{01} \cup (U(\omega_{01}) \vec{R}_{12})$ where $U(\theta)$ is the rotation matrix by the angle $\theta$. Furthermore, $\vec{T}[012] = (1-U(2\pi/M))^{-1}(1-U(\omega_{01}))\vec{R}_2$.} The first component physically corresponds to the total magnetic flux density. Here $A_{XY}$ is a quadratic form in $\vec{R}$ and is physically interpreted as an area element: the quantity $\frac{1}{2\pi}\int_W A_{XY}$ is identified with the number of unit cells in $W$. Therefore $\int_W d \delta A + \frac{\phi}{2\pi} A_{XY}$ measures the total magnetic flux in $W$, including the uniform background (proportional to $A_{XY}$) and the excess part (given by $d\delta A$). This can equivalently be written as $dA$, where $A$ is the usual vector potential. Note that the Lagrangian will always be written in terms of $A$ and not $\delta A$ (see Eq.~\eqref{eq:mainresponse}). 

The second component of the flux, denoted $\vec{T}$, is interpreted as the torsion, which physically corresponds to the total Burgers vector of defects within a given region (multiplied by $2\pi$) . The third component $d\omega$ is interpreted as disclination flux: $\int_W d\omega$ is identified with the total disclination angle of defects in $W$. Finally, it is convenient to view the quantity $\frac{1}{2\pi} A_{XY}$ as an `area flux' that counts the number of unit cells. Then, the different terms in the topological response theory can be physically interpreted as assigning either $\U(1)^f$ charge, linear or angular momentum to the four gauge fluxes introduced above.\footnote{In fact, for a flat gauge field $B$, the four gauge fluxes introduced here correspond (up to a normalization) to the four generators of the group $\H^2(G_b,\Z)$ pulled back using $B$. Here $G_b = G_f/\Z_2^f$ is the bosonic symmetry group.}

\subsection{Math conventions}

We denote by $\ZZ_N \equiv\ZZ/N\ZZ$ the integers modulo $N$. Furthermore, when we say that a real number $x \in a\cdot \ZZ_N = (a\ZZ)/(N\ZZ)$ for $a \in \RR$, we mean that $x$ is an integer multiple of $a$ that is only defined modulo $N$. $a|b$ means for integers $a$ and $b$, $a$ divides $b$. When we have a real number $x$ and a positive number $n$, we denote by $[x]_n$ the residue of $x$ after dividing by $n$.\footnote{In other words, $[x]_n\in [0, n)$ and $[x]_n= x\mod n$. } We denote by $\gcd(m,n)$ the greatest common divisor between the positive integers $m$ and $n$.

\section{Classification from topological field theory}\label{sec:FT}

\begin{table*}
\centering
    \begin{tabular}{l|l|p{0.35\textwidth}|p{0.15\textwidth}}
       Coefficient  & Quantization & Physical interpretation & Comments  \\ \hline
        $C$ & $C = c_- + 8 k_1, \quad c_-, k_1 \in \Z$ & Hall conductivity & \\ 
         $\SO$ & $\SO -\frac{c_-}{2} \in \Z_M$ & Discrete shift & $\SO = \frac{C}{2} \mod 1$ \\ 
          $\ell_{\OO}$ & $\ell_{\OO} - \frac{c_-}{4} \in \Z_M$ & Related to expectation value of ground state under partial rotations (Eq.~\eqref{eq:ells_Theta})& $\ell_{\OO} = \frac{C}{4} \mod 1$\\ 
        $\PO$ &$
 \begin{cases}
     M=2: \frac{1}{2} (\overline{\mathscr{P}}_{\text{o},x},\overline{\mathscr{P}}_{\text{o},y}), &(\overline{\mathscr{P}}_{\text{o},x},\overline{\mathscr{P}}_{\text{o},y}) \in \Z_2 \times \Z_2\\
 M=4: \overline{\mathscr{P}}_{\text{o}} \frac{1}{2} (1,1), &\quad \overline{\mathscr{P}}_{\text{o}} \in \Z_2\\
 M=3: \overline{\mathscr{P}}_{\text{o}} \frac{1}{3} (1,2), &\quad \overline{\mathscr{P}}_{\text{o}} \in \Z_3\\
 M=6: (0,0). &\\
\end{cases}$ &Defines quantized charge polarization $\vec{P}_{\OO} = - \PO \times \hat{z}$  & $\PO$ is defined modulo integer vectors\\ 
        $\vec{\mathscr{P}}_{s,\OO}$ & See Eqs.~\eqref{eq:FTresult_C4},~\eqref{eq:FTresult_C2},~\eqref{eq:FTresult_C3} & Polarization of angular momentum &Also defined modulo integer vectors \\ 
        $\kappa$ & $\kappa \in \Z$ & Charge per unit cell at zero magnetic field & $\nu = C \phi/2\pi + \kappa$ \\ 
        $\nu_s$ & $\nu_s \in \Z_M$ & Angular momentum per unit cell & \\ 
    \end{tabular}
    \caption{Summary of coefficients in the topological field theory for invertible fermionic phases with $G_f = \U(1)^f\times_{\phi} [\Z^2 \rtimes \Z_M]$. We assume that the origin $\OO$ is symmetric under $M$-fold rotations. These coefficients are not necessarily generators of the overall classification. }\label{tab:FT_summary}
\end{table*}

\subsection{General form of topological action}
Fix an $\OO$ for which $\MO = M$, that is, $\OO$ has the highest possible symmetry. Let $\omega$ be a $\Z_M$ background gauge field corresponding to the subgroup of $G_f$ generated by $\Cmop$. The response theory defined with respect to $\Cmop$ reads 
\begin{align}\label{eq:mainresponse}
    \mathcal{L} &= \frac{C}{4\pi} A \wedge dA + \frac{\SO}{2\pi} A \wedge d\omega +\frac{\tilde{\ell}_{\OO}}{4\pi} \omega \wedge d\omega +\frac{\PO}{2\pi} \cdot A \wedge \vec{T} \nonumber \\ & +\frac{\vec{\mathscr{P}}_{s,\OO}}{2\pi} \cdot \omega \wedge \vec{T} +\frac{\kappa}{2\pi} A \wedge A_{XY} +\frac{\nu_s}{2\pi} \omega \wedge A_{XY}.
\end{align}
The response coefficients can be understood as follows; see Table~\ref{tab:FT_summary} for a summary. $C$ is the Chern number, which sets the Hall conductance $\sigma_H = C e^2/h$. $\kappa$ is an integer invariant related to the filling $\nu$ as $\nu = C \phi/2\pi + \kappa$ (therefore $\kappa$ is the filling at zero magnetic field). $\SO$ is the `discrete shift' measured with respect to $\OO$, and was studied in detail in Refs.~\cite{zhang2022fractional,zhang2022pol}. In particular, $\SO$ is defined mod $\MO$ and specifies a fractionally quantized contribution to the charge around a lattice disclination created at $\OO$, as well as the angular momentum of magnetic flux. $\PO$ is a quantized charge polarization with respect to $\OO$ (see \cite{zhang2022pol} for a detailed study). It specifies a fractionally quantized contribution to the charge around a lattice \textit{dislocation} defined wih respect to $\OO$, as well as the \textit{linear} momentum of magnetic flux. $\PO$ can be determined fully if we know $\kappa$ and $\mathscr{S}_{\OO'}$ for each high symmetry point $\OO'$; see Table~\ref{tab:PvsS} for the relevant equations, which were derived in \cite{zhang2022pol}.

\begin{table}[t]
    \centering
    \begin{tabular}{c|c|l}
       $\MO$  & $\OO$  & Equation\\ \hline
        2 & $\alpha$ & $\vec{\mathscr{P}}_{\alpha} = \frac{(1,1)}{2}(\mathscr{S}_{\alpha}+\kappa) + \frac{1}{2}(\mathscr{S}_{\delta},\mathscr{S}_{\gamma})$\\
         \hline
         3 & $\alpha$ & $\vec{\mathscr{P}}_{\alpha} = \frac{(1,2)}{3}(\mathscr{S}_{\alpha}-\mathscr{S}_{\beta}-\kappa)$ \\
          \hline
          \multirow{2}{*}{4} & $\alpha$ & $\vec{\mathscr{P}}_{\alpha} = \frac{(1,1)}{2}(\mathscr{S}_{\beta}-\mathscr{S}_{\alpha}+\kappa)$ \\
          & $\beta$ & $\vec{\mathscr{P}}_{\beta} = \frac{(1,1)}{2}(\mathscr{S}_{\beta}-\mathscr{S}_{\alpha}-\kappa)$ \\ \hline
    \end{tabular}
    \caption{Expressions for $\vec{\mathscr{P}}_{\OO}$ in terms of differences of $\SO$ and $\kappa$. For $\MO = 2,3$, the relations for $\OO \ne \alpha$ are obtained by translating each point in the above relations by a fixed vector. Note that $\PO$ is trivial for $\MO = 6$. See Ref.~\cite{zhang2022pol} for the derivation.}
    \label{tab:PvsS}
\end{table}

The quantity $\tilde{\ell}_{\OO} = c_-/12 + \ell_{\OO}$ has two contributions. The contribution proportional to $c_-$ arises due to the framing anomaly \cite{witten1989,Gromov2015} and needs to be added by hand because it does not appear in the TQFT-category theoretic derivation. The other contribution $\ell_{\OO}$ is defined mod $\MO/2$ if $\MO$ is even, and mod $\MO$ if $\MO$ is odd. $\ell_{\OO}$ can be extracted from the phase of the expectation value of a given ground state under a partial rotation operator restricted to an open region around $\OO$. This was explained in detail in Ref.~\cite{zhang2023complete}; we also provide a brief discussion in Sec.~\ref{sec:RSI}. 

Finally, although we do not have a direct understanding of $\vec{\mathscr{P}}_{s,\OO}$ and $\nu_s$, we can understand them as linear combinations of $\ell_{\OO}$ over different choices of $\OO$. $\vec{\mathscr{P}}_{s,\OO}$ is interpreted as an angular momentum polarization and is related to $\ell_{\OO}$ in much the same way as $\PO$ is related to $\SO$. Furthermore, $\nu_s$ is a $\Z_M$ invariant obtained by taking a weighted sum of partial rotation expectation values over the different high symmetry points $\OO$ in the unit cell. The TQFT suggests that it be understood as an overall measure of angular momentum per unit cell.

The derivation of Eq.~\eqref{eq:mainresponse} uses techniques from group cohomology and $G$-crossed braided tensor category theory, and is based on a general theory of invertible fermionic states with symmetry in (2+1) dimensions proposed in Ref.~\cite{barkeshli2021invertible} (together with the fCEP). The basic idea is that each invertible state with symmetry $G_f$ and bosonic symmetry group $G_b = G_f/\Z_2^f$ can be specified by a set of data $(c_-, n_1, n_2, \nu_3)$ with $n_1 \in \H^1(G_b,\Z_2),n_2 \in C^2(G_b,\Z_2),\nu_3 \in C^3(G_b,\U(1))$; these fix the properties of $G_f$ symmetry defects in the theory. Here $C^n(G_b,M)$ is the set of $n$-variable functions from $G_b$ to $M$, while $\H^1(G_b,\Z_2)$ is the group of functions $\alpha_1:G_b \rightarrow \Z_2$ satisfying $d \alpha_1 = 0$. A useful introduction to the group cohomology notation used here can be found in the appendices of Refs.~\cite{Chen2013,barkeshli2021invertible}.

The data are constrained by a set of consistency equations, whose derivation can be found in Ref.~\cite{barkeshli2021invertible}. In particular, we have an obstruction relation
\begin{equation}\label{eq:O4}
    d \nu_3 = \mathcal{O}_4[c_-,n_1,n_2]
\end{equation}
which is the constraint on $\nu_3$. The topological effective action $\mathcal{L}$ is then obtained by pulling back $\nu_3$ using $B$:
\begin{equation}
    \mathcal{L} = 2\pi B^* \nu_3.
\end{equation}
$[\mathcal{O}_4]$ is sometimes referred to as the 't Hooft anomaly for the invertible state. To have a well-defined effective action in (2+1)D, this obstruction needs to vanish. This constraint forces the response coefficients in the action to take specific quantized values.

In the special case of bosonic SPT states (which can be described by setting $c_-=0 ,n_1=0, n_2=0$), Eq.~\eqref{eq:O4} reduces to 
\begin{equation}
    d\nu_{3,\text{bSPT}} = 0.
\end{equation}
Thus we recover the known fact that bosonic SPTs are classified by representative cocycles of the group $\H^3(G_b,\U(1))$ (modulo coboundaries) \cite{Chen2013}. Since the obstruction is more complicated for invertible fermionic states, the response coefficients in the resulting effective action are quantized differently compared to those in bSPTs (generally in smaller fractions). Moreover, two fermionic states with the same $\mathcal{O}_4$ obstruction differ by a term of the form $\nu_{3,\text{bSPT}}$. We will see several examples of this below.

The parameters $n_1$ and $n_2$ specify intrinsically fermionic states. When $c_-$ is an integer as we have here, $n_1$ specifies whether unpaired Majorana zero modes can exist at symmetry defects. But in a system with $\U(1)^f$ charge conservation symmetry (and no particle-hole symmetry), this is not possible \cite{manjunath2022mzm}, therefore we set $n_1 = 0$ throughout. The parameter $n_2$ can be nonzero and affects the quantization of the response theory, as mentioned above.

There are some useful transformations of the response theory, which we briefly comment on. The above response theory was defined with respect to the operator $\Cmop$. The response theory defined with respect to $\Cmom$ can be obtained from Eq.~\eqref{eq:mainresponse} after relabelling $A \rightarrow A + \omega/2$. More generally, the response theory with respect to $\tilde{C}_{M_{\OO},\chi}^{+}$ can be obtained by taking $A \to A +\chi \omega $ for $\chi \in \frac{1}{2}\ZZ$. Note that under the above transformation, the coefficients $\SO, \ell_{\OO},\vec{\mathscr{P}}_{s,\OO}$ and $\nu_s$ all transform. Therefore, strictly speaking, we should use $+$ and $-$ superscripts for these coefficients to indicate their dependence on $\Cmop$ and $\Cmom$ respectively. However, in this paper we will primarily use the coefficients with superscript $+$, and therefore we drop the superscript notation. Whenever we refer to the coefficients with $-$ superscripts, we will make it explicit in the notation.

\subsection{General results for $\MO=1,2,3,4,6$}
In this section, we list some predictions of the topological action that apply to any origin with $\MO \ge 1$. The derivations are all contained in Appendix~\ref{app:DerTopAct}.

\begin{table}[t]
    \centering
    \begin{tabular}{c||c|l||l|l|l|l|l}
    
         $M$ & $s_{\OO}$& $\vec{t}_{\OO}$ & $k_1$&$k_{2,\OO}$ & $k_{3,\OO}$& $k_{4,\OO}$& $k_{5,\OO}$    \\ \hline
         1 & -& - & $\Z$ & - & -& -& -    \\ 
         2 & $\Z_2$& $\Z_2^2$ & $\Z$&$\Z_2$ & $\Z_2$& $\Z_2^2$& $\Z_2^2$    \\ 
         3 &-& - & $\Z$&$\Z_3$ & $\Z_3$& $\Z_3$& $\Z_3$   \\ 
         4 & $\Z_2$& $\Z_2$ & $\Z$&$\Z_4$ & $\Z_4$& $\Z_2$& $\Z_2$    \\ 
         6 & $\Z_2$& - & $\Z$&$\Z_6$ & $\Z_6$& -& -   \\ 
          
    \end{tabular}
    \caption{Quantization of the parameters which fix the response coefficients in Eqs.~\eqref{eq:FTresult_C1},~\eqref{eq:coeffs_C4},~\eqref{eq:coeffs_C2},~\eqref{eq:coeffs_C3},~\eqref{eq:coeffs_C6}.}
    \label{tab:FT_quantization}
\end{table}

There are three independent integer invariants which are well-defined even in the absence of point group rotation symmetry:
\begin{align}\label{eq:FTresult_C1}
    c_- &\in \Z  \nonumber \\
    C &= c_- + 8 k_1\nonumber \\
    \kappa &\in \Z.
\end{align}
$c_-$ is the chiral central charge of the invertible state; in general it can be an integer or a half-integer, but since we assume $\U(1)^f$ charge conservation symmetry, $c_-$ is forced to be an integer \cite{manjunath2022mzm}. $C$ is the Chern number and sets the Hall conductivity $\sigma_H = C \frac{e^2}{h}$. In non-interacting fermion systems of charge $\pm 1$, we have $C = c_-$, but in general they differ by a multiple of 8. The difference can be understood in terms of bosonic integer quantum Hall (BIQH) states, which have $C = 8 k_1,c_- = 0$ for some integer $k_1$, and are intrinsically interacting states.\footnote{From e.g. Ref.~\cite{senthil2013} we know that BIQH states have an even Hall conductance $2k_1$ in units of $e_b^2/h$ where $e_b$ is the elementary boson charge and $k_1$ is some integer. Since $e_b = 2 e$, the BIQH Hall conductance is actually $8 k_1$ in units of $e^2/h$.} The quantization of the parameters which specify the response coefficients is summarized in Table~\ref{tab:FT_quantization}.

The integer $\kappa$ is related to the $\U(1)$ filling, which we denote by $\nu$. If the system has flux $\phi$ per unit cell, we have $\nu = C \phi/2\pi + \kappa$, therefore $\kappa$ equals the filling at zero magnetic field. To measure $C$ and $\kappa$ in a Chern insulator defined on a spatial torus, we change $\phi$ by inserting additional magnetic flux and find the corresponding values of $\nu$ for which the system remains gapped. Then, $C$ and $\kappa$ are given by the slope and intercept of the plot of $\nu$ versus $\phi/2\pi$. This relation between $\nu,C,\kappa$ also holds for invertible states with interactions; this was argued using flux-insertion arguments in Ref.~\cite{Lu2017fillingenforced} and using TQFT arguments in Ref.~\cite{Manjunath2020fqh}. We present a short derivation based on TQFT in App.~\ref{app:phinonzero}.

The remaining field theory coefficients all depend on the point group symmetry. The basic quantization and physical interpretation of these coefficients is given in Table~\ref{tab:FT_summary}; below we explain how they fit into the overall classification of invertible states. An important prediction of the field theory is that the coefficients $\SO,\ell_{\OO}$ are partially fixed by the Chern number, for $M = 2,3,4,6$:
\begin{align}
    \SO &= \frac{C}{2} \mod 1, \nonumber \\
    \ell_{\OO} &= \frac{C}{4} \mod \gcd(M,2).
\end{align}
The quantization of $\SO$ was checked numerically in Hofstadter models for $M = 4$ in Ref.~\cite{zhang2022fractional}, and for $M=2,3,6$ in Ref.~\cite{zhang2022pol}, by explicitly computing the charge at lattice disclinations and the angular momentum of magnetic flux. The quantization of $\ell_{\OO}$ was checked only for $M=4$, in Ref.~\cite{zhang2023complete}, using a method based on partial rotation operators that we will discuss in Sec.~\ref{sec:RSI}.

As we will see below, a set of independent invariants can be constructed by taking suitable linear combinations of the response coefficients that appear in Eq.~\eqref{eq:mainresponse}. There are some common invariants for all $M$: choosing $\OO$ so that $\MO = M$, we define 
\begin{equation}
    \begin{split}
        I_1 &:= \SO- \ell_{\OO}-\frac{c_-}{4}\in 
        \ZZ_{\gcd(M,2) M} \\
        I_2 &:= \nu_s \in \Z_M. 
    \end{split}
\end{equation}
We also define 
\begin{align}
    \delta \ell_{\OO} &= \ell_{\OO}-\frac{c_-}{4}  \in 2\ZZ_{M}\cong \ZZ_{M/\gcd(M,2)} \nonumber \\
     \delta \PO &= 2(\PO - 2 \vec{\mathscr{P}}_{s,\OO}) \in \Z_M \times \Z_M;\,\, M=2,4 .
\end{align}
These can be trivial for specific $M$ and do not have a common structure. In terms of the above, we summarize a set of independent integer valued invariants (aside from $\kappa,C , (C-c_-)/8$) for $M=2,4,6$ in Table~\ref{tab:indep_invts}, where we have chosen $\OO = \alpha$ by convention. For $M=3$, the invariants are $I_1, I_2, \mathscr{P}_{\alpha,x}, \mathscr{P}_{s,\alpha,x}, \delta \ell_{\alpha}/2$, and each generates a $\ZZ_3$ subgroup of the classification. 
\begin{table}[h!]
    \centering
    \begin{equation*}
        \begin{array}{c||c|cc|c|c}
            M & I_1 & \delta \mathscr{P}_{\alpha;x} & \delta \mathscr{P}_{\alpha;y} & I_2 & \delta\ell_{\alpha} /2 \\ \hline
            2 & {\red 4} & {\red 4} & {\red 4} & 2 \\ \hline 
            4 & {\red 8} & {\red 4} & & 4 & 2\\\hline 
            6 &{\red 12} &  &  &6 &3 \\ \hline 
        \end{array}
    \end{equation*}
    \caption{List of independent generators of the classification. Each integer denotes the order of the cyclic group generated by the corresponding invariant. Missing entries in each row are either trivial or can be determined from the listed generators. The integers in {\red red} mean that the odd elements are intrinsically fermionic. }
    \label{tab:indep_invts}
\end{table}

\subsection{$M=4$}
We consider some $\OO$ for which $\MO = 4$ and define the response theory with respect to the operator $\Cmop$. (This is the same convention that was used in Refs.~\cite{zhang2022fractional,zhang2022pol}). In this case a group cohomology calculation, outlined in App.~\ref{app:DerTopAct}, gives the quantization conditions in two main steps. The first step is to determine a partial quantization of the coefficients, without including certain equivalences that arise in the computation. The second step is to include these equivalences. 

After the first step, we find that  
\begin{align}
    C &= c_- + 8 k_1, \nonumber \\
    \SO &= \frac{c_-}{2} + [s_{\OO}]_2 + 2 k_{2,\OO} \mod 8,  \nonumber\\
    \ell_{\OO} &= \frac{c_-}{4} + 2k_{3,\OO} \mod 8,  \nonumber\\
    \PO &= \frac{([t_{\OO}]_2 + 2 k_{4,\OO})}{2} (1,1) \mod (2\Z)^2, \nonumber\\
    \vec{\mathscr{P}}_{s,\OO} &= \frac{k_{5,\OO}}{2}(1,1) \mod \Z^2,  \nonumber\\
    \kappa &\in \ZZ \nonumber\\
    \nu_s &\in \ZZ_4. \label{eq:coeffs_C4}
\end{align}
See Table~\ref{tab:FT_quantization} for the quantization of $s_{\OO},\vec{t}_{\OO}$ and the integers $k_i$. The quantities $s_{\OO},t_{\OO}$ are `intrinsically fermionic' contributions which must be zero in any invertible bosonic state; they are related to the data $n_2$ introduced previously. Interestingly, these parameters also appear in the classification of topologically ordered states with charge conservation and crystalline symmetries, as pointed out in Refs.~\cite{manjunath2021cgt,Manjunath2020fqh}. In that context, $s_{\OO}$ and $\vec{t}_{\OO}$ are referred to as the `discrete spin vector' and the `discrete torsion vector' respectively; they specify the fractional angular and linear momentum of anyons in the theory. 

In the second step, we find two equivalences that reduce the classification. These correspond to relabelling disclination and dislocation defects by fermions; such relabellings can modify the coefficients but should be considered trivial, since fermions can be created by local operators \footnote{Another way to say this is that $\nu_3$, modulo the correct relations, encodes properties of `equivalence classes of defects' which each include a representative defect and the same defect with a fermion attached. Therefore, two defects that differ only by the quantum numbers of a fermion are in the same equivalence class.}. For $a,b \in \Z_2$, these relabellings imply that
\begin{align}\label{eq:relabel_C4}
    (k_{2,\OO},k_{3,\OO}) \simeq (k_{2,\OO} + 2a, k_{3,\OO} + 2a) \nonumber \\
    (k_{4,\OO}, k_{5,\OO}) \simeq (k_{4,\OO} + b, k_{5,\OO} + b)
\end{align}
where all other coefficients remain unchanged. After modding out by the equivalences, a complete and independent set of (integer quantized) invariants is given by 
\begin{align}\label{eq:FTresult_C4}
    & c_-,\kappa,k_1 \in \ZZ \nonumber \\
    I_1 &:= \SO- \ell_{\OO}-\frac{c_-}{4} \in \ZZ_8 \nonumber \\
    I_2 &:= \nu_s \in \ZZ_4 \nonumber\\
    I_3 &:= \frac{1}{2}\left(\ell_{\OO} - \frac{c_-}{4}\right) \in \ZZ_2  \nonumber \\
   I_4 &:= 2 (\PO - 2 \vec{\mathscr{P}}_{s,\OO}) \cdot (1,0) \in \ZZ_4 .
\end{align}
Thus the full classification is $\Z^3 \times (\Z_8 \times \Z_2) \times \Z_4 \times \Z_4$. If we set $c_-=0$ and $k_1 = 0$, we obtain the classification in Ref.~\cite{zhang2022realspace}.\footnote{The possibly confusing combination $\PO - 2 \vec{\mathscr{P}}_{s,\OO}$ appears because of the definition of the gauge fields: $\PO$ is associated with a $\U(1)^f$ gauge field $A$, while $\vec{\mathscr{P}}_{s,\OO}$ is associated with a bosonic gauge field $\omega$.} 

Note that $\SO$ does not give a $\Z_8$ generator of the classification unless we also know $\ell_{\OO}$, since after including equivalences it can only take 4 distinct values for a fixed $c_-$. Similarly, $\PO$ and $\vec{\mathscr{P}}_{s,\OO}$ are individually not generators of the classification.

\subsection{$M=2$}

We fix an origin $\OO$ with $\MO = 2$. In this case, the coefficients appearing in Eq.~\eqref{eq:mainresponse} can be parametrized as follows:
\begin{align}\label{eq:coeffs_C2}
    C &= c_- + 8 k_1,  \nonumber \\
    \SO &= \frac{c_-}{2} + [s_{\OO}]_2 + 2 k_{2,\OO} \mod 4,  \nonumber \\
    \ell_{\OO} &= \frac{c_-}{4} + 2k_{3,\OO} \mod 4,  \nonumber\\
    \PO &= \frac{([\vec{t}_{\OO}]_2 + 2 \vec{k}_{4,\OO})}{2},  \nonumber\\
    \vec{\mathscr{P}}_{s,\OO} &= \frac{\vec{k}_{5,\OO}}{2} \nonumber\\
    \kappa &\in \ZZ \nonumber\\
    \nu_s &\in \ZZ_2.
\end{align}

There are two equivalences that correspond to relabelling disclination and dislocation defects by fermions; such relabellings should be considered trivial as the fermion is a local operator. For $a,b_x,b_y \in \Z_2$, these relabellings change the coefficients as follows:
\begin{align}\label{eq:relabel_C2}
    (k_{2,\OO},k_{3,\OO}) \simeq (k_{2,\OO} + a, k_{3,\OO} + a) \nonumber \\
    (\vec{k}_{4,\OO}, \vec{k}_{5,\OO}) \simeq (\vec{k}_{4,\OO} + \vec{b}, \vec{k}_{5,\OO} + \vec{b})
\end{align}
where we assume all other coefficients remain unchanged. After modding out by the equivalences, a complete and independent set of invariants for a fixed $\OO$ is given by
\begin{align}\label{eq:FTresult_C2}
    & c_-,\kappa,k_1 \in \ZZ \nonumber \\
    I_1 &:= \SO - \ell_{\OO}-\frac{c_-}{4} \in \ZZ_4 \nonumber \\
    I_2 &:= \nu_s \in \ZZ_2 \nonumber\\
   (I_{3},I_{4}) &:= 2 (\PO - 2 \vec{\mathscr{P}}_{s,\OO}) \in \ZZ_4\times \Z_4.
\end{align}
Thus the full classification is $\Z^3 \times \Z_4^3 \times \Z_2$. The invariant associated to $\ell_{\OO}$ alone turns out to be trivial in this case, and does not appear above.

\subsection{$M=3$}

Here we can pick any high symmetry point $\OO$, since $\MO$ is always equal to 3. In this case the coefficients can be parametrized as follows:
\begin{align}\label{eq:coeffs_C3}
    C &= c_- + 8 k_1 \nonumber \\
    \SO &= \frac{c_-}{2} + 2 k_{2,\OO} \mod 3\nonumber \\
    \ell_{\OO} &= \frac{c_-}{4} + 2k_{3,\OO} \mod 3 \nonumber\\
    \PO &= \frac{k_{4,\OO}}{3} (1,2) \mod \Z^2 \nonumber\\
    \vec{\mathscr{P}}_{s,\OO} &= \frac{k_{5,\OO}}{3}(1,2) \mod \Z^2 \nonumber\\
    \kappa &\in \ZZ \nonumber\\
    \nu_s &\in \ZZ_3.
\end{align}

The $c_-$ dependence and mod 3 quantization of $\SO$ and $\ell_{\OO}$ for $M=3$ are somewhat subtle points which are explained in App.~\ref{app:cdep_M=3}. In this case there are no further equivalences, so a complete set of invariants is given by 
\begin{align}\label{eq:FTresult_C3}
    & c_-,\kappa,k_1 \in \ZZ \nonumber \\
    I_1 &:= \SO- \ell_{\OO} -\frac{c_-}{4} \nonumber \\ 
    I_2 &:= \nu_s \nonumber \\
    I_3 &:= \ell_{\OO} - \frac{c_-}{4} \nonumber \\
    I_{4} &:= 3\PO \cdot (1,0)\nonumber \\
    I_{5} &:= 3  \vec{\mathscr{P}}_{s,\OO} \cdot (1,0),
\end{align}
where $I_1, \dots , I_5$ are all $\Z_3$ valued. Thus the full classification is $\Z^3 \times \Z_3^5$.

\subsection{$M=6$}

There is only one high symmetry point with $M=6$, which we denote as $\alpha$ in Fig.~\ref{fig:wyckoff}. Therefore we must pick $\OO = \alpha$. In this case the coefficients can be parametrized as follows:
\begin{align}\label{eq:coeffs_C6}
    C &= c_- + 8 k_1 \nonumber \\
    \SO &= \frac{c_-}{2} + [s_{\OO}]_2 + 2 k_{2,\OO} \mod 12 \nonumber \\
    \ell_{\OO} &= \frac{c_-}{4} + 2 k_{3,\OO} \mod 12 \nonumber\\
    \kappa &\in \ZZ \nonumber\\
    \nu_s &\in \ZZ_6.
\end{align}

In this case we have the further equivalence 
\begin{equation}\label{eq:relabel_C6}
    (k_{2,\OO}, k_{3,\OO}) \simeq (k_{2,\OO} + 3 a, k_{3,\OO} + 3 a), \quad a \in \Z_2.
\end{equation}
A complete set of invariants is given by
\begin{align}\label{eq:FTresult_C6}
    & c_-,\kappa,k_1 \in \ZZ \nonumber \\
    I_1 &:= \SO-\ell_{\OO}-\frac{c_-}{4}  \in \Z_{12} \nonumber \\ 
     I_2 &:= \nu_s \in \ZZ_6\nonumber \\
    I_3 &:= \frac{1}{2}\left(\ell_{\OO}-\frac{c_-}{4}\right) \in \Z_3.
\end{align}
Thus the full classification is $\Z^3 \times \Z_{12} \times \Z_3 \times \Z_6$.

\section{Classification using real-space invariants}\label{sec:RSI}

In this section, we consider interacting invertible states with the same symmetry as in the previous section, but present a completely different approach to classify and characterize them, based on real-space invariants. We first assume $c_- = C = 0$; in this limit we can explicitly construct a fixed-point wavefunction for each invertible phase that appears in the TQFT, but in terms of a set of real-space invariants $\{n_{\OO},m_{\OO}\}$ defined for each high symmetry point $\OO$. Following this, we introduce a set of real-space invariants $\Theta^{\pm}_{\OO}$ for each $\OO$, which generalize $\{n_{\OO},m_{\OO}\}$ and are well-defined even when $c_-,C \ne 0$. After discussing the basic quantization and physical meaning of $\Theta^{\pm}_{\OO}$, we relate them to the field theory coefficients of Sec.~\ref{sec:FT}. We explain how, along with $c_-, C, \kappa$, they fully generate the classification given in Sec.~\ref{sec:FT}.

\subsection{Real-space construction when $c_- = 0=C$}

Throughout this section, we set $c_- = 0 = C$. All the details in this section have appeared previously in Ref.~\cite{zhang2022realspace}, and are stated here for completeness. Corresponding to each invertible phase, we construct a representative state in which the degrees of freedom are exponentially localized at the high symmetry points of the unit cell (see Fig.~\ref{fig:wyckoff}). In this limit, the topological invariants can be identified straightforwardly.

\begin{figure}
    \centering
    \includegraphics[width=0.4\textwidth]{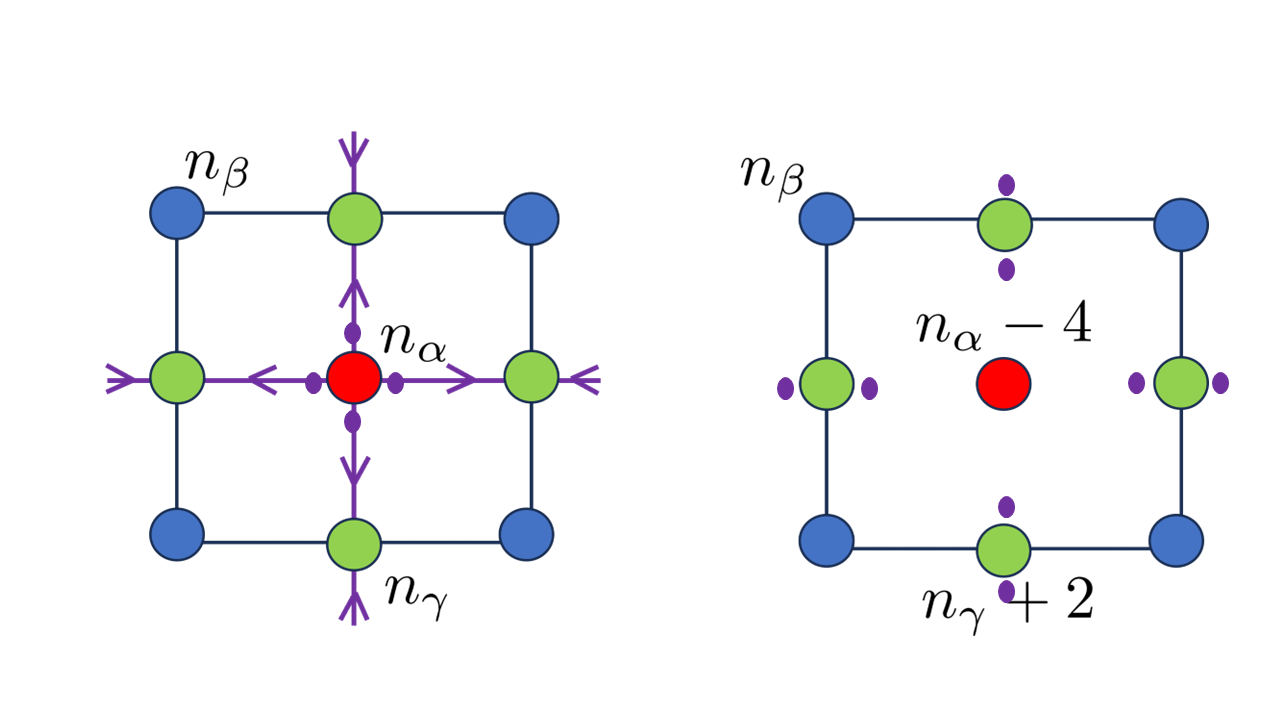}
    \caption{Equivalence for the real-space $c_- = 0=C$ invariants on a square lattice. Each purple dot represents one filled fermion orbital moved from $\alpha$ to $\gamma$ as indicated by arrows. Since the process shifts an entire orbit from $\alpha$ to $\gamma$, both $\{n_{\OO}\}$ and $\{m_{\OO}\}$ change.}
    \label{fig:movement}
\end{figure}

\subsubsection{$M=4$}

We begin with the square lattice. The construction proceeds by placing at each high symmetry point $\OO$ a charge under the subgroup that leaves $\OO$ fixed. In our case this is always $\U(1)^f \times \Z_{\MO}$. Accordingly, we place a $\U(1)^f$ charge $n_{\OO} \in \ZZ$ at $\OO=\alpha,\beta,\gamma$ as well as a $\ZZ_4$ charge $m_{\OO}$ at $\OO=\alpha,\beta$, and a $\Z_2$ charge $m_{\gamma}$ at $\gamma$. By `$\Z_4$ charge of $m_{\alpha}$ around $\alpha$' we specifically mean a set of filled orbitals localized at $\alpha$ that together transform under $\tilde{C}_{M_{\alpha}}$ as a nontrivial representation of $\Z_4$ with total eigenvalue $e^{i \frac{2\pi}{M_{\alpha}} m_{\alpha}}$.

Each tuple $(n_{\alpha},n_{\beta},n_{\gamma}|m_{\alpha},m_{\beta},m_{\gamma})$ characterizes an invertible state. But different tuples can be equivalent. In particular, as shown in Fig.~\ref{fig:movement}, consider a set of four symmetric filled orbitals localized at $\alpha$ which transform into each other under $\tilde{C}_{M_{\alpha}}$. We symmetrically shift these orbitals in space so that they end up localized around $\gamma$. The $\U(1)^f$ charge at each $\alpha$ point thus reduces by 4, while the total $\U(1)^f$ charge at each $\gamma$ point increases by 2. Moreover, the original orbitals at $\alpha$ form a closed orbit under rotations about their common origin, with eigenvalues $\pm 1, \pm i$ (the fourth roots of unity). Thus, moving them to $\gamma$ changes the total rotation eigenvalue at $\alpha$ by the product of these phases, which is -1; equivalently, $m_{\alpha} \rightarrow m_{\alpha} + 2 \mod 4$. Similarly, the new orbitals that are localized at each $\gamma$ point form a closed orbit with rotation eigenvalues $+1$ and -1, and $m_{\alpha}$ changes by the product of these eigenvalues, which is -1. Therefore $m_{\gamma} \rightarrow m_{\gamma} + 1 \mod 2$. 

From this, we have the equivalence
\begin{align}
 & (n_{\alpha},n_{\beta},n_{\gamma}|m_{\alpha},m_{\beta},m_{\gamma}) \nonumber \\ \simeq &(n_{\alpha}-4,n_{\beta},n_{\gamma}+2|m_{\alpha}+2,m_{\beta},m_{\gamma}+1)  \nonumber \\
 \implies &(0,0,0|0,0,0) \simeq (-4,0,2|2,0,1).
\end{align}

By analogously moving fermions between $\beta$ and $\gamma$, we get a second equivalence 
\begin{align}
    (0,0,0|0,0,0) \simeq (0,4,-2|0,2,1).
\end{align}
We then mod out the different configurations by the above equivalences. We find that a convenient basis for the invariant quantities in this problem is
\begin{align}
    n_{\alpha}+n_{\beta}+ 2n_{\gamma} &\in \Z \nonumber \\
    n_{\alpha} + 2 m_{\alpha} &\in \Z_8 \nonumber \\
    m_{\alpha} &\in \Z_2 \nonumber \\
    n_{\gamma} + 2 m_{\gamma} &\in \Z_4 \nonumber \\
    m_{\alpha}+m_{\beta}+ 2m_{\gamma}&\in \Z_4.
\end{align}
The relationship between these quantities and the response coefficients in the above limit (which we will call the atomic insulator or AI limit) is the following:
\begin{align}
   \kappa &=n_{\alpha}+n_{\beta}+ 2n_{\gamma}  \nonumber \\
   \nu_{s,\AI} &= m_{\alpha}+m_{\beta}+ 2m_{\gamma} \mod 4 \nonumber \\
   \mathscr{S}_{\OO,\AI} &= n_{\OO} \mod \MO \nonumber\\
   \vec{\mathscr{P}}_{\alpha,\AI} &= \frac{n_{\beta}+n_{\gamma}}{2}(1,1) \nonumber \\
   \vec{\mathscr{P}}_{\beta,\AI} &= \frac{n_{\alpha}+n_{\gamma}}{2}(1,1) \nonumber \\
   \vec{\mathscr{P}}_{s,\alpha,\AI} &= \frac{m_{\beta}+m_{\gamma}}{2}(1,1) \nonumber \\
   \vec{\mathscr{P}}_{s,\beta,\AI} &= \frac{m_{\alpha}+m_{\gamma}}{2}(1,1) \nonumber \\
   \ell_{\OO,\AI} &= 2m_{\OO} \mod \MO.
\end{align}
Let us explain these relations.  Intuitively, the equation for $\kappa$ holds because the filling is a weighted sum of $n_{\OO}$ at each high symmetry point. The equation for $\SO$ holds because the excess fractional charge at a disclination of angle $2\pi/\MO$ centred at $\OO$ is $\SO/\MO \mod 1$ from field theory, but should also equal $n_{\OO}/\MO$ if we consider the same disclination in the real space construction. $\PO$ is related to the dipole moment of $n_{\OO}$ within the unit cell, and Ref.~\cite{zhang2022pol} shows how to compute this dipole moment for general $M$. These relations for $\SO, \PO$ have also been extensively checked in Refs.~\cite{zhang2022fractional,zhang2022pol}. 

Finally, the relations for $\vec{\mathscr{P}}_{s,\OO}$,$\ell_{\OO}$ and $\nu_s$ have not been checked numerically in prior work but we are confident in these expressions, for the following reasons. $\vec{\mathscr{P}}_{s,\OO}$ is a polarization of angular momentum with the same quantization as $\PO$, therefore it is reasonable that the expression for $\vec{\mathscr{P}}_{s,\OO}$ should resemble that of $\PO$, with the $\U(1)^f$ charge replaced by a $\Z_{\MO}$ charge. Furthermore, in a fermion SPT state, the SPT invariant, which is normalized as $\ell_{\OO}/2$ in field theory, should give the angular momentum at the rotation center $\OO$, as we know from real-space constructions \cite{cheng2018rotation,Rasmussen2020HOSPT,zhang2022realspace}. But this is simply $m_{\OO}$. Finally, the expression for $\nu_s$ should depend only on $m_{\OO}$, should be defined mod 4, and should be invariant under a shift of origin. The expression $m_{\alpha} + m_{\beta} + 2m_{\gamma}$ is the simplest such expression, and is identical to the expression for $\kappa$ if we replace $m_{\OO}$ with $n_{\OO}$. Therefore if we interpret $\nu_s$ as the angular momentum per unit cell, this is the most natural expression for it. 

Note, finally, that the above construction alone does not guarantee that the real-space classification is complete, however we have confidence in its completeness because it reproduces the predictions of TQFT after we set $C = 0 = c_-$. 

\subsubsection{$M=2$}

We use the p2 unit cell as shown in Fig.~\ref{fig:wyckoff}. The real space construction proceeds by placing $\U(1)$ charge $n_i \in \ZZ$ as well as a $\ZZ_2$ charge $m_i$ at 
$i=\alpha,\beta,\gamma,\delta$. Each invertible phase is characterized by a set $(n_{\alpha},n_{\beta},n_{\gamma},n_{\delta}|m_{\alpha},m_{\beta},m_{\gamma},m_{\delta})$. But we can symmetrically move a pair of fermion orbitals between any two high symmetry points, and relate the different allowed configurations. From an argument similar to the $M=4$ case above, this leads to the equivalences 
\begin{align}
    &(0,0,0,0|0,0,0,0) \simeq (-2,2,0,0|1,1,0,0) \nonumber \\ \simeq &(-2,0,2,0|1,0,1,0) \simeq (-2,0,0,2|1,0,0,1).
\end{align}
A complete set of invariants can be identified as
\begin{align}
    \kappa &:= n_{\alpha}+n_{\beta}+n_{\gamma}+n_{\delta}  \in \Z \nonumber \\
    \nu_{s,\AI} &= m_{\alpha}+m_{\beta}+m_{\gamma}+m_{\delta}  \in \Z_2 \nonumber \\
    \mathscr{S}_{\OO,\AI}- \ell_{\OO,\AI} &:= n_{\OO} + 2 m_{\OO}  \in \Z_4, \quad \OO = \alpha,\beta,\gamma.
\end{align}
The other response coefficients can be written as follows (we only show the relations for $\OO = \alpha$, but the other relations can be obtained by a constant overall shift of origin):
\begin{align}
   \vec{\mathscr{P}}_{\alpha,\AI} &= \frac{1}{2}(n_{\beta}+n_{\gamma},n_{\beta}+n_{\delta}) \nonumber \\
   \vec{\mathscr{P}}_{s,\alpha,\AI} &= \frac{1}{2}(m_{\beta}+m_{\gamma},m_{\beta}+m_{\delta}) \nonumber \\
   \ell_{\alpha,\AI} &= 2m_{\alpha} \mod 2.
\end{align}

\subsubsection{$M=3$}

The real space construction proceeds by placing $\U(1)$ charge $n_{\OO} \in \ZZ$ and $\ZZ_3$ charge $m_{\OO}$ at $\OO=\alpha,\beta,\gamma$. Each invertible phase is characterized by a set $(n_{\alpha},n_{\beta},n_{\gamma}|m_{\alpha},m_{\beta},m_{\gamma})$. But we can symmetrically move fermions from $\alpha$ to $\beta$, $\beta$ to $\gamma$, and so on and adiabatically relate different allowed configurations. This leads to the equivalences 
\begin{align}
    (0,0,0|0,0,0) \simeq (3,-3,0|0,0,0) \simeq (3,0,-3|0,0,0).
\end{align}
This calculation differs slightly from the case with $M$ even. If we move 3 fermions from, say, $\alpha$ to $\beta$, they form a closed orbit under rotations with eigenvalues $1,e^{i 2\pi/3}, e^{i 4\pi/3}$ whose product is 1. Therefore the value of $m_{\OO}$ does not change under the reconfiguration.

A convenient basis for the invariant quantities in this problem is
\begin{align}
    \kappa &= n_{\alpha} + n_{\beta} + n_{\gamma} &\in \Z \nonumber \\
    \mathscr{S}_{\OO,\AI} &= n_{\OO} \mod 3 &\in \Z_3 \nonumber \\
    \ell_{\OO,\AI} &= 2 m_{\OO} \mod 3 &\in \Z_3 \nonumber \\
    3\mathscr{P}_{\alpha,x,\AI} &= n_{\beta} + 2 n_{\gamma} &\in \Z_3 \nonumber \\
    3\mathscr{P}_{s,\alpha,x,\AI} &= m_{\beta} + 2 m_{\gamma} &\in \Z_3 \nonumber \\
    \nu_{s,\AI} &=  m_{\alpha} + m_{\beta} + m_{\gamma}  &\in \Z_3.
\end{align}
\subsubsection{$M=6$}

Here the construction proceeds by placing $\U(1)$ charge $n_{\OO} \in \ZZ$ for $\OO = a,b,c$, $\ZZ_6$ charge $m_{\alpha}$ at $\alpha$, a $\ZZ_3$ charge $m_{\beta}$ at $\beta$, and a $\ZZ_2$ charge $m_{\gamma}$ at $\gamma$. Each invertible phase is characterized by a set $(n_{\alpha},n_{\beta},n_{\gamma}|m_{\alpha},m_{\beta},m_{\gamma})$. But we can symmetrically move fermions from $\alpha$ to $\beta$, $\beta$ to $\gamma$, and so on and relate different allowed configurations. This leads to the equivalences 
\begin{align}
    (0,0,0,0,0,0) \simeq (-6,3,0,3,0,0) \simeq (-6,0,2,3,0,1).
\end{align}
A convenient basis for the invariant quantities in this problem is
\begin{align}
    \kappa &= n_{\alpha} + 2 n_{\beta} + 3 n_{\gamma} &\in \Z \nonumber \\
    \mathscr{S}_{\alpha,\AI} - \ell_{\alpha,\AI} &= n_{\alpha} - 2 m_{\alpha} &\in \Z_{12}\nonumber \\
    \ell_{\alpha,\AI} &= m_{\alpha} &\in \Z_3 \nonumber \\
    \nu_{s,\AI} &= m_{\alpha} + 2 m_{\beta} + 3 m_{\gamma} &\in \Z_6.
\end{align}

\subsection{Real-space invariants for $c_- \ne 0$}

From the previous section, it is natural to argue that if we allow for general $c_-$ and $C$, the classification as a group should simply acquire two factors of $\Z$ in addition to those obtained above. However, this is not the whole story, because the exponentially localized Wannier limit does not exist when $c_- \ne 0$, and therefore we first need to give an alternative definition of the real-space invariants that remains valid in this case. 

In this section we address this issue and define a pair of invariants $\Theta^+_{\OO},\Theta^-_{\OO}$ for each high symmetry point $\OO$ of the unit cell. The main result of this section is that these invariants are well-defined for general $c_-$ and $C$, and that the classification of invertible states can be fully captured if $\Theta^+_{\OO},\Theta^-_{\OO}$ are known for each $\OO$. We also derive a set of relations between $\Theta^+_{\OO},\Theta^-_{\OO}$ and the response coefficients discussed in Sec.~\ref{sec:FT}. 

\subsubsection{Definition of invariants}
Consider an invertible state $\ket{\Psi}$ on a torus or an open disk with symmetry operators $\Cmopm$ about an origin $\OO$, as defined in Sec.~\ref{sec:Defs}. We define $\Cmopm|_D$ to be the restriction of $\Cmopm$ to some symmetric open region $D$ centered at $\OO$. The results in this section appeared previously in Ref.~\cite{zhang2023complete}.

We define
\begin{align}\label{eq:K_def}
    \bra{\Psi} \Cmopm|_D \ket{\Psi} &= e^{- \gamma |\partial D| + i \frac{2\pi}{\MO} K^{\pm}_{\OO}} (1 + O(e^{- \epsilon |\partial D|})).
\end{align}
$\gamma$ sets the amplitude of the expectation value, while $\epsilon$ is some positive number that captures subleading contributions. Numerically we have found for $M=4$ \cite{zhang2023complete} that this equation is obeyed with a quantized value of $K^{\pm}_{\OO}$, and moreover the quantities
\begin{align}\label{eq:Theta_def_K}       
    \Theta^+_{\OO} &:= \begin{cases} K^+_{\OO} \mod \frac{\MO}{2} & \MO \text{ even}\\ K^+_{\OO} \mod \MO & \MO \text{ odd}\end{cases}, \nonumber \\
    \Theta^-_{\OO} &:=  \begin{cases} K^-_{\OO} \mod \MO & \MO \text{ even}\\ K^-_{\OO} \mod \frac{\MO}{2} & \MO \text{ odd}\end{cases}
\end{align}
are invariants, that is, they remain constant within a given invertible phase. The reason $\Theta^{\pm}_{\OO}$ are defined by taking a modular reduction of $K^{\pm}_{\OO}$ is that the values of $K_{\OO}^{+}$ for $\MO$ even, and $K^-_{\OO}$ for $\MO$ odd, can jump by multiples of $\MO/2$ for a fixed $\ket{\Psi}$ as the size of $D$ is changed. This behavior was seen numerically in Ref.~\cite{zhang2023complete} and is explained in App.~\ref{app:rel-Theta-K}.

We do not have a mathematical proof that $\Theta^{\pm}_{\OO}$ is indeed a quantized many-body invariant of invertible states. However, we can justify Eq.~\eqref{eq:Theta_def_K} and also show that $\Theta^{\pm}_{\OO}$ is quantized for different $\MO$, if we make the assumption that the density matrix $\rho_D$ of the system restricted to $D$ equals $\rho_{\text{CFT}}$, the density matrix of the conformal field theory that lives on the edge of $D$. This assumption has been discussed in the literature of topological phases previously \cite{Haldane2008entanglement,Qi2012entanglement}. The calculation that justifies the quantization was done in Ref.~\cite{zhang2023complete}; it uses conformal field theory and is also briefly outlined around Eq.~\eqref{eq:ells_Theta}. Moreover, from the previous section we know that when $c_-=0=C$, the arguments of such partial rotation expectation values are expected to be invariants because they simply measure the $\U(1)^f$ and $\Z_{\MO}$ charges localized at $\OO$, and these are quantized by construction. 

In contrast to the invariants in the previous section, $\Theta^{\pm}_{\OO}$ are defined when $c_-$ and $C$ are both nonzero. For a given $c_-$, the possible values of $\Theta^+_{\OO}$ always differ by integers, while the possible values of $\Theta^-_{\OO}$ always differ by half-integers. Note that when $\MO$ is even, $\Theta^+_{\OO}$ is defined mod $\MO/2$ and can take $\MO/2$ different values, while $\Theta^-_{\OO}$ is defined mod $\MO$ and can take $2\MO$ different values. On the other hand, when $\MO$ is odd, $\Theta^+_{\OO}$ is defined mod $\MO$ while $\Theta^-_{\OO}$ is defined mod $\MO/2$; both can take $\MO$ different values in this case. 

More generally, if we consider the operators $\tilde{C}^{\pm}_{\MO,\chi}$, we can write $ \bra{\Psi} \tilde{C}^{\pm}_{\MO,\chi}|_D \ket{\Psi} \simeq e^{- \gamma |\partial D| +i \frac{2\pi}{\MO} l^\pm_{D,\OO,\chi}}$ to leading order, and we can show that 
\begin{equation}\label{eq:angular_momentum}
    l^\pm_{D,\OO,\chi}=\frac{C\chi^2}{2}+\mathscr{S}^{\pm}_{\OO}\chi + K^{\pm}_{\OO} \mod \MO.
\end{equation}
This equation was derived analytically, and also verified numerically when $M = 2,4$, in Ref.~\cite{zhang2023complete}. It shows how $\Theta^{\pm}_{\OO}$ (which equals $l^{\pm}_{D,\OO,0}$ up to modular reductions) transforms under redefining the rotation operators, and can be derived in two steps. The first step is to relate $\Theta^{\pm}_{\OO}$ to the TQFT coefficient $\ell_{\OO}$, using Eq.~\eqref{eq:ells_Theta}. This is itself an involved computation which we explain when we introduce that equation. Next, as we stated in Sec.~\ref{sec:FT}, we know how the full TQFT transforms under the above redefinition of rotation operators. Therefore we can calculate the transformation of $\ell_{\OO}$ from field theory, and from this recover the transformation of $\Theta^{\pm}_{\OO}$.

It is possible to write a much more explicit form for $\Theta^{\pm}_{\OO}$ in two separate limits. First, when $c_-=0 = C$, $\Theta^{\pm}_{\OO}$ can be related to the parameters $\{n_{\OO},m_{\OO}\}$ of an exponentially localized Wannier limit of the invertible state, as introduced above. Next, for each $c_- > 0$, $\Theta^{\pm}_{\OO}$ can be analytically computed in the Landau level limit for a $c_- = 1$ integer quantum Hall state.
We now explain how to use these two limits to write an expression for $\Theta^{\pm}_{\OO}$ in an arbitrary invertible state.

\subsubsection{$c_-=0 = C$: Wannier limit}

Assuming $c_- = 0 = C$, let $n_{\OO}$ be the $\U(1)^f$ charge at $\OO$, and $m_{\OO} \mod \MO$ be the charge at $\OO$ under the operator $\Cmop$ which generates the subgroup $\Z_{\MO}$. Since the real space construction in this case pertains to atomic insulators, we use the subscript $\AI$ to refer to the corresponding invariants.  Then, we define
\begin{align}\label{eq:Theta_def_c=0}
    \Theta^+_{\OO,\AI} &:= \begin{cases} m_{\OO} \mod \frac{\MO}{2} & \MO \text{ even}\\
    m_{\OO} \mod \MO & \MO \text{ odd} \end{cases} \nonumber \\
    \Theta^-_{\OO,\AI} &:= \begin{cases}
        \frac{n_{\OO}}{2} + m_{\OO} \mod \MO & \MO \text{ even}\\
        \frac{n_{\OO}}{2} + m_{\OO} \mod \frac{\MO}{2} & \MO \text{ odd}.
    \end{cases}
\end{align}
Note that $\Theta^+_{\OO,\AI}$ and $2\Theta^-_{\OO,\AI}$ are quantized to integer values. This Wannier limit representation of $\Theta^{\pm}_{\OO,\AI}$ is valid for arbitrary $\phi$.

\subsubsection{Landau level limit}

In order to define $\Theta^{\pm}_{\OO}$ for all invertible states, we need to define them for at least one state with $c_- \ne 0$. Below we give their definition for the $c_- = 1$ integer quantum Hall state, where we only consider the wallpaper subgroup of the continuous translational and rotational symmetry group. 

First assume $c_- > 0$. Integer quantum Hall states in the continuum (that is, with continuous translation and rotation symmetries) have their $c_-$ lowest Landau levels filled by fermions of charge +1. In the TQFT picture, a Chern-Simons theory calculation (see App.~\ref{app:RSI}) shows that these states have the following nonzero response coefficients:
\begin{align}\label{eq:invts_LL}
C = c_- ; \quad
\mathscr{S}_{\text{LL}} = \frac{c_-^2}{2}; \quad
\ell_{\text{LL}} = \frac{4 c_-^3 - c_-}{12}.
\end{align}
By considering the time-reversed partners of the above states, we can also obtain states with $c_-<0$ that satisfy the same formulas as in Eq.~\eqref{eq:invts_LL}.

In the continuum, these invariants are well-defined as rational numbers, with no modular reduction and no origin dependence. If the system instead has a crystalline symmetry $G_{\text{space}}$, the last two invariants in general acquire a dependence on the chosen orign $\OO$, and also get reduced mod $\MO$. Moreover, from Sec.~\ref{sec:FT} we see that we need to consider the additional invariants $\kappa, \PO, \vec{\mathscr{P}}_{s,\OO},\nu_s$. However, even in the crystalline setting the TQFT classification predicts a specific state with $c_- = 1$, with the invariants
\begin{align}
C &= c_- = 1; \quad 
\mathscr{S}_{\OO,\text{LLL}} = \frac{1}{2} \mod \MO ; \nonumber \\
\ell_{\OO,\text{LLL}} &= \frac{1}{4} \mod \MO,
\end{align}
where additionally $\kappa, \PO, \vec{\mathscr{P}}_{s,\OO},\nu_s$ all vanish. 
We refer to this state as the `lowest Landau level limit', or LLL limit, because its topological invariants are identical to those of the $c_- = 1$ IQH state. It also satisfies the relation $\nu = C \phi/2\pi$, which can be shown from TQFT. The LLL limit is completely well-defined assuming the TQFT classification from Sec.~\ref{sec:FT} is complete. Note that a concrete realization of this limit exists in the region with $\phi \rightarrow 0^+, \nu \rightarrow 0, C=1$ of Hofstadter models defined on the appropriate crystalline lattices.  

Importantly, in this limit we can analytically show (subject to the assumption $\rho_D = \rho_{\text{CFT}}$) that 
\begin{align}\label{eq:Theta_def_LL}
    \Theta^+_{\OO,\text{LLL}} &= \begin{cases}
        \frac{4-\MO^2}{24} \mod \frac{\MO}{2}, & \MO \text{ even} \\
        \frac{1-\MO^2}{24} \mod \MO, & \MO \text{ odd}
    \end{cases} \\
    \Theta^-_{\OO,\text{LLL}} &= \begin{cases}
        \frac{10-\MO^2}{24} \mod \MO, & \MO \text{ even} \\
         \frac{13-\MO^2}{24} \mod \frac{\MO}{2}, & \MO \text{ odd}.
    \end{cases}
\end{align}
There are two main steps. The first step again involves showing Eq.~\eqref{eq:ells_Theta}, the general relation between $\Theta^{\pm}_{\OO}$ and $\ell^{\pm}_{\OO}$, which we discuss further below. The second step is to compute $\ell^{\pm}_{\OO}$ for a system of $c_-$ filled LLs in the continuum; this can be done using Chern-Simons theory \cite{Wen1992shift}, as we show in Appendix~\ref{app:RSI}. For convenience, we summarize the relevant values of $\Theta_{\OO,\text{LLL}}^{\pm}$ in Table ~\ref{tab:Theta_relevant}.

\subsubsection{General formula}
First note that any invertible state which is adiabatically connected to a tensor product of an atomic insulator and $c_-$ copies of the LLL state must satisfy the linearity relation
\begin{equation}\label{eq:Theta_def_gen}
    \Theta^{\pm}_{\OO} = c_-\Theta^{\pm}_{\OO,\text{LLL}} + \Theta^{\pm}_{\OO,\AI}.
\end{equation}
In Eq.~\eqref{eq:Theta_def_gen}, the parameters $\{n_{\OO},m_{\OO}\}$ specify how a given invertible state differs from $c_-$ copies of the LLL limit. In the TQFT classification this accounts for all invertible states with $C = c_-$.

The TQFT classification includes one additional integer invariant $k_1 = (C - c_-)/8$, which also describes the bosonic integer quantum Hall (BIQH) states. The derivation of $\Theta^{\pm}_{\OO}$ from conformal field theory \cite{zhang2023complete} shows that they are however independent of $k_1$. Therefore, assuming the TQFT classification is complete, we conclude that Eq.~\eqref{eq:Theta_def_gen} gives the most general quantization of $\Theta^{\pm}_{\OO}$ for invertible fermionic states in (2+1) dimensions.

\subsection{Relation between $\Theta^{\pm}_{\OO}$ and response coefficients}\label{sec:rel-Theta-resp}
We now state several relations between $\Theta^{\pm}_{\OO}$ and the response coefficients discussed in Sec.~\ref{sec:FT}. The derivations will be given in App.~\ref{app:RSI}. In most cases, the derivation is simply to check the relation in the $c_- = 0$ limit and in the LLL limit, and then use the linearity of the field theory parameters as well as the real-space invariants under stacking.

\begin{table}[t]
    \centering
    \begin{tabular}{c|c|c}
    \hline 
    \multicolumn{3}{c}{$\ell_{\OO}^{\pm} = 2 \Theta^{\pm}_{\OO} + t_{\pm} c_- \mod \MO$ } \\ \hline
       $\MO$  &  $t_+$ & $t_-$  \\ \hline
        even & $\frac{\MO^2-1}{12}$& $\frac{\MO^2+2}{12}$\\
        odd & $\frac{\MO^2+2}{12}$ & $\frac{\MO^2-1}{12}$\\
        \hline
    \end{tabular}
    \caption{Relation between the real space invariants $\Theta^{\pm}_{\OO}$ and the field theory parameters $\ell_{\OO}^{\pm}$, derived from conformal field theory. The quantity we usually write as $\ell_{\OO}$ is referred to as $\ell^+_{\OO}$ here.}
    \label{tab:Theta_vs_ells}
\end{table}

\begin{table}[t]
    \centering
    \begin{tabular}{c|c|c|c|c}
    \hline 
    $M_{\OO}$ & 2 & 4 & 3 & 6\\ \hline
    $\Theta_{\OO,\text{LLL}}^{+}$& 0 & $-\frac{1}{2}$ & $-\frac{1}{3}$ & $-\frac{4}{3}$\\
    $\Theta_{\OO,\text{LLL}}^{-}$ & $+\frac{1}{4}$ & $-\frac{1}{4}$ & $+\frac{1}{6}$&  $ -\frac{13}{12}$\\
        \hline
    \end{tabular}
    \caption{Value of the real space invariants for the lowest Landau level, $\Theta_{\OO,\text{LLL}}^{\pm}$, relevant to two dimensional crystalline systems. }
    \label{tab:Theta_relevant}
\end{table}

The most important relation is
\begin{equation}\label{eq:ells_Theta}
    \ell_{\OO}^{\pm} = 2 \Theta^{\pm}_{\OO} + t_{\pm} c_- \mod \MO,
\end{equation}
where $t_{\pm}$ depends on $\MO$ and is given in Table~\ref{tab:Theta_vs_ells}. This relation can be derived using a conformal field theory argument that was previously given in Ref.~\cite{zhang2023complete}. 

To derive this we start from the definitions, Eqs.~\eqref{eq:K_def} and~\eqref{eq:Theta_def_K}.
The expectation value of $\Cmopm|_D$ can be approximated by making the assumption that $\rho_D$, the density matrix of the state $\ket{\Psi}$ restricted to $D$, equals $\rho_{\text{CFT}}$, the density matrix of the conformal field theory that lives on the edge of $D$:
\begin{equation}
    \rho_D = \rho_{\text{CFT}}.
\end{equation}
This relationship was originally conjectured by Li and Haldane \cite{Haldane2008entanglement} for fractional quantum Hall states, and argued to hold more generally for gapped topological phases in Ref.~\cite{Qi2012entanglement}. Using this relation, a conformal field theory calculation performed in Ref.~\cite{zhang2023complete} shows that
\begin{align}\label{eq:CFTresult}
    \bra{\Psi} \Cmopm|_D \ket{\Psi} & \propto e^{-\frac{2\pi i c_-}{24} t_{\pm}} \times \mathcal{I}_{\MO}^{\pm}, 
\end{align}
for $t_{\pm}$ as defined in Table~\ref{tab:Theta_vs_ells}. The quantity $\mathcal{I}_{\MO}^{\pm}$ is directly related to the $G$-crossed modular $T$ matrix of the invertible state; it appears because the partial rotation expectation value equals the trace of the rotation operator acting in the CFT over the topological defect sectors of the CFT/TQFT, which in turn equals (to leading order) an expression involving only the $G$-crossed modular $T$ matrices. By evaluating $\mathcal{I}_{\MO}^{\pm}$ using the TQFT data, we additionally find that
\begin{equation}\label{eq:IM_vs_ells}
    \mathcal{I}_{\MO}^{\pm} = e^{\frac{\pi i}{\MO} \ell^{\pm}_{\OO}}.
\end{equation}
Upon combining Eqs.~\eqref{eq:Theta_def_K},~\eqref{eq:CFTresult} and~\eqref{eq:IM_vs_ells}, we obtain Eq.~\eqref{eq:ells_Theta}. The main utility of Eq.~\eqref{eq:ells_Theta} is that we can calculate $\ell^{\pm}_{\OO}$ analytically for a system of Landau levels even though we cannot directly compute $\Theta^{\pm}_{\OO}$. This gives us a simple way to obtain the $c_-$ dependence of $\Theta^{\pm}_{\OO}$.

There are other important relations. $\Theta^{\pm}_{\OO}$ are related to the discrete shift $\SO$ as follows:
\begin{equation}
    \SO = \begin{cases}
    2 (\Theta^-_{\OO} - \Theta^+_{\OO}) \mod \MO & \MO \text{ even} \\
    2 (\Theta^-_{\OO} - \Theta^+_{\OO}) - \frac{C}{2} \mod \MO & \MO \text{ odd}.
\end{cases}
\end{equation}
This can be shown using a relabelling of the field theory, together with Eq.~\eqref{eq:ells_Theta}. Using this, and the relation between $\SO$ and $\PO$ (see Table~\ref{tab:PvsS}), we can also deduce relations between $\Theta^{\pm}_{\OO}$ and $\PO$. 

For completeness, note that we can calculate $\vec{\mathscr{P}}_{\OO}$ directly in two steps. When $c_- = 0$, $\PO$ is the dipole moment at $\OO$ within the unit cell, which can be written as a linear combination of $n_{\OO}$ (see App.~B of Ref.~\cite{zhang2022pol}). In the LL limit, where the same system can be thought of as having a continuous rotational symmetry, $\PO$ vanishes because it must be invariant under arbitrarily small rotations and therefore cannot take nontrivial quantized values. We can similarly define $\vec{\mathscr{P}}_{s,\OO}$ if in the above discussion we replace $n_{\OO}$ with $m_{\OO}$.

Finally, $\Theta^{\pm}_{\OO}$ are related to the angular momentum per unit cell $\nu_s$ as follows. When $c_- = 0 = C$, $\nu_s$ takes the value $\sum_{\OO} m_{\OO} W_{\OO} \mod M$ where $W_{\OO}$ be the total number of points in the unit cell belonging to the maximal Wyckoff position containing $\OO$. (For example, in the square lattice, $W_{\alpha} = W_{\beta} = 1$ and $W_{\gamma} = 2$, and $\nu_s = m_{\alpha} + m_{\beta} + 2 m_{\gamma}$.) This implies that
\begin{equation}
    \nu_{s,\AI}  =\begin{cases}
        - \frac{\kappa}{2} + \sum_{\OO} \Theta^-_{\OO,\AI} W_{\OO} \mod M & M \text{ even}, \\
        \sum_{\OO} \Theta^+_{\OO,\AI} W_{\OO} \mod M, & M \text{ odd}.
    \end{cases}
\end{equation}
The reason we define $\nu_{s,\AI}$ using $\Theta^-_{\OO,\AI}$ when $M$ is even, and not $\Theta^+_{\OO,\AI}$, is because we require an invariant mod $M$, but $\Theta^+_{\OO}$ are only defined mod $M/2$ in this case. Next, note that $\nu_{s,\text{LLL}} = 0$ because it is an invariant associated to discrete translation symmetries, whereas the Landau level system has a continuum limit in which this invariant must vanish. By combining these results, we obtain 
\begin{equation}
    \nu_{s}  =\begin{cases}
        - \frac{\kappa}{2} + \sum_{\OO} (\Theta^-_{\OO} - c_-\Theta^-_{\OO,\text{LLL}}) W_{\OO} \mod M, & M \text{ even}, \\
        \sum_{\OO} (\Theta^+_{\OO}- c_-\Theta^+_{\OO,\text{LLL}}) W_{\OO} \mod M, & M \text{ odd}.
    \end{cases}
\end{equation}
Here we used Eq.~\eqref{eq:Theta_def_gen}. Note that $\nu_s$ is well defined modulo $M$, as required by the field theory. Additionally, when $M$ is even, we have
\begin{equation}
    \nu_s = \sum_{\OO}(\Theta^+_{\OO}- c_-\Theta^+_{\OO,\text{LLL}}) W_{\OO} \mod \frac{M}{2}.
\end{equation}
From these results, we see that for each $M$ $\nu_s$ is indeed a weighted sum of partial rotation expectation values at different high symmetry points (suitably corrected to vanish for Landau levels). This matches our intuition that $\nu_s$ describes an angular momentum per unit cell. Interestingly, these formulas also imply a relationship between $\Theta^+_{\OO}$ and $\Theta^-_{\OO}$ which was noticed in Ref.~\cite{zhang2023complete} but not proven there. 

We can also write down explicit formulas which express the generators of the TQFT classification (see Eqs.~\eqref{eq:FTresult_C4},~\eqref{eq:FTresult_C2},~\eqref{eq:FTresult_C3},~\eqref{eq:FTresult_C6}) in terms of $\Theta^{\pm}_{\OO}, C, c_-$ and $\kappa$. These formulas are summarized in App.~\ref{app:Gens-RSI-FT}.

\section{Non-interacting band Insulators as invertible phases}\label{sec:FImap}
\def\Gm{\boldsymbol{\Gamma}}
\def\Xm{\boldsymbol{X}}
\def\Mm{\boldsymbol{M}}
\def\Ym{\boldsymbol{Y}}
\def\Km{\boldsymbol{K}}
\def\bKm{\bar{\boldsymbol{K}}}

This section derives the free-to-interacting map between the classification of non-interacting fermions at zero flux to the interacting classification with $G_f = \text{pM} \times \U(1)^f$. We start by reviewing the non-interacting classification in Sec.~\ref{sec:ShortReview}. In Sec.~\ref{sec:ManyBodyInv-nIInv}, we express the real-space invariants of Sec.~\ref{sec:RSI} for any band insulator in terms of their non-interacting invariants. A similar calculation was carried out in Ref.~\cite{Li2020disc} but for a different set of interacting invariants.

\subsection{Short review of non-interacting band insulators }\label{sec:ShortReview}

\begin{figure}
    \centering
    \includegraphics[width=0.5\textwidth]{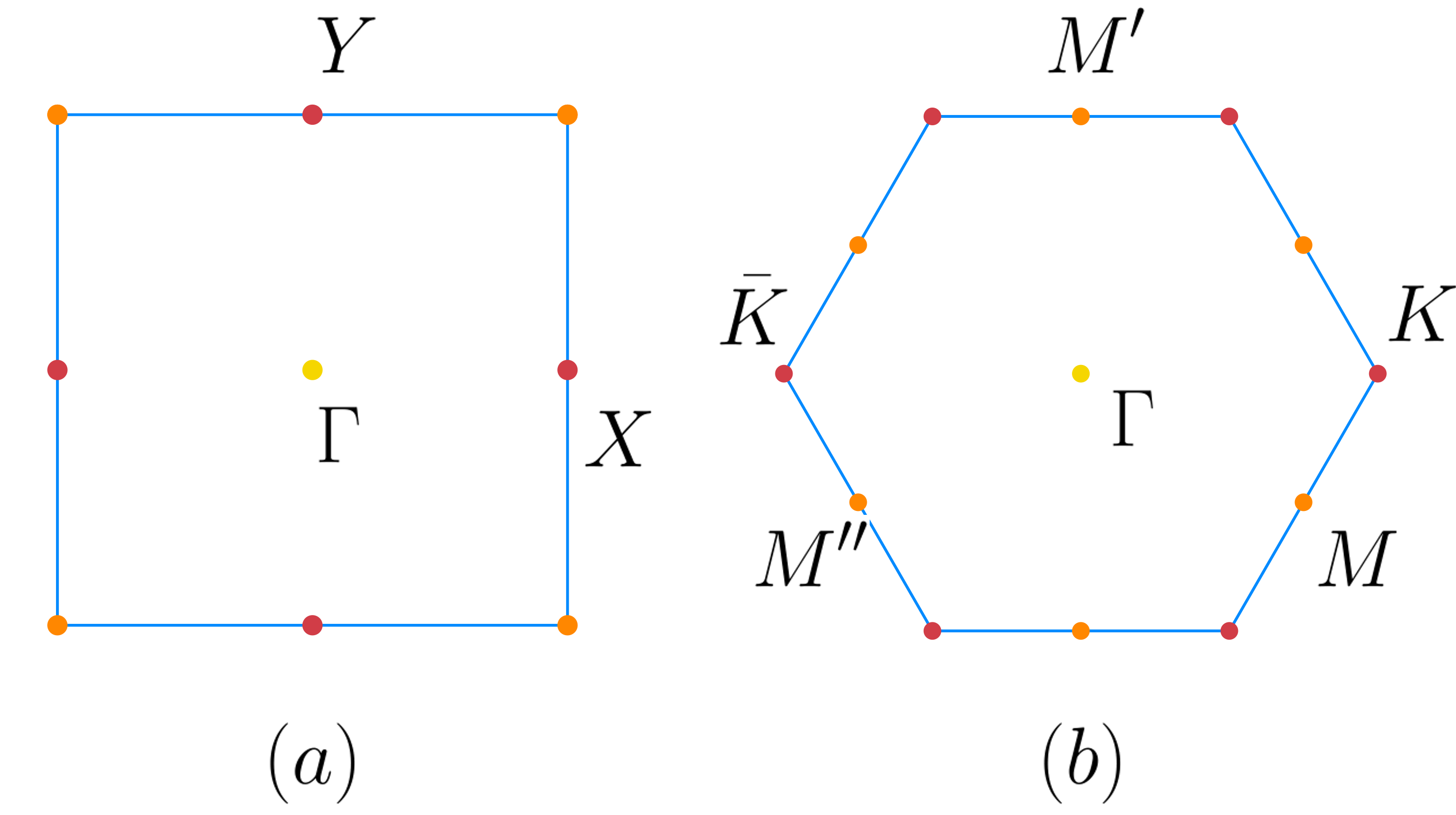}
    \caption{Brillouin zones for systems with (a) $M=2,4$ , (b) $M=3,6$.}
    \label{fig:BZ}
\end{figure}

Non-interacting band insulators (nIBIs) with orientation-preserving wallpaper groups and zero flux ($\phi = 0$) have been mathematically classified by the formalism of K-theory,  e.g. \cite{Shiozaki2018AH, stehouwer2018classification, cornfeld2021tenfold}. However, the same results can be phrased physically in terms of the (total) Chern number and the eigenvalues (counted with multiplicities) of the filled bands under a (fixed) rotation operator \cite{Kruthoff2017TIBandComb}. For the sake of completeness and to fix conventions for the free-to-interacting map in Sec.~\ref{sec:ManyBodyInv-nIInv}, we review the results of the non-interacting classification in App.~\ref{app:BandInsulators} and provide a summary of the known results below. Note that much less is known for the case of non-zero flux $\phi\neq 0$; see \cite{herzog2020Hofstadter,herzog2022Hofstadter, Fang2023SymIndRationalFlux} for the current understanding.

\subsubsection{Reciprocal space notation}
Let us first establish some notation for each $\Gspace=$p1,p2,p3,p4,p6. For p2 and p4, we use a square Brillouin zone (BZ). For p3 and p6, we use a hexagonal BZ. Both BZs are illustrated in Fig.~\ref{fig:BZ} with the high symmetry momenta ($\HSM$) highlighted. Upon setting the lattice constants to one, these momenta have coordinates $\Gm=(0,0)$, $\Xm=(\pi,0)$, $\Ym = (0,\pi)$, $\Km = (4\pi/\sqrt{3},0)$, $\bKm=(-4\pi/\sqrt{3},0)$. For the square BZ, $\Mm = (\pi,\pi)$, while for the hexagonal BZ, $\Mm =(\pi,-\pi/\sqrt{3})$. \footnote{The free fermion literature usually replaces $(\Xm,\Ym) \to (\Xm_1,\Xm_2)$ for $M=4$ and $(\Km,\Mm) \to (\Km_1,\Mm_1)$ for $M=6$, but we avoid the subscript notation to make our later equations simpler.}

We denote by $\HSM^{\star}$ the set of orbits of the high symmetry momenta ($\HSM)$ under point group rotations, and by $\HSMsp$ the set $\HSM^{\star}$ with the orbit of the origin $\Gamma = (0,0)$ excluded. $\kbs^{\star}$ denotes the orbit with representative $\kbs$. See Table ~\ref{tab:SummaryHSMstar} for $\HSM^{\star}$ for $M=2,3,4,6$.
Given a momentum $\kbs$, we denote by $M_{\kbs}$ the maximal order of a rotation which maps $\kbs$ to itself.\footnote{For example, when $M=4$, this gives $M_{\Gm}= M_{\Mm}=4$ and $M_{\Xm}=2$.}

\begin{table}[t]
        \centering
       \begin{equation*}
    \begin{array}{|c|l|}\hline
       M  & \HSM^{\star} \\\hline
        2 & \{\Gm\}, \{\Xm\}, \{\Ym\}, \{\Mm\} \\
        4 & \{\Gm\}, \{\Xm,\Ym\},  \{\Mm\} \\
        3 & \{\Gm\}, \{\Km\}, \{\bKm\} \\
        6 & \{\Gm\}, \{\Km,\bKm\}, \{\Mm,\Mm',\Mm''\} \\
        \hline
    \end{array}
\end{equation*}
        \caption{
        Orbits of the high symmetry momenta, denoted as $\HSM^{\star}$. We take the first element of each orbit as the representative. 
        }
        \label{tab:SummaryHSMstar}
\end{table}

\subsubsection{Non-interacting classification}
\def\CmopNI{\Cmop}

Using the many-body rotation operator $\Cmop$ discussed above, we can deduce how it acts on the single-particle eigenstates. We choose $\OO$ such that $M_{\OO}=M$. 
Single particle eigenstates are labeled by their crystal momentum $\kbs$ and their orbital wave function. A rotation that maps $\kbs$ to itself also maps occupied single-particle eigenstates to themselves; this allows us to assign eigenvalues under the rotation to occupied states. Let $\#\kbs_{j,\OO}^{(m)}$, with $j=1,2,\dots,m$, be the degeneracy among the occupied bands of the eigenvalue $\lambda=e^{2\pi \ii (j-1)/m}$ under $[\CmopNI]^{\MO/m}$, a rotation of order $m$ around $\OO$. (See App.~\ref{app:BandComb:DefI} for more details.)

A non-interacting band insulator has the following three types of invariants: 
\begin{enumerate}
    \item The total Chern number $\Ch$ of the filled bands, 
    \begin{equation}\label{eq:ChernNumberDef}
        \Ch =\sum_{n \in \text{Occ.}} \int_{\BZ} \eps_{ij}\partial_i\left[\mel{u_{\kbs}^{(n)}}{\partial_j}{u_{\kbs}^{(n)}}\right]\frac{\dd^2{\kbs}}{2\pi\ii};
    \end{equation}
    where $\ket{u_{\kbs}^{(n)}}$ are the orbital wave-functions of the occupied bands; $\partial_j = \pdv{\,}{\kbs_k}$; and $\eps$ is the Levi-Civita tensor ($\eps_{12}=-\eps_{21}=1$).
    \item The integers $\#\Gm_{j,\OO}^{(M)}$, with $ j=1,\dots,M$. $\Gm= (0,0)$ is the origin of the BZ. 
    \item For each orbit $\kbs^{\star}\in\HSM'^{\star}$, the rotation invariants from a representative $\kbs$, defined as
    \begin{equation}
    [\kbs_{j,\OO}^{(M_{\kbs)}}]= \# \kbs_{j,\OO}^{(M_{\kbs)}} - \#\Gm_{j,\OO}^{(M_{\kbs)}};\, j = 1,\dots, M_{\kbs}. 
    \end{equation}
 
\end{enumerate}

Even though the above set of invariants is complete, it is partially redundant for two reasons. 1) nIBIs have fully filled bands so the total number of eigenvalues is constant as a function of $\kbs$ and equal to the filling: 
\begin{equation}\label{eq:KappaConstrain}
    \kappa = \sum_{j=1}^{M_{\kbs}} \#\kbs_{j,\OO}^{(M_{\kbs})}; \,\,\forall \kbs \in \BZ.
\end{equation}
2) Rotation invariants determine the Chern number modulo $M$\footnote{We derive this expression in App.~\ref{app:DefInv:OrPreserving}  based on Ref.~\cite{Fang2012PGS}. }:
\begin{equation}\label{eq:ChQuantizedRot}
  \frac{\Ch}{M}+  \sum_{\kbs^{\star}\in \HSM'^{\star}} \sum_{j=1}^{M_{\kbs}}\frac{(j-1)}{M_{\kbs}}[\kbs_{j,\OO}^{(M_{\kbs})}]  = 0  \mod 1 .
\end{equation}
There are no more constraints as one can obtain explicit models realizing any set of invariants subject to the above constraints (See App.~\ref{app:GeneratingSet}). A complete set of invariants for nIBI is given in Tab.~\ref{tab:basisNonInteracting}. We denote the invariants as $J = (J_1,J_2,\dots)$, and the invariants calculated for a state $\psi$ are denoted by $J[\psi]$. Note that the second constraint above implies that there is a relation modulo $M$ between these invariants.  

\def\psiCoeff{a}
Within the non-interacting classification, any nIBI $\psi$ can be formally written as a linear combination of a small, finite set of basis states:
\begin{equation}\label{eq:ZDecomposition}
    \psi \sim_{S} \sum_{k=1}^{D_M} \psiCoeff_k\Psi_k,
\end{equation}
where $\psiCoeff_{k}\in \ZZ$. $D_M$ is an integer that is equal to the exponent of $\ZZ$ in the ``free classification" column of Table~\ref{tab:Classif_Summary}. The explicit set of states can be found in Table~\ref{tab:basisComplete} in App.~\ref{app:GeneratingSet}. We take $\Psi_1$ to be a specific Chern insulator with $\Ch=+1$: it is the Qi-Wu-Zhang model for $M=2,4$ \cite{Qi2006QSHE_SquareLattice} and the Haldane model for $M=3,6$ (see App.~\ref{app:ChernInsulatorsExamples}). The remaining basis states are atomic insulators induced from maximal Wyckoff positions (see App.~\ref{app:GeneratingAI}.)

There are two subtleties in Eq.~\eqref{eq:ZDecomposition}. First, the symbol $\sim_S$ denotes ``stable equivalence". This means that there exists some ancilla tensor product state $\psi'$ that allows us to find a path connecting the LHS and RHS, {i.e,}:
\begin{equation}\label{eq:AdiabaticPath}
    \psi' + \psi  \sim   \psi' +  \sum_{k=1}^{D_M} \psiCoeff_k\Psi_k^{}
\end{equation}
where the $+$ symbol here corresponds to the tensor product (``stacking") of many-body states and $n \Psi_{k}^{} $ is a shorthand for stacking $n$ copies of state $\Psi_{k}$. The $\sim$ symbol means that there is an adiabatic path in the space of free fermion Hamiltonians connecting the state on the LHS to the state on the RHS.

Second, to interpret Eq.~\eqref{eq:ZDecomposition} when there are negative coefficients $\psi_k$, we need to first move these negative coefficients to the other side before applying Eq.~\eqref{eq:AdiabaticPath}. That is, Eq.~\eqref{eq:ZDecomposition} becomes
\begin{equation}\label{eq:ZDecompositionPositive}
     \psi +  \sum_{k=1}^{D_M} [\psiCoeff_k]_-\Psi_k \sim_{S} \sum_{k=1}^{D_M} [\psiCoeff_k]_+\Psi_k.
\end{equation}
$[a]_{+}=a$ if $a\geq 0$ and zero, otherwise. Similarly, $[a]_{-}=-a$ if $a\leq 0$ and zero, otherwise.

\begingroup
\renewcommand{\arraystretch}{1.5}
\begin{table}[t]
        \centering
       \begin{equation*}
    \begin{array}{|c|l|}\hline
       M  & \text{Basis} \\\hline\hline
        1 & \Ch, \#\Gm_1^{(1)} \\ \hline
        2 & \Ch, \#\Gm_1^{(2)}, \#\Gm_2^{(2)}, [\Mm_{2}^{(2)}], [\Xm_2^{(2)}], [\Ym_2^{(2)}]\\\hline
        3 & \Ch, \#\Gm_1^{(3)}, \#\Gm_2^{(3)}, \#\Gm_3^{(3)}, [\Km_{2}^{(3)}], [\Km_{3}^{(3)}], [\bKm_{2}^{(3)}], [\bKm_{3}^{(3)}]\\\hline
        4 & \Ch, \#\Gm_1^{(4)}, \#\Gm_2^{(4)},\#\Gm_3^{(4)},\#\Gm_4^{(4)}, [\Mm_2^{4}], [\Mm_3^{4}], [\Mm_4^{4}], [\Xm_2^{(2)}]\\ \hline
        \multirow{2}{*}{6}& \Ch, \#\Gm_1^{(6)}, \#\Gm_2^{(6)}, \#\Gm_3^{(6)}, \#\Gm_4^{(6)}, \#\Gm_5^{(6)},\#\Gm_6^{(6)},\\
        & [\Km_{2}^{(3)}], 
        [\Km_{3}^{(3)}], [\Mm_{2}^{(2)}]\\
        \hline
    \end{array}
\end{equation*}
        \caption{
        Basis choice for the free invariants in the non-interacting band insulator classification.
        }
        \label{tab:basisNonInteracting}
\end{table}
\endgroup
\begin{table}[t]
        \centering
       \begin{equation*}
    \begin{array}{|c|l|}\hline
       M  & \text{Invariants} \\\hline
        1 & C, c_-, \kappa \\
        2 & C, c_-, \kappa, \Theta_{\alpha}^{-}, \Theta_{\beta}^{-}, \Theta_{\gamma}^{-}, \Theta_{\delta}^{-} \\
        4 & C, c_-, \kappa, \Theta_{\alpha}^{+}, \Theta_{\beta}^{+}, 
        \Theta_{\alpha}^{-}, \Theta_{\beta}^{-}, \Theta_{\gamma}^{-} \\
        3 & C, c_-, \kappa, \Theta_{\alpha}^{+}, \Theta_{\beta}^{+}, 
        \Theta_{\gamma}^{+}, 
        \Theta_{\gamma}^{-},
        \Theta_{\beta}^{-}, \Theta_{\gamma}^{-} \\
        6 & C, c_-, \kappa, \Theta_{\alpha}^{+}, 
        \Theta_{\beta}^{+}, 
        \Theta_{\alpha}^{-},
        \Theta_{\beta}^{-}, 
        \Theta_{\gamma}^{-}\\ 
        \hline
    \end{array}
\end{equation*}
        \caption{
        Choice for a complete set of real-space interacting invariants used to derive the free-to-interacting map. 
        \label{tab:basisInteracting}
        }
\end{table}

\subsection{Many-body invariants from non-interacting invariants}\label{sec:ManyBodyInv-nIInv}
\def\Qmc{\mathcal{Q}}
\def\Imc{\mathcal{I}}
\def\AI{\mathsf{AI}}
\def\ai{\mathsf{ai}}

As we learnt in Sec.~\ref{sec:RSI}, a complete set of interacting real-space invariants is given by $c_-,C,\kappa$ and the various $\Theta_{\OO}^{\pm}$. We summarize these invariants in Table ~\ref{tab:basisInteracting} and denote them by $\Qmc = (\Qmc_{1},\Qmc_2,\dots)$ following the order of the table.

Given a decomposition of $\psi$ as in Eq.~\eqref{eq:AdiabaticPath} and the fact that topological invariants are linear under stacking, we can write
\begin{equation}\label{eq:Zlinearity}
    \Imc[\psi] = \sum_{k=1}^{D_M} \psiCoeff_k \Imc[\Psi_{k}]
\end{equation}
where $\Imc$ can now be either $J_j$ (a free fermion band invariant from Table~\ref{tab:basisNonInteracting}) or $\Qmc_j$ (an interacting invariant from Table~\ref{tab:basisInteracting}). 

We can compactly express the free fermion invariants in terms of a matrix $O$, defined as $(O)_{jk}=J_{j}[\Psi_k]$, whose inverse $O^{-1}$ exists\footnote{This is where choosing the right set of states $\{\Psi_1,\Psi_2,\dots\}$ to use a basis is important. We present the explicit $O$ matrices in Eqs.~\ref{eq:Omatrix_M2}, \ref{eq:Omatrix_M4}, \ref{eq:Omatrix_M3} and \ref{eq:Omatrix_M6} in App.~\ref{app:GeneratingSet}.}. This allows us to express  $\psiCoeff_k = (O^{-1})_{kj}J_{j}[\psi]$. In other words, once we know the band invariants of $\Psi_k$ and $\psi$, we can recover the precise linear combination of $\Psi_k$ which determines $\psi$.

Now we reapply Eq.~\eqref{eq:Zlinearity} with $\mathcal{I} = \Qmc_i$ and with $\psiCoeff_k$ as above. This gives
\begin{equation}\label{eq:ZlinearQ}
    \Qmc_i[\psi] = \sum_{k=1}^{D_M} \Qmc_i[\Psi_k] (O^{-1})_{kj}J_{j}[\psi].
\end{equation}
This equation gives the interacting invariants of $\psi$ in terms of its free fermion invariants and the matrix $O$. Since this equation holds for any $\psi$, we have the relation $\Qmc_i = \sum_{j=1}^{D_M} \Mmc_{ij }J_{j}$ with $ \Mmc_{ij} =\sum_{k=1}^{D_M} \Qmc_i[\Psi_k] (O^{-1})_{kj}$. Therefore, once we know $J[\Psi_k]$ and $\Qmc[\Psi_k]$, we can calculate the matrix $\Mmc$, which is the desired free-to-interacting map. 

In order to evaluate the invariants above, we need to make choices for Hamiltonians and rotation operators. All the models we consider are invariant under the translations $f_{\rbs} \to f_{\rbs+\Rbs}$ where $\Rbs$ is a lattice vector and $\rbs$ is the real space position of the orbital. This allows us to define the rotation operators around $\OO\neq \alpha$ in terms of $\tilde{C}^{+}_{M_{\alpha}}$. For example, for $M=4$ we can define the rotation around $\beta$ as $\tilde{C}^{+}_{M_{\beta}} =  T_{x}\tilde{C}^{+}_{M_{\alpha}}$ where $T_x$ is the above translation by one unit cell in the $x$ direction. The explicit choices for the Hamiltonian and rotation operators for the states $\Psi_1$ can be found in App.~\ref{app:ChernInsulatorsExamples}. The evaluation of the invariants can be found in the same Appendix. The calculation of invariants for the other $\Psi_k$ is straightforward as these are atomic insulators. Descriptions of these models can be found in App.~\ref{app:AI}.
Using these results, we evaluate $\Mmc$ and present our results below. We omit expressions for $C$ and $c_-$ because they are always equal to the $C$ of the non-interacting model.

\begin{widetext}
The free-to-interacting map is
\begingroup
\renewcommand{\arraystretch}{1.5}
\begin{itemize}
    \item $M=2:$
\begin{equation}\label{eq:FImap-C2}
     \begin{pmatrix}
    \kappa\\
    \Theta_{\alpha}^{-}\\
    \Theta_{\beta}^{-}\\
    \Theta_{\gamma}^{-}\\
    \Theta_{\delta}^{-}\\
\end{pmatrix}=  \begin{pmatrix}
  0 & 1 & 1 & 0 & 0 & 0 \\
 1 & \frac{1}{2} & \frac{3}{2} & \frac{3}{4} & \frac{3}{4} & \frac{3}{4} \\
 1 & 0 & 0 & \frac{3}{4} & \frac{5}{4} & \frac{5}{4} \\
 1 & 0 & 0 & \frac{5}{4} & \frac{3}{4} & \frac{5}{4} \\
 1 & 0 & 0 & \frac{5}{4} & \frac{5}{4} & \frac{3}{4} \\
\end{pmatrix}
\begin{pmatrix}
    \Ch\\
    \#\Gm_{1,\alpha}^{(2)}\\
    \#\Gm_{2,\alpha}^{(2)}\\ 
    [\Mm_{2,\alpha}^{(2)}]\\
    [\Xm_{2,\alpha}^{(2)}]\\
    [\Ym_{2,\alpha}^{(2)}]
\end{pmatrix}
\end{equation}
\item $M=4:$
\begin{equation}\label{eq:FImap-C4}
   \begin{pmatrix}
    \kappa\\
    \Theta_{\alpha}^{+}\\
    \Theta_{\beta}^{+}\\
    \Theta_{\alpha}^{-}\\
    \Theta_{\beta}^{-}\\
    \Theta_{\gamma}^{-}\\
\end{pmatrix}= \begin{pmatrix}
  0 & 1 & 1 & 1 & 1 & 0 & 0 & 0 & 0 \\
 0 & 0 & 1 & 0 & 1 & \frac{1}{2} & 0 & \frac{3}{2} & 0 \\
 0 & 0 & 0 & 0 & 0 & \frac{3}{2} & 0 & \frac{1}{2} & 0 \\
 -1 & \frac{1}{2} & \frac{3}{2} & \frac{5}{2} & \frac{7}{2} & \frac{13}{4} & 1 & \frac{1}{4}
   & \frac{5}{2} \\
 -1 & 0 & 0 & 0 & 0 & \frac{13}{4} & 3 & \frac{9}{4} & \frac{3}{2} \\
 -1 & 0 & 0 & 0 & 0 & \frac{5}{4} & 0 & \frac{5}{4} & 0 \\
\end{pmatrix}
\begin{pmatrix}
    \Ch\\
    \#\Gm_{1,\alpha}^{(4)}\\
    \#\Gm_{2,\alpha}^{(4)}\\ 
    \#\Gm_{3,\alpha}^{(4)}\\
    \#\Gm_{4,\alpha}^{(4)}\\
    [\Mm_{2,\alpha}^{(4)}]\\
    [\Mm_{3,\alpha}^{(4)}]\\
    [\Mm_{4,\alpha}^{(4)}]\\
    [\Xm_{2,\alpha}^{(2)}]\\
\end{pmatrix}
\end{equation}
\item $M=3:$
\begin{equation}\label{eq:FImap-C3}
\begin{pmatrix}
    \kappa\\
    \Theta_{\alpha}^{+}\\
    \Theta_{\beta}^{+}\\
    \Theta_{\gamma}^{+}\\
    \Theta_{\alpha}^{-}\\
    \Theta_{\beta}^{-}\\
    \Theta_{\gamma}^{-}\\
\end{pmatrix}=\begin{pmatrix}
 0 & 1 & 1 & 1 & 0 & 0 & 0 & 0 \\
 1 & 0 & 1 & 2 & \frac{4}{3} & \frac{5}{3} & \frac{4}{3} & \frac{5}{3} \\
 1 & 0 & 0 & 0 & \frac{4}{3} & \frac{8}{3} & \frac{1}{3} & \frac{5}{3} \\
 1 & 0 & 0 & 0 & \frac{1}{3} & \frac{5}{3} & \frac{4}{3} & \frac{8}{3} \\
 -1 & \frac{1}{2} & 0 & 1 & \frac{1}{3} & \frac{2}{3} & \frac{1}{3} & \frac{2}{3} \\
 -1 & 0 & 0 & 0 & \frac{5}{6} & \frac{7}{6} & \frac{1}{3} & \frac{7}{6} \\
 -1 & 0 & 0 & 0 & \frac{1}{3} & \frac{7}{6} & \frac{5}{6} & \frac{7}{6} \\
\end{pmatrix}
\begin{pmatrix}
    \Ch\\
    \#\Gm_{1,\alpha}^{(3)}\\
    \#\Gm_{2,\alpha}^{(3)}\\ 
    \#\Gm_{2,\alpha}^{(3)}\\
    [\Km_{2,\alpha}^{(3)}]\\
    [\Km_{3,\alpha}^{(3)}]\\
    [\bKm_{2,\alpha}^{(3)}]\\
    [\bKm_{3,\alpha}^{(3)}] 
\end{pmatrix}
\end{equation}
\item $M=6:$

\begin{equation}\label{eq:FImap-C6}
\begin{pmatrix}
    \kappa\\
    \Theta_{\alpha}^{+}\\
    \Theta_{\beta}^{+}\\
    \Theta_{\alpha}^{-}\\
    \Theta_{\beta}^{-}\\
    \Theta_{\gamma}^{-}\\
\end{pmatrix}=
\begin{pmatrix}
0 & 1 & 1 & 1 & 1 & 1 & 1 & 0 & 0 & 0 \\
 \frac{3}{2} & 0 & 1 & 2 & 0 & 1 & 2 & \frac{8}{3} & \frac{1}{3} & \frac{3}{2} \\
 1 & 0 & 0 & 0 & 0 & 0 & 0 & \frac{5}{3} & \frac{4}{3} & 0 \\
 0 & \frac{1}{2} & \frac{3}{2} & \frac{5}{2} & \frac{7}{2} & \frac{9}{2} & \frac{11}{2} &
   \frac{2}{3} & \frac{4}{3} & \frac{21}{4} \\
 \frac{1}{2} & 0 & 0 & 0 & 0 & 0 & 0 & \frac{7}{6} & \frac{5}{6} & 0 \\
 1 & 0 & 0 & 0 & 0 & 0 & 0 & 0 & 0 & \frac{5}{4} \\
\end{pmatrix}\begin{pmatrix}
    \Ch\\
    \#\Gm_{1,\alpha}^{(6)}\\
    \#\Gm_{2,\alpha}^{(6)}\\
    \#\Gm_{3,\alpha}^{(6)}\\
    \#\Gm_{4,\alpha}^{(6)}\\
    \#\Gm_{5,\alpha}^{(6)}\\
    \#\Gm_{6,\alpha}^{(6)}\\
    [\Km_{2,\alpha}^{(3)}]\\
    [\Km_{3,\alpha}^{(3)}]\\
    [\Mm_{2,\alpha}^{(2)}]
\end{pmatrix}
\end{equation}
\end{itemize}
\endgroup
\end{widetext}

\section{Discussion}\label{sec:Disc}

In this paper we considered the symmetry group $G_f = \U(1)^f \times_{\phi} G_{\text{space}}$, where $G_{\text{space}}$ is an orientation preserving wallpaper group in two space dimensions with magnetic flux $\phi$ per unit cell. We discussed several distinct approaches to classify (2+1) dimensional invertible fermionic topological states with this symmetry, as summarized in Table~\ref{tab:Classif_Summary}. First, in Sec.~\ref{sec:FT}, we derived a topological effective action for many-body invertible states (Eq.~\eqref{eq:mainresponse}) which expresses the desired topological invariants in terms of the response of the system to inserting symmetry defects. Next, in Sec.~\ref{sec:RSI}, we considered a physically distinct real-space classification, and defined a pair of invariants $\Theta^+_{\OO},\Theta^-_{\OO}$ for each high-symmetry point in the real-space unit cell. We showed that this real-space classification reproduces all the invariants in the above topological field theory, except for an integer invariant describing bosonic integer quantum Hall states. The above results hold for a general chiral central charge $c_-$ as well as an arbitrary flux $\phi$. Finally, in Sec.~\ref{sec:FImap}, we specialized to $\phi = 0$ and derived expressions relating $\Theta^{\pm}_{\OO}$ to the band invariants that arise in the free fermion classification of invertible states. This provides an explicit map between the free and interacting classifications.

Several of our results are based on some underlying assumptions, which we elaborate on below. 

In the TQFT-based derivation of Sec.~\ref{sec:FT}, we assumed the fermionic crystalline equivalence principle (fCEP), which states that the classification of invertible fermionic states with a spatial symmetry $G_f$ is in one-to-one correspondence with that of invertible fermionic states with an \textit{internal} symmetry $G_f^{\text{eff}}$. We used a version of the fCEP stated in Ref.~\cite{manjunath2022mzm}, which also states an explicit algebraic formula relating $G_f$ and $G_f^{\text{eff}}$. Specifically, we have $G_f/\Z_2^f \cong G_f^{\text{eff}}/\Z_2^f := G_b$, but the central extension of $G_b$ by the fermion parity $\Z_2^f$ is different in each case. The algebraic formula stated there was not derived, but justified heuristically based on a comparison of several known examples of topological insulators and superconductors with the same $G_b$, which in some cases acted as a spatial symmetry and in others as an internal symmetry. Furthermore, while Ref.~\cite{manjunath2022mzm} and prior work on the fCEP \cite{Thorngren2018,zhang2020realspace,debray2021invertible} stated it as a correspondence between the groups that classify invertible states with symmetries $G_f$ and $G_f^{\text{eff}}$, we make a stronger assumption here, namely that the coefficients in a given topological action for invertible states with spatial symmetry $G_f$ have the same quantization and mathematical properties as those in a corresponding topological action for invertible states with internal symmetry $G_f^{\text{eff}}$. Although there is no formal derivation of this stronger correspondence, we are confident in the predictions derived from the fCEP in our case, because they satisfy a number of numerical checks that were performed previously in Refs.~\cite{zhang2022fractional,zhang2022pol,zhang2023complete}, and they also give results consistent with the real-space classification.

In Sec.~\ref{sec:RSI} we defined the quantities $\Theta^{\pm}_{\OO}$ in terms of the argument of the expectation value of the ground state under partial rotation operations, see Eqs.~\eqref{eq:K_def},~\eqref{eq:Theta_def_K}. Then we stated that these are quantized many-body invariants, and explained how this quantization depends on the topological invariants that arise in the TQFT of Sec.~\ref{sec:FT}. The theoretical analysis that predicts quantization of these terms relies on the assumption $\rho_D \approx \rho_{\text{CFT}}$, that is, the density matrix within the partial rotation region $D$ can be approximated by the conformal field theory defined on the boundary of $D$. Alternatively, the assumption states that the entanglement spectrum of the density matrix is equivalent to that of its edge CFT. Assuming this correspondence allows us to compute $\Theta^{\pm}_{\OO}$ explicitly using conformal field theory, and the resulting predictions have been verified in detail numerically for the square lattice \cite{zhang2023complete}.

This assumption was discussed previously in Refs.~\cite{Haldane2008entanglement,Qi2012entanglement}, which gave heuristic arguments supporting the correspondence but did not formally derive it. In fact, it has known exceptions, for example the `spurious' topological entanglement entropy (TEE) \cite{zou2016spuriousTEE,Cano2015entanglement} in topologically ordered states, where depending on the choice of entanglement cut the value of the TEE calculated from the entanglement Hamiltonian may not equal the expected universal value $\log \mathcal{D}$, which is the total quantum dimension of anyons in the edge CFT. These exceptions seem non-generic and somewhat fine-tuned, so the correspondence is still of great practical value. Nevertheless, it would be of interest to have a more systematic understanding of when this general correspondence is violated and its implications for invariants obtained from partial rotations. 

We conclude by noting some further directions of interest. 
The free-to-interacting map we have derived here specifies that certain free fermion states can in principle be adiabatically connected by turning on suitable interactions, but is silent on precisely what types of interactions are required. It would be interesting to refine the map by studying the minimal interaction term (4-body, 6-body, etc.) required to connect two specific free fermion states, and clarifying how the answer depends on their symmetry data. For example, the 4-body case was studied for some crystalline topological states in Ref.~\cite{Morimoto2015}.

Finally, it would be very interesting to explore whether the partial rotation methods discussed here to measure real-space crystalline topological invariants can be implemented in current experiments. More generally, the results in this paper emphasize that measuring a single invariant of an invertible state, such as the Chern number in Chern insulators, is not enough to experimentally characterize it or conclusively distinguish it from another invertible state. A full characterization instead requires us to measure all the invariants predicted by either the free or interacting classifications discussed here, whichever is most relevant. Developing experimental protocols to measure them (in particular the many-body invariants, which have been much less explored) is therefore an important future direction.

\section{Acknowledgements}

We thank Yuxuan Zhang, Gautam Nambiar and Ryohei Kobayashi for collaboration on related work. We are grateful to Ryohei Kobayashi for sharing CFT calculations that were unpublished at the time of writing. N.M. and V.C. contributed equally to this work.

This work is supported by NSF
CAREER grant (DMR- 1753240), and the Laboratory for Physical Sciences through the Condensed Matter Theory Center (NM, MB). Research at Perimeter Institute is supported in part by the Government of Canada through
the Department of Innovation, Science and Economic Development and by the Province of
Ontario through the Ministry of Colleges and Universities.

\newpage
\appendix
\begin{widetext}

\section{Derivation of topological action}\label{app:DerTopAct}
\subsection{Details for general $M$}\label{app:details_genM}
Here we summarize the steps of the TQFT derivation, keeping $M$ general. We fix an origin $\OO$ which is symmetric under $M$-fold rotations, that is, $\MO = M$. The calculations specific to $M=2,3,4,6$ will be given in the following subsections. Note that Apps~\ref{app:details_genM},~\ref{app:spM} assume $\phi = 0$. The extension to $\phi \ne 0$ will be discussed in App.~\ref{app:phinonzero}.

\subsubsection{Overview and definition of $G_f$}

We have $G_f = \U(1)^f\times [\Z^2\rtimes\Z_M]$ (that is, we assume $\phi = 0$), and $G_b = G_f/\Z_2^f = \U(1)\times [\Z^2\rtimes \Z_M]$. Let $[\omega_2] \in \H^2(G_b,\Z_2)$ determine $G_f$ as a group extension of $G_b$ by $\Z_2^f$. We assume that the group $\Z_M$ is generated by the operator $\Cmop$ defined in the main text. The general topological action $\mathcal{L}$ for $G_f$ symmetric invertible fermion phases will be derived in two main steps. First we use the fermionic crystalline equivalence principle (fCEP), which states that the desired classification and topological action for $G_f$ invertible phases is in one-to-one correspondence with those for invertible phases with an \textit{effective internal} symmetry $G_f^{\text{eff}}$. (In our case it will turn out that $G_f^{\text{eff}} \cong G_f$.) Next, we use the general theory of invertible fermionic phases with internal symmetries developed in Ref.~\cite{barkeshli2021invertible} to convert the problem of finding $\mathcal{L}$ into a calculation in group cohomology.

\subsubsection{Group cohomology definitions}
The following calculations make liberal use of results from group cohomology. An introduction aimed at applications to topological phases of matter can be found, for example, in the appendices of Refs.~\cite{Chen2013} (general introduction and many examples),~\cite{manjunath2021cgt} (computation of several cohomology groups relevant to this work),~\cite{Manjunath2020fqh} (several general results, especially a development of spectral sequences),~\cite{barkeshli2021invertible} (cup products) and~\cite{manjunath2022mzm} (further computational tricks). 
 
 We begin by stating some results about the group cohomology of $G_b = \U(1)\times [\Z^2\rtimes\Z_M]$. First we have the result
\begin{align}
    \H^2(G_b,\Z) &\cong \H^2(\U(1),\Z) \times \H^2(\Z_M,\Z) \times \H^1(\Z_M,\H^1(\Z^2,\Z)) \times \H^0(\Z_M,\H^2(\Z^2,\Z)) \\
    &\cong \Z \times \Z_{M}\times K_M \times \Z
\end{align}
where $K_M := \frac{\Z^2}{(1-U({\bf h}))\Z^2}$ and ${\bf h}$ is the generator of $\Z_M$, which rotates vectors in $\Z^2$ as specified by the matrix $U({\bf h})$. Thus, elements of $K_M$ are given by integer vectors modulo an equivalence relation. We find that $K_2 \cong \Z_2\times \Z_2, K_3 \cong \Z_3, K_4 \cong \Z_2, K_6 \cong \Z_1$. This group cohomology result can be proven using a spectral sequence decomposition, as argued in Ref.~\cite{manjunath2021cgt}, and can be verified numerically using the GAP program. 

We denote the generators of this group as 
$$[\zz_2]\in \H^2(\U(1),\Z), \quad [\hh_2]\in \H^2(\Z_M,\Z), \quad [\vec{\tt}_2] = ([\tt_{2,x}],[\tt_{2,y}]) \in \H^1(\Z_M,\H^1(\Z^2,\Z)), \quad [\aa_2]\in \H^0(\Z_M,\H^2(\Z^2,\Z)).$$
(Here $[x]$ is the cohomology class corresponding to the cocycle representative $x$.)
Note that in this case, $\H^2(G_b,\Z_2) \cong \H^2(G_b,\Z) \otimes \Z_2 \cong \Z_2 \times \Z_{(M,2)} \times (K_M \otimes \Z_2) \times \Z_2$. This means that any $\Z_n$ or $\Z$ factor of $\H^2(G_b,\Z)$ reduces to a $\Z_{(n,2)}$ or $\Z_2$ factor of $\H^2(G_b,\Z_2)$, respectively. The symbol $(m,n)$ denotes the greatest common divisor of $m$ and $n$.  

For our choice of $G_f$, we moreover have
\begin{equation}
    \omega_2 = \zz_2 \mod 2,
\end{equation}
indicating that the $\U(1)$ subgroup of $G_b$ nontrivially extends fermion parity, but the spatial part of $G_b$ does not. This corresponds to the convention that for every $\OO$ in the unit cell, $(\Cmop)^{\MO} = +1$.  

Next, we discuss cohomology in degree 4 with integer coefficients. Note that
\begin{align}
    \H^4(G_b,\Z) &\cong \H^4(\U(1),\Z) \times \H^2(\Z^2\rtimes\Z_M,\Z) \times \H^4(\Z^2\rtimes\Z_M,\Z) \\
    &\cong \Z \times [\Z \times \Z_M \times K_M] \times [\Z_M^2 \times K_M].
\end{align}
We consider the three factors in the decomposition separately. The generators of this group can be expressed in terms of cup products of the generators of $\H^2(G_b,\Z_2)$ discussed above (see App. B of Ref.~\cite{barkeshli2021invertible} for a throrough introduction to cup products and their properties). The first factor is generated by the cocycle representative $\zz_2 \cup \zz_2 \equiv \zz_2^2$. The second factor is generated by the cocycles $\zz_2 \aa_2 ~(\Z), \zz_2 \hh_2 ~(\Z_M), \{\zz_2\tt_{2,x},\zz_2\tt_{2,y}\} ~(K_M)$. The third factor is generated by the cocycles $\hh_2^2~ (\Z_M), \aa_2 \hh_2 ~(\Z_M), \{\hh_2\tt_{2,x},\hh_2\tt_{2,y}\} (K_M)$. 

Note that $\H^3(G_b,\U(1)) \cong \H^4(G_b,\Z)$. In Ref.~\cite{manjunath2021cgt}, a concrete parametrization was obtained for representative cocycles $[\lambda_3] \in \H^3(G_b,\U(1))$, and we rewrite it here for convenience. First define $\zz_1,\hh_1$ as the generators of $\H^1(\U(1),\U(1)),\H^1(\Z_M,\U(1))$ respectively. Assume that the $\U(1)$ coefficient module is defined as $\R/\Z$. Then, we have \cite{manjunath2021cgt}
\begin{equation}\label{eq:lambda3}
    \lambda_3 = k_1 \zz_1 \zz_2 + k_2 \hh_1 \zz_2 + k_3 \hh_1 \hh_2 + \vec{k}_4 \cdot \zz_1 \vec{\tt}_2 + \vec{k}_5 \cdot \hh_1 \vec{\tt}_2 + k_6 \zz_1 \aa_2 + k_7 \hh_1 \aa_2 \mod 1.
\end{equation}
Here we have suppressed the dependence of the coefficients on the origin $\OO$, although this dependence will be made explicit in the final result. The fact that $d\lambda_3 = 0 \mod 1$ can be seen by observing that $d\zz_1 = \zz_2$ and $d\hh_1 = \hh_2$; these follow from the fact that the coboundary $d$ in this case is the Bockstein map for the short exact sequence
$$ 1 \rightarrow \U(1) \rightarrow \R \rightarrow \Z \rightarrow 1.$$

Finally, it will be useful to define cocycle representatives for the group $\H^1(G_b,\Z_2)$. This group is isomorphic to $K_M \times \Z_{2}$ for even $M$, and is trivial for $M=3$. When $M$ is even, the first factor is generated by $2\tt_{1,x} \mod 2,2\tt_{1,y} \mod 2$, where $d\vec{\tt}_1 = \vec{\tt}_2 \mod 2$ by definition. The second factor is generated by $M \hh_1 \mod 2$. 

\subsubsection{Fermionic crystalline equivalence principle}\label{app:fcep}

Let us discuss the first step in finding $\mathcal{L}$, using the formulation of the fCEP in Ref.~\cite{manjunath2022mzm}. Applied to our case, that result simply states that $G_f^{\text{eff}}$ is also a group extension of $G_b$ by $\Z_2^f$ corresponding to a cocycle $[\omega^{\text{eff}}_2] \in \H^2(G_b,\Z_2)$, and in particular
\begin{equation}\label{eq:omeffdef}
    \omega^{\text{eff}}_2 = \om + \hh_2 \mod 2 = \zz_2 + \hh_2 \mod 2.
\end{equation}
A heuristic justification of the principle can be found in Ref.~\cite{manjunath2022mzm}. Although the fCEP has been discussed in several prior works and verified in many specific examples, we are not aware of a general proof. 

\subsubsection{Definition of $G_b$ gauge field}

In Refs.~\cite{manjunath2021cgt,Manjunath2020fqh,zhang2022fractional}, the $G_b$ gauge field was defined as $B = (A_b,\vec{R},\omega)$, where $A_b \sim A_b + 2\pi, \omega \sim \omega + 2\pi$. Here we will instead define $\tilde{X} := X/2\pi$ for $X = B, A_b,\vec{R},\omega$. We use tilded gauge fields in this appendix so that we have a more convenient normalization for $\nu_3$ and $\mathcal{L}$. But we will stick to the usual condensed matter notation without tildes in the main text.

Since our system has $\U(1)^f$ charge conservation symmetry, it cannot host unpaired Majorana zero modes \cite{manjunath2022mzm}. In this case we can obtain $\mathcal{L}$ by setting 
$$\mathcal{L} = 2\pi \tilde{B}^*\nu_3$$
where $*$ is the pullback operation and we are viewing the background gauge field $\tilde{B}$ here as a map from the space-time manifold to the classifying space $BG_b$. The parameter $\nu_3 \in C^3(G_b,\mathbb{R}/\Z)$ is a 3-cochain which will be determined in terms of group cohomology data. Note that it is unclear if and to what extent this relation holds when the system has unpaired Majorana zero modes, that is when $n_1 \ne 0$ in the formalism of Ref.~\cite{barkeshli2021invertible}.

Let us parameterize elements of $G_b$ as ${\bf g} = (z,\vec{r},h)$ where $z \in \RR/\Z \cong \U(1),\vec{r} \in \Z^2, h \in \Z/M\Z \cong \Z_M$. On a 3-manifold $\mathcal{M}^3$ with a triangulation, we then define the flat $G_b$ gauge field
\begin{equation}
    \tilde{B} = (\tilde{A}_b,\vec{\tilde{R}},\tilde{\omega}).
\end{equation}
The component $\tilde{A}_b$ is real valued, with $\tilde{A}_b \sim \tilde{A}_b + 1$. It is a gauge field for the bosonic $\U(1)$ symmetry, which is defined as $\U(1)^f/\Z_2^f$ (note that the fermion has charge 1/2 under this group). $\vec{\tilde{R}} = (\tilde{X},\tilde{Y})$ is integer valued. Finally $\tilde{\omega}$ is valued in multiples of $1/M$, with $\tilde{\omega} \sim \tilde{\omega} + 1$. 

For a flat gauge field $\tilde{B}$, the fluxes of $\tilde{B}$ can be obtained by pulling back representative cocycles of the the previously defined cohomology classes that generate $\H^2(G_b,\Z)$: 
\begin{equation}
    d\tilde{A}_b = \tilde{B}^* \zz_2; \quad d\tilde{\omega} = \tilde{B}^* \hh_2; \quad (1-U\left(2\pi/M\right))^{-1} d\tilde{\vec{R}} = \vec{\tt}_2; \quad \tilde{A}_{XY} = \tilde{B}^* \aa_2.
\end{equation}
Here $2\pi d\tilde{A}_b$ is interpreted as magnetic flux. $2\pi d\tilde{\omega}$ is interpreted as disclination flux: the integral of  $2\pi d\tilde{\omega}$ over a region $W$ measures the total disclination angle of defects within $W$. $(1-U\left(\frac{2\pi}{M}\right))^{-1} d\vec{\tilde{R}}$ is interpreted as disclocation flux: the integral of  this quantity within $W$ measures the total Burgers vector of defects within $W$, up to conjugation by elements in $G_b$ \cite{manjunath2021cgt}. $\tilde{A}_{XY}$ has the physical interpretation of an area element, whose integral over a region $W$ counts the number of unit cells in $W$. 

\def\Pmc{\mathcal{P}}
\subsubsection{General condition on $\mathcal{L}$}
The condition below is derived using the general theory of invertible fermionic phases developed in Ref.~\cite{barkeshli2021invertible}. We wish to write down a topological action $\mathcal{L}$ for invertible fermionic phases with (unitary) symmetry $G_f^{\text{eff}}$ and chiral central charge $c_-$. Since the symmetries we consider contain $\U(1)^f$ charge conservation as a subgroup, a number of simplifications occur:
\begin{enumerate}
    \item The chiral central charge $c_-$ must be an integer.
    \item A system with $\U(1)^f$ charge conservation symmetry cannot host unpaired Majorana zero modes at symmetry defects. 
\end{enumerate}
If we work out the theory with these simplifications, we find that 
\begin{align}
     \mathcal{L} &= 2\pi\tilde{B}^* \nu_3, \\
    d\nu_3 &= \mathcal{O}_4 = \frac{1}{2} n_2 \cup (n_2 + \omega_2^{\text{eff}}) + \frac{c_-}{8} \omega_2^{\text{eff}} \cup \omega_2^{\text{eff}}\mod 1 \label{eq:O4eq}
\end{align}
where $n_2$ is a general element of $\H^2(G_b,\Z_2)$, that is, $dn_2 = 0 \mod 2$.\footnote{In the general form of Eq.~\eqref{eq:O4eq}, there is an extra term $\frac{c_-}{8} \omega_2^{\text{eff}} \cup_1 d\omega_2^{\text{eff}}$. Moreover, the equation for $n_2$ is more generally $dn_2 = c_- \omega_2^{\text{eff}} \cup_1 \omega_2^{\text{eff}} \mod 2$. However, for the symmetries we consider, the terms with $\cup_1$ products turn out to vanish. Moreover, when we write $\frac{c_-}{8} \omega_2^{\text{eff}} \cup \omega_2^{\text{eff}}$ in Eq.~\eqref{eq:O4eq}, what we mean by $\omega_2^{\text{eff}}$ is strictly speaking a class in $\H^2(G_b,\Z)$ whose mod 2 reduction equals $\omega_2^{\text{eff}}$.} In terms of the generators of $\H^2(G_b,\Z_2)$, the most general form of $n_2$ is
\begin{equation}\label{eq:n2def}
    n_2 = q \zz_2 + s \hh_2 + \vec{t} \cdot \vec{\tt}_2 + m \aa_2.
\end{equation}
Note that in the general theory, $n_2$ and $n_2 + \omega_2^{\text{eff}}$ are equivalent \cite{barkeshli2021invertible}; this reflects the fact that we can relabel the fermion parity defects without changing any physical properties. Using this, we set $q=0$ without loss of generality. By substituting Eqs.~\eqref{eq:omeffdef},~\eqref{eq:n2def} into Eq.~\eqref{eq:O4eq}, and simplifying, we find
\begin{equation}\label{eq:dnu3}
    d\nu_3 = \frac{1}{2}( t_x \tt_{2,x}^2 + t_y \tt_{2,y}^2 + m \aa_2^2 + \hh_2 (t_x \tt_{2,x} + t_y \tt_{2,y} + m \aa_2)) + \frac{1}{2}\zz_2(s \hh_2 + t_x \tt_{2,x} + t_y \tt_{2,y} + m \aa_2)+\frac{c_-}{8} (\zz_2^2 + 2 \zz_2 \hh_2 + \hh_2^2).
\end{equation}
Here we have suppressed the cup product notation for conciseness. We also used the fact that for 2-cocycles $\alpha_2,\beta_2$, $\frac{1}{2}\alpha_2 \beta_2 = \frac{1}{2}\beta_2 \alpha_2 \mod 1$ up to coboundaries; this simplifies many of the mixed terms. 

Eq.~\eqref{eq:dnu3} is the main equation we will consider separately for $M = 2,3,4,6$. It can be further simplified using relations between 4-cocycles that we will specify below; this means that the various terms in Eq.~\eqref{eq:dnu3} are not independent, but rather there exist relations between them which depend on the value of $M$.

We can integrate this equation to find $\nu_3$. Note that there is a constant of integration $\lambda_3$ satisfying $d\lambda_3 = 0 \mod 1$; this corresponds precisely to a $G_b$ bosonic SPT phase classified by the group $\H^3(G_b,\U(1))$. Therefore the expression for $\nu_3$ should include a generic contribution from a $G_b$ bosonic SPT, represented by the cocycle in Eq.~\eqref{eq:lambda3}. We will see this explicitly below.

Using the definition $\mathcal{L} = 2\pi \tilde{B}^* \nu_3$, we obtain an expression in terms of $\tilde{A}_b$ and the other gauge field components. However, this is unsatisfactory because the physical system is described more naturally by the $\U(1)^f$ gauge field $\tilde{A} = \tilde{A}_b/2$, under which the fermion has charge 1. (Note that the correctly normalized vector potential for the system is $A = 2\pi\tilde{A}$.) The final step is thus to replace $\tilde{A}_b \rightarrow 2\tilde{A}$. 

\subsection{Calculations for specific $M$}\label{app:spM}

\subsubsection{$M=4$}
When $M=4$, we find that $[\tt_{2,x}] = [\tt_{2,y}] = [\tt_2]$. Moreover, one can show the following relations:
\begin{equation}
    [\aa_2^2] = [\hh_2 \aa_2] \mod 2, \quad [\tt_{2}^2] = [\tt_2 \hh_2] \mod 2.
\end{equation}
These can either be formally derived using properties of the cohomology ring of $G_b$, or explicitly verified using Mathematica by calculating a finite number of cohomology invariants associated to $\H^4(G_b,\Z_2)$. Now, defining $t = t_x + t_y \mod 2$, we can simplify Eq.~\eqref{eq:dnu3} as follows:
\begin{equation}
    d\nu_3 = \frac{1}{2} \zz_2 (s \hh_2 + t \tt_2 + m \aa_2) + \frac{c_-}{8}(\zz_2^2 + \zz_2 \hh_2 + \hh_2^2) \mod 1.
\end{equation}
The full expression obtained after integrating $d\nu_3$ and simplifying is
\begin{align}\label{eq:nu3_C4}
    \nu_3 &= \left(\frac{c_-}{8}+k_1\right)\zz_1 \zz_2 + \left(\frac{c_-}{4}+\frac{[s]_2}{2}+k_2\right)\hh_1\zz_2 +  \left(\frac{c_-}{8}+k_3\right)\hh_1\hh_2 \nonumber \\ &\quad+  \left(\frac{[t]_2}{2}+k_4\right)\zz_1 \tt_2 + k_5\hh_1\tt_2 + \left(\frac{[m]_2}{2}+k_6\right)\zz_1 \aa_2 + k_7\hh_1\aa_2 \mod 1,
\end{align}
where
\begin{equation}\label{eq:kidef_C4}
    k_1, k_6 \in \Z; \quad k_2, k_3, k_7 \in \Z_4; \quad k_4, k_5 \in \Z_2.
\end{equation}
The coefficients $s,t,m, k_i$ all in principle have a dependence on the origin $\OO$, which we have suppressed here but will restore at the end of the calculation. Note that $\tilde{B}^*\zz_1 = \tilde{A}_b, \tilde{B}^*\hh_1 = \tilde{\omega},\tilde{B}^*\tt_2 = \frac{1}{2}(1,1) \cdot d\tilde{\vec{R}}$. Therefore
\begin{align}
    &\tilde{B}^*\nu_3 = \left(\frac{c_-}{8}+k_1\right)\tilde{A}_b \cup d\tilde{A}_b + \left(\frac{c_-}{4}+\frac{[s]_2}{2}+k_2\right)\tilde{\omega} \cup d\tilde{A}_b +  \left(\frac{c_-}{8}+k_3\right)\tilde{\omega} \cup d\tilde{\omega} \nonumber \\ &\quad+  \frac{(1,1)}{2} \cdot \left(\frac{[t]_2}{2}+k_4\right)\tilde{A}_b \cup d\vec{\tilde{R}} + \frac{(1,1)}{2} \cdot k_5\tilde{\omega} \cup d\vec{\tilde{R}} + \left(\frac{[m]_2}{2}+k_6\right)\tilde{A}_b \cup  \tilde{A}_{XY} + k_7\tilde{\omega} \cup \tilde{A}_{XY} \mod 1.
\end{align}
Next, as explained above, we replace $\tilde{A}_b = 2\tilde{A}$, where $\tilde{A}$ is a $\U(1)^f$ gauge field, to obtain 
\begin{align}
    &\tilde{B}^*\nu_3 = \frac{c_- + 8 k_1}{2}\tilde{A} \cup d\tilde{A} + \left(\frac{c_-}{2}+[s]_2+2k_2\right)\tilde{\omega} \cup d\tilde{A}_b +  \left(\frac{c_-}{8}+k_3\right)\tilde{\omega} \cup d\tilde{\omega} \nonumber \\ &\quad+  \frac{(1,1)}{2} \cdot ([t]_2 + 2 k_4)\tilde{A} \cup d\vec{\tilde{R}} + \frac{(1,1)}{2} \cdot k_5\tilde{\omega} \cup d\vec{\tilde{R}} + ([m]_2 + 2k_6)\tilde{A} \cup  \tilde{A}_{XY} + k_7\tilde{\omega} \cup \tilde{A}_{XY} \mod 1.
\end{align}
Finally, we multiply by $2\pi$ to obtain $\mathcal{L} = 2\pi \tilde{B}^* \nu_3$ in terms of the tilded gauge fields, and then replace each $\tilde{X}$ with $X/2\pi$. The resulting action (with $\OO$ subscripts restored) is
\begin{align}
    &\mathcal{L} = \frac{c_- + 8 k_1}{4\pi}A \cup dA + \left(\frac{c_-}{2}+[s_{\OO}]_2+2k_{2,\OO}\right)\frac{\omega \cup dA}{2\pi} +  \left(\frac{c_-}{4}+2k_{3,\OO}\right)\frac{\omega \cup d\omega}{4\pi} \nonumber \\ &\quad+  \frac{(1,1)}{2} \cdot ([t_{\OO}]_2 + 2 k_{4,\OO})\frac{A \cup d\vec{R}}{2\pi} + \frac{(1,1)}{2} \cdot k_{5,\OO}\frac{\omega \cup d\vec{R}}{2\pi} + ([m]_2 + 2k_6)\frac{A \cup A_{XY}}{2\pi} + k_7\frac{\omega \cup A_{XY}}{2\pi}  \mod 2\pi.
\end{align}
$m,k_6, k_7$ set the charge and angular momentum per unit cell, and are independent of $\OO$.

Using the quantization of $k_i$ obtained in Eq.~\eqref{eq:kidef_C4}, we can define the coefficients of the action as given in Eq.~\eqref{eq:coeffs_C4}. This completes the derivation of the effective action prior to modding out by equivalences.

From the general theory outlined in Ref.~\cite{barkeshli2021invertible}, once $n_2$ is gauge fixed (by setting $q=0$), the only remaining equivalences are on $\nu_3$ and are given by
\begin{equation}
    \nu_3 \simeq \nu_3 + \frac{1}{2}\chi_1 \cup \omega_2^{\text{eff}},
\end{equation}
where $\chi_1$ runs over the group $\H^1(G_b,\Z_2)$. These equivalences correspond to relabelling $G_b$ defects (in this case, disclination and dislocation defects as determined by the choice of $\chi_1$) by fermions. A general form for $\chi_1$ is
\begin{equation}
    \chi_1 = a (4\hh_1) + b (2\tt_1) \mod 2
\end{equation}
where we define $\tt_1 \in C^1(G_b,\R/\Z)$ so that $d\tt_1 = \tt_2 \mod 2$. Therefore,
\begin{align}
    \nu_3 &\simeq \nu_3 + \frac{1}{2}(4 a \hh_1 + 2 b \tt_1) (\zz_2 + \hh_2) \nonumber \\
    &\simeq \nu_3 + 2 a \hh_1 (\zz_2 + \hh_2) + b (\zz_1+\hh_1) \tt_2 \mod 1.
\end{align}
Here we used the fact that $\zz_1 \tt_2 = \tt_1 \zz_2$ and $\hh_1 \tt_2 = \tt_1 \hh_2$ up to coboundaries. If we now attribute the change in $\nu_3$ to a change in the coefficients $k_i$ (which is reasonable because the relabelling does not change any fermion quantum numbers, and can be thought of as stacking a bosonic SPT), we obtain Eq.~\eqref{eq:relabel_C4}.

\subsubsection{$M=2$}
When $M=2$, we verify using Mathematica that
\begin{equation}
    [\aa_2^2] = [\hh_2 \aa_2], \quad [\tt_{2,i}^2] = [\tt_{2,i} \hh_2], \quad i = x,y.
\end{equation}
We can simplify Eq.~\eqref{eq:dnu3} as follows:
\begin{equation}
    d\nu_3 = \frac{1}{2} \zz_2 (s \hh_2 + \vec{t} \cdot \vec{\tt}_2 + m \aa_2) + \frac{c_-}{8}(\zz_2^2 + \zz_2 \hh_2 + \hh_2^2) \mod 1.
\end{equation}
The full expression for $\nu_3$ is
\begin{align}\label{eq:nu3_C2}
    \nu_3 &= \left(\frac{c_-}{8}+k_1\right)\zz_1 \zz_2 + \left(\frac{c_-}{4}+\frac{[s]_2}{2}+k_2\right)\hh_1\zz_2 +  \left(\frac{c_-}{8}+k_3\right)\hh_1\hh_2 \nonumber \\ &\quad+  \left(\frac{[\vec{t}]_2}{2}+\vec{k}_4\right) \cdot \zz_1 \vec{\tt}_2 + \vec{k}_5 \cdot \hh_1\vec{\tt}_2 + \left(\frac{[m]_2}{2}+k_6\right)\zz_1 \aa_2 + k_7\hh_1\aa_2 \mod 1,
\end{align}
where
\begin{equation}\label{eq:kidef_C2}
    k_1, k_6 \in \Z; \quad k_2, k_3, k_{4,i}, k_{5,i}, k_7 \in \Z_2 .
\end{equation}
The coefficients $s,t,m, k_i$ all in principle have a dependence on the origin $\OO$, which we have suppressed here but will restore at the end of the calculation. Note that $\tilde{B}^*\zz_1 = \tilde{A}_b. \tilde{B}^*\hh_1 = \tilde{\omega},\tilde{B}^*\vec{\tt}_2 = \frac{1}{2}\cdot d\tilde{\vec{R}}$. Therefore
\begin{align}
    &\tilde{B}^*\nu_3 = \left(\frac{c_-}{8}+k_1\right)\tilde{A}_b \cup d\tilde{A}_b + \left(\frac{c_-}{4}+\frac{[s]_2}{2}+k_2\right)\tilde{\omega} \cup d\tilde{A}_b +  \left(\frac{c_-}{8}+k_3\right)\tilde{\omega} \cup d\tilde{\omega} \nonumber \\ &\quad+  \frac{1}{2}\left(\frac{[\vec{t}]_2}{2}+\vec{k}_4\right) \cdot \tilde{A}_b \cup d\vec{\tilde{R}} + \frac{1}{2} \vec{k}_5 \cdot\tilde{\omega} \cup d\vec{\tilde{R}} + \left(\frac{[m]_2}{2}+k_6\right)\tilde{A}_b \cup  \tilde{A}_{XY} + k_7\tilde{\omega} \cup \tilde{A}_{XY} \mod 1.
\end{align}
Replacing $\tilde{A}_b = 2\tilde{A}$ gives 
\begin{align}
    &\tilde{B}^*\nu_3 = \frac{c_- + 8 k_1}{2}\tilde{A} \cup d\tilde{A} + \left(\frac{c_-}{2}+[s]_2+2k_2\right)\tilde{\omega} \cup d\tilde{A}_b +  \left(\frac{c_-}{8}+k_3\right)\tilde{\omega} \cup d\tilde{\omega} \nonumber \\ &\quad+  \frac{1}{2} ([\vec{t}]_2 + 2\vec{k}_4) \cdot \tilde{A} \cup d\vec{\tilde{R}} + \frac{1}{2} \vec{k}_5 \cdot \tilde{\omega} \cup d\vec{\tilde{R}} + ([m]_2 + 2k_6)\tilde{A} \cup  \tilde{A}_{XY} + k_7\tilde{\omega} \cup \tilde{A}_{XY} \mod 1.
\end{align}
Finally, we replace each $\tilde{X}$ with $X/2\pi$ and write the resulting action (with $\OO$ subscripts restored):
\begin{align}
    &\mathcal{L} = \frac{c_- + 8 k_1}{4\pi}A \cup dA + \left(\frac{c_-}{2}+[s_{\OO}]_2+2k_{2,\OO}\right)\frac{\omega \cup dA}{2\pi} +  \left(\frac{c_-}{4}+2k_{3,\OO}\right)\frac{\omega \cup d\omega}{4\pi} \nonumber \\ &\quad+  \frac{1}{2} (\vec{t}_{\OO}]_2 + 2 \vec{k}_{4,\OO}) \cdot \frac{A \cup d\vec{R}}{2\pi} + \frac{1}{2} \vec{k}_{5,\OO} \cdot \frac{\omega \cup d\vec{R}}{2\pi} + ([m]_2 + 2k_6)\frac{A \cup A_{XY}}{2\pi} + k_7\frac{\omega \cup A_{XY}}{2\pi}  \mod 2\pi.
\end{align}
Here $m,k_6, k_7$ are independent of $\OO$. Using Eq.~\eqref{eq:kidef_C2}, we can define the coefficients of the action as given in Eq.~\eqref{eq:coeffs_C2}. 

This completes the derivation of the effective action prior to modding out by equivalences. The remaining equivalences on $\nu_3$ are given by
\begin{equation}
    \nu_3 \simeq \nu_3 + \frac{1}{2}\chi_1 \cup \omega_2^{\text{eff}},
\end{equation}
where $\chi_1$ runs over $\H^1(G_b,\Z_2)$. A general form for $\chi_1$ is
\begin{equation}
    \chi_1 = a (2\hh_1) + \vec{b} \cdot (2\vec{tt}_1) \mod 2
\end{equation}
where we define $\tt_{1,i} \in C^1(G_b,\R/\Z)$ for $i=x,y$ so that $d\vec{\tt}_1 = \vec{\tt}_2 \mod 2$. Therefore,
\begin{align}
    \nu_3 &\simeq \nu_3 + \frac{1}{2}(2 a \hh_1 + 2 b \tt_1) (\zz_2 + \hh_2) \nonumber \\
    &\simeq \nu_3 + a \hh_1 (\zz_2 + \hh_2) + \vec{b} \cdot (\zz_1+\hh_1) \vec{\tt}_2 \mod 1.
\end{align}
Attributing this change in $\nu_3$ to the coefficients $k_i$ as before, we obtain Eq.~\eqref{eq:relabel_C2}.

\subsubsection{$M=3$}\label{app:cdep_M=3}

When $M=3$, we need to set $\vec{\tt}_2 \mod 2$ to be trivial in $\H^2(G_b,\Z_2)$, although $\vec{\tt}_2$ generates a $\Z_3$ factor in $\H^2(G_b,\Z)$. We also need to set $s = 0$, because $\hh_2 \mod 2$ is trivial in $\H^2(G_b,\Z_2)$. Furthermore, we can verify using Mathematica that
\begin{equation}
    [\aa_2^2] = [\hh_2 \aa_2].
\end{equation}
Therefore we can simplify Eq.~\eqref{eq:dnu3} as follows:
\begin{equation}
    d\nu_3 = \frac{1}{2} \zz_2 ( m \aa_2) + \frac{c_-}{8}(\zz_2^2 + \zz_2 \hh_2 + \hh_2^2) \mod 1.
\end{equation}
There is a subtle point regarding the term proportional to $c_-$. We have $\omega_2^{\text{eff}} = \zz_2 + \hh_2 \mod 2$, and the last term is of the form $\frac{c_-}{8} \omega_2^{\text{eff}} \cup \omega_2^{\text{eff}}$. But we stated above that when $M=3$, the cocycle $\hh_2 \mod 2$ is trivial in $\H^2(G_b,\Z_2)$. Therefore it may seem that we can simply use $\omega_2^{\text{eff}} = \zz_2 \mod 2$ instead and rewrite the last term as $\frac{c_-}{8} \zz_2^2$. However, this is not correct, for multiple reasons. First, as we noted above, when we write $\frac{c_-}{8} \omega_2^{\text{eff}} \cup \omega_2^{\text{eff}}$, we really mean an integral lift of $\omega_2^{\text{eff}}$. In this case, the correct integral lift is not just $\zz_2$, but $\zz_2 + \hh_2$, as a result of the fCEP. A more physical reason is that a system with $M=6$ (see below) is a special case of a system with $M=3$, and therefore if the coefficients $\SO, \ell_{\OO}$ depend on $c_-$ when $M=6$, they should have a similar dependence on $c_-$ when $M=3$. The only way to achieve this is to take the integral lift of $\omega_2^{\text{eff}}$ to be $\zz_2 + \hh_2$ and not $\zz_2$.

The full expression for $\nu_3$ is
\begin{align}\label{eq:nu3_C3}
    \nu_3 &= \left(\frac{c_-}{8}+k_1\right)\zz_1 \zz_2 + \left(\frac{c_-}{4}+k_2\right)\hh_1\zz_2 +  \left(\frac{c_-}{8}+k_3\right)\hh_1\hh_2 \nonumber \\ &\quad+  \vec{k}_4 \cdot \zz_1 \vec{\tt}_2 + \vec{k}_5 \cdot \hh_1\vec{\tt}_2 + \left(\frac{[m]_2}{2}+k_6\right)\zz_1 \aa_2 + k_7\hh_1\aa_2 \mod 1,
\end{align}
where
\begin{equation}\label{eq:kidef_C3}
    k_1, k_6 \in \Z; \quad  \vec{k}_{4} := k_4(1,2);  \vec{k}_{5} := k_5(1,2); \quad k_2, k_3, k_4,k_5,k_7 \in \Z_3.
\end{equation}
The coefficients $k_i$ all in principle have a dependence on the origin $\OO$, which we have suppressed here but will restore at the end of the calculation. Now using $\tilde{B}^*\zz_1 = \tilde{A}_b,\tilde{B}^*\hh_1 = \tilde{\omega},\tilde{B}^*\vec{\tt}_2 = \frac{1}{3} d\tilde{\vec{R}}$. Therefore
\begin{align}
    &\tilde{B}^*\nu_3 = \left(\frac{c_-}{8}+k_1\right)\tilde{A}_b \cup d\tilde{A}_b + \left(\frac{c_-}{4}+k_2\right)\tilde{\omega} \cup d\tilde{A}_b +  \left(\frac{c_-}{8}+k_3\right)\tilde{\omega} \cup d\tilde{\omega} \nonumber \\ &\quad+  \frac{1}{3}(1,2) \cdot\ k_4 \tilde{A}_b \cup d\vec{\tilde{R}} + \frac{1}{3}(1,2)\cdot k_5 \tilde{\omega} \cup d\vec{\tilde{R}} + \left(\frac{[m]_2}{2}+k_6\right)\tilde{A}_b \cup  \tilde{A}_{XY} + k_7\tilde{\omega} \cup \tilde{A}_{XY} \mod 1.
\end{align}
Replacing $\tilde{A}_b = 2\tilde{A}$ gives 
\begin{align}
    &\tilde{B}^*\nu_3 = \frac{c_- + 8 k_1}{2}\tilde{A} \cup d\tilde{A} + \left(\frac{c_-}{2}+2k_2\right)\tilde{\omega} \cup d\tilde{A}_b +  \left(\frac{c_-}{8}+k_3\right)\tilde{\omega} \cup d\tilde{\omega} \nonumber \\ &\quad+  \frac{k_4}{3}(1,2) \cdot  \tilde{A} \cup d\vec{\tilde{R}} +\frac{k_5}{3}(1,2)\cdot  \tilde{\omega} \cup d\vec{\tilde{R}} + ([m]_2 + 2k_6)\tilde{A} \cup  \tilde{A}_{XY} + k_7\tilde{\omega} \cup \tilde{A}_{XY} \mod 1.
\end{align}
Finally, we replace each $\tilde{X}$ with $X/2\pi$ and write the resulting action (with $\OO$ subscripts restored):
\begin{align}
    &\mathcal{L} = \frac{c_- + 8 k_1}{4\pi}A \cup dA + \left(\frac{c_-}{2}+2k_{2,\OO}\right)\frac{\omega \cup dA}{2\pi} +  \left(\frac{c_-}{4}+2k_{3,\OO}\right)\frac{\omega \cup d\omega}{4\pi} \nonumber \\ &\quad+  \frac{k_4}{3}(1,2) \cdot  \frac{A \cup d\vec{R}}{2\pi} + \frac{k_5}{3}(1,2)\cdot \frac{\omega \cup d\vec{R}}{2\pi} + ([m]_2 + 2k_6)\frac{A \cup A_{XY}}{2\pi} + k_7\frac{\omega \cup A_{XY}}{2\pi}  \mod 2\pi.
\end{align}
Here $m,k_6, k_7$ are independent of $\OO$. Using Eq.~\eqref{eq:kidef_C3}, we can define the coefficients of the action as given in Eq.~\eqref{eq:coeffs_C3}. 

We need to make a final point about the quantization of $\SO$ and $\ell_{\OO}$. Note that the terms $\frac{M}{2\pi} \omega dA$ and $\frac{M}{4\pi} \omega d\omega$ give a trivial partition function on any closed 3-manifold. A quick check is to see that the integrals $\int \frac{dA}{2\pi},\int \frac{d\omega}{4\pi}$ are integer quantized over any closed 2-manifold (due to the quantization of flux and spatial curvature respectively), while $\int M \omega$ is a multiple of $2\pi$ on any 1-manifold. Physically, we can see that these terms are trivial because they assign integer charge and angular momentum to an elementary disclination. Therefore in Eq.~\eqref{eq:coeffs_C3}, $\SO$ and $\ell_{\OO}$ should be defined mod $3$ and not mod 6. 

The remaining equivalences on $\nu_3$ are given by
\begin{equation}
    \nu_3 \simeq \nu_3 + \frac{1}{2}\chi_1 \cup \omega_2^{\text{eff}},
\end{equation}
where $\chi_1$ runs over $\H^1(G_b,\Z_2)$. However, $\chi_1$ is trivial in this case, so we directly obtain Eq.~\eqref{eq:FTresult_C3}.
\subsubsection{$M=6$}

When $M=6$, we need to set $\vec{\tt}_2$ to be trivial in $\H^2(G_b,\Z)$. Therefore torsional terms depending on $d\vec{\tilde{R}}$ do not appear in the result at all. We also have
\begin{equation}
    [\aa_2^2] = [\hh_2 \aa_2],
\end{equation}
which can be verified using Mathematica. Therefore we can simplify Eq.~\eqref{eq:dnu3} as follows:
\begin{equation}
    d\nu_3 = \frac{1}{2} \zz_2 (s \hh_2 + m \aa_2) + \frac{c_-}{8}(\zz_2^2 + \zz_2 \hh_2 + \hh_2^2) \mod 1.
\end{equation}
The full expression for $\nu_3$ is
\begin{align}\label{eq:nu3_C6}
    \nu_3 &= \left(\frac{c_-}{8}+k_1\right)\zz_1 \zz_2 + \left(\frac{c_-}{4}+\frac{[s]_2}{2}+k_2\right)\hh_1\zz_2 +  \left(\frac{c_-}{8}+k_3\right)\hh_1\hh_2 \nonumber \\ &\quad + \left(\frac{[m]_2}{2}+k_6\right)\zz_1 \aa_2 + k_7\hh_1\aa_2 \mod 1,
\end{align}
where
\begin{equation}\label{eq:kidef_C6}
    k_1, k_6 \in \Z; \quad  \quad k_2, k_3, k_7 \in \Z_6.
\end{equation}
The coefficients $s, k_i$ all in principle have a dependence on the origin $\OO$, which we have suppressed here but will restore at the end of the calculation. Now using $\tilde{B}^*\zz_1 = \tilde{A}_b,\tilde{B}^*\hh_1 = \tilde{\omega},\tilde{B}^*\vec{\tt}_2 = \frac{1}{3} d\tilde{\vec{R}}$, we write
\begin{align}
    &\tilde{B}^*\nu_3 = \left(\frac{c_-}{8}+k_1\right)\tilde{A}_b \cup d\tilde{A}_b + \left(\frac{c_-}{4}+\frac{[s]_2}{2}+k_2\right)\tilde{\omega} \cup d\tilde{A}_b +  \left(\frac{c_-}{8}+k_3\right)\tilde{\omega} \cup d\tilde{\omega} \nonumber \\ &\quad+ \left(\frac{[m]_2}{2}+k_6\right)\tilde{A}_b \cup  \tilde{A}_{XY} + k_7\tilde{\omega} \cup \tilde{A}_{XY} \mod 1.
\end{align}
Replacing $\tilde{A}_b = 2\tilde{A}$ gives 
\begin{align}
    &\tilde{B}^*\nu_3 = \frac{c_- + 8 k_1}{2}\tilde{A} \cup d\tilde{A} + \left(\frac{c_-}{2}+[s]_2+2k_2\right)\tilde{\omega} \cup d\tilde{A}_b +  \left(\frac{c_-}{8}+k_3\right)\tilde{\omega} \cup d\tilde{\omega} \nonumber \\ &\quad + ([m]_2 + 2k_6)\tilde{A} \cup  \tilde{A}_{XY} + k_7\tilde{\omega} \cup \tilde{A}_{XY} \mod 1.
\end{align}
Finally, we replace each $\tilde{X}$ with $X/2\pi$ and write the resulting action (with $\OO$ subscripts restored):
\begin{align}
    &\mathcal{L} = \frac{c_- + 8 k_1}{4\pi}A \cup dA + \left(\frac{c_-}{2}+[s_{\OO}]_2+2k_{2,\OO}\right)\frac{\omega \cup dA}{2\pi} +  \left(\frac{c_-}{4}+2k_{3,\OO}\right)\frac{\omega \cup d\omega}{4\pi} \nonumber \\ &\quad + ([m]_2 + 2k_6)\frac{A \cup A_{XY}}{2\pi} + k_7\frac{\omega \cup A_{XY}}{2\pi}  \mod 2\pi.
\end{align}
Here $m,k_6, k_7$ are independent of $\OO$. Using Eq.~\eqref{eq:kidef_C6}, we can define the coefficients of the action as given in Eq.~\eqref{eq:coeffs_C6}. This completes the derivation of the effective action prior to modding out by equivalences.

The remaining equivalences on $\nu_3$ are given by
\begin{equation}
    \nu_3 \simeq \nu_3 + \frac{1}{2}\chi_1 \cup \omega_2^{\text{eff}},
\end{equation}
where $\chi_1$ runs over $\H^1(G_b,\Z_2)$. A general form for $\chi_1$ is
\begin{equation}
    \chi_1 = a (6\hh_1) \mod 2.
\end{equation}
Therefore,
\begin{align}
    \nu_3 &\simeq \nu_3 + a \hh_1 (\zz_2 + \hh_2) \mod 1.
\end{align}
This gives Eq.~\eqref{eq:relabel_C6}.
\subsection{The case with $\phi \ne 0$}\label{app:phinonzero}
Extending the derivation to any $\phi \ne 0$ turns out to be straightforward. The final result is that for any $M$,  the complete topological action is still given by Eq.~\eqref{eq:mainresponse} with the coefficients still quantized according to Eqs.~\eqref{eq:FTresult_C4},~\eqref{eq:FTresult_C2},~\eqref{eq:FTresult_C3},~\eqref{eq:FTresult_C6}. Here $A$ is understood as the full vector potential, so that $\int_W dA = \phi \int_W A_{XY} + \int_W \delta A$ gives the full magnetic flux within the system, including a uniform background contribution proportional to $\phi$, and an excess flux, as specified by $\delta A$. We explain this below.

We have $G_b = \U(1) \times_{2\phi} [\Z^2\rtimes \Z_M]$. The notation $\times_{2\phi}$ is used because the elementary boson, which is a bound state of two fermions, sees $2\phi$ flux per unit cell. Proceeding as in App.~\ref{app:details_genM}, we define a $G_b$ gauge field $\tilde{B} = (\delta \tilde{A}_b,\tilde{\vec{R}},\tilde{\omega})$. Crucially, $\delta \tilde{A}_b$ is not the full vector potential for the bosonic $\U(1)$ symmetry; it is the deviation of $\tilde{A}_b$ from a reference gauge field that assigns uniform flux $2\phi$ per unit cell. Since the fermion sees half the total flux seen by an elementary boson, the full vector potential $\tilde{A}$ is related to $\delta \tilde{A}_b$ as follows:
\begin{equation}\label{eq:dA_phinonzero}
    d\tilde{A} = \frac{1}{2} d\tilde{A}_b = \frac{1}{2}\left( d\delta \tilde{A}_b + \frac{2\phi}{2\pi} \tilde{A}_{XY}\right)
\end{equation}

The main steps for $\phi \ne 0$ only differ in terms of the definitions of $n_2,\om$. Note that $\zz_2 = d \zz_1$ is not longer an element of $\H^2(G_b,\Z)$ when $\phi \ne 0$; we need to replace this with the quantity $d\zz_1 + \frac{2\phi}{2\pi} \aa_2$, which is a valid cocycle representative. The other generators of $\H^2(G_b,\Z)$ remain unchanged.

If we then go through the calculation as before, we eventually obtain Eq.~\eqref{eq:dnu3} (the $\phi = 0$ result) with $\zz_2$ replaced by
$d\zz_1 + \frac{2\phi}{2\pi} \aa_2$. Therefore in the resulting field theory we should replace $\tilde{B}^* \zz_2 = d\delta \tilde{A}_b$ with
$$\tilde{B}^*(\zz_2 + \frac{2\phi}{2\pi} \aa_2) = d\delta \tilde{A}_b + \frac{2\phi}{2\pi} A_{XY}.$$ But from Eq.~\eqref{eq:dA_phinonzero}, this just equals $\tilde{B}^* (2d \tilde{A})$. Therefore, the dependence of $\mathcal{L}$ on $A$ is exactly the same for $\phi = 0$ and $\phi \ne 0$. Thus we have proven our main claim that Eq.~\eqref{eq:mainresponse} continues to hold, but $A$ is now understood as the full vector potential for the system. 

It is also straightforward to verify that the quantization of the response coefficients does not depend on $\phi$. Note, however, that some important properties which do not directly appear in Eq.~\eqref{eq:mainresponse} do acquire a $\phi$-dependence. For example, consider the filling per unit cell $\nu$. On a spatial torus $T^2$ with $N$ unit cells, the total charge predicted by the field theory is
    \begin{align}
        \nu N &= \int_{T^2} \frac{\delta \mathcal{L}}{ \delta A} = \frac{C}{2\pi} \int_{T^2} dA + \frac{\kappa}{2\pi} \int_{T^2} A_{XY} \\
        &= \left(\frac{C \phi }{2\pi} + \kappa\right)N.
    \end{align}
    Here we used $\int_{T^2} dA = \phi N$ $\int_{T^2} A_{XY} = 2\pi N$. Since $N$ is arbitrary, $\kappa$ must be related to the filling $\nu$ as follows:
    \begin{equation}
        \nu = \frac{C \phi }{2\pi} + \kappa.
    \end{equation}
    This relation was previously derived for Chern insulators using flux-threading arguments \cite{Lu2017fillingenforced}, in which the role of symmetry was not completely clear. The above derivation shows that it is a direct consequence of the magnetic translation symmetry in the problem.

\section{Computations involving $\Theta^{\pm}_{\OO}$}\label{app:RSI}

\subsection{Relation between $\Theta^{\pm}_{\OO}$ and $K_{\OO}^{\pm}$}\label{app:rel-Theta-K}

Here we explain why $K_{\OO}^{\pm}$, as defined by Eq.~\eqref{eq:Theta_def_K}, is not necessarily invariant upon changing the size of the partial rotation region $D$, but taking a suitable modular reduction gives a well-defined invariant $\Theta^{\pm}_{\OO}$.

First consider $\Theta^+_{\OO}$. Suppose we symmetrically increase the size of $D$ to include a new set of $\MO$ additional points which form a closed orbit under rotations about $\OO$. Since $(\Cmop)^{\MO} = +1$, the states defined at these points span a closed subspace under $\Cmop$ with rotation eigenvalues $e^{\frac{2\pi i j}{\MO}}$ for $j = 0, 1, \dots , \MO-1$. The phase of the expectation value of $\Cmop|_D$ will therefore change by the product of these eigenvalues, which equals $-1$ when $\MO$ is even and $+1$ when $\MO$ is odd. As a result, $K_{\OO}^+ \rightarrow K_{\OO}^+ + \frac{\MO}{2}$ for even $\MO$, and therefore only $K_{\OO}^+ \mod \frac{\MO}{2}$ is invariant.

Next consider $\Theta^-_{\OO}$. Again, increasing the size of $D$ introduces a new orbit of size $\MO$. However, since $(\Cmom)^{\MO} = (-1)^F$, the states defined at these points have eigenvalues $e^{\frac{2\pi i }{\MO}(j + \frac{1}{2})}$ for $j = 0, 1, \dots , \MO-1$. Now the product of these eigenvalues equals $+1$ when $\MO$ is even and $-1$ when $\MO$ is odd. Therefore, when $\MO$ is odd $K_{\OO}^- \mod \frac{\MO}{2}$ and $K^+_{\OO} \mod \MO$ are the right invariant quantities. This explains the modular reduction taken in Eq.~\eqref{eq:Theta_def_K}.

\subsection{Equations for $\Theta^{\pm}_{\OO,\text{LL}}$}
Here we explain how to derive Eq.~\eqref{eq:Theta_def_LL} for $c_->0$ and $c_-<0$. We use the subscript LL to refer to general $c_-$; for the lowest Landau level, as in the main text, we set $c_- = 1$. First we compute $\ell_{\OO,\text{LL}}^{\pm}$ in these two cases. Let us begin with the case $c_->0$. In this case, it is known that the topological effective action for $c_-$ lowest filled LLs in the continuum is given by
\begin{equation}\label{eq:EA_LL}
    \mathcal{L} = \sum\limits_{i=1}^{c_-} \left((A + s_i \omega) \wedge d(A + s_i \omega) - \frac{1}{48\pi} \omega \wedge d\omega\right),
\end{equation}
where $A$ is the usual vector potential, $\omega$ is an $\text{SO}(2)$ spin connection \cite{Gromov2014,Gromov2015}, and $s_i= i-\frac{1}{2}$. The coefficient of $\frac{1}{2\pi} A \wedge d\omega$ equals $c_-^2/2$, and is identified with $\mathscr{S}^+_{\text{LL}}$. The coefficient of $\frac{1}{4\pi} \omega \wedge d\omega$ which comes from the first term alone is identified with $\ell_{\OO,\text{LL}}^+$ (the term $-\frac{c_-}{48\pi} \omega d \omega$ is related to the framing anomaly and is therefore not identified with $\ell_{\OO,\text{LL}}^+$). We find that
\begin{equation}\label{eq:ells+_LL}
    \ell_{\OO,\text{LL}}^+ = \frac{4c_-^3 - c_-}{12}.
\end{equation}
To find $\ell_{\OO,\text{LL}}^-$, we shift $A \rightarrow A + \omega/2$ in the above action and again compute the coefficient of $\frac{1}{4\pi} \omega \wedge d\omega$ coming from the first term alone. Then we obtain
\begin{equation}\label{eq:ells-_LL}
    \ell_{\OO,\text{LL}}^- = \frac{2c_-^3 + 3 c_-^2 + c_-}{6}.
\end{equation}
The above equations were derived assuming $c_->0$. When $c_-<0$, we define the state corresponding to the `LL limit' as the time-reversed partner of the usual LL state with $|c_-|$ filled LLs. Now under time-reversal, $\omega_0 \rightarrow - \omega_0$ and $\omega_i \rightarrow \omega_i$ for $i = 1,2$. Therefore we must also have $\ell_{\OO} \rightarrow - \ell_{\OO}$. But since Eq.~\eqref{eq:ells+_LL} is odd in $c_-$, this means that it holds for $c_-<0$ as well. 

To compute $\ell_{\OO,\text{LL}}^-$ for $c_-<0$, we consider Eq.~\eqref{eq:EA_LL}, apply a time-reversal operation, and then take $A \rightarrow A + \omega/2$. The time-reversal operation changes $\omega$ as stated above, but also takes $A_0 \rightarrow A_0, A_i \rightarrow - A_i$. As a result we obtain
\begin{equation}
    \mathcal{L}_{\text{TR}} = - \frac{|c_-|}{4\pi} A \wedge dA + \frac{c_-^2/2}{2\pi} A \wedge d\omega + \frac{\ell_{\text{LL}}^+}{4\pi} \omega \wedge d\omega + \frac{c_-}{48\pi} \omega \wedge d\omega.
\end{equation}
If we now take $A \rightarrow A + \omega/2$ and compute $\ell_{\text{LL}}^-$, we find that Eq.~\eqref{eq:ells-_LL} also holds when $c_-<0$.

\subsection{Strategy for remaining derivations}
In the main text we sketched the derivation of Eq.~\eqref{eq:ells_Theta}, which is the main relation between the real-space invariants and the TQFT. The other relations in this section are derived using a common idea, which assumes the linearity of both $\Theta^{\pm}_{\OO}$ and the field theory coefficients under stacking. Each relation is first verified for $c_-=0$, and then for $c_- \ne 0$ in the LLL limit. The assumption of linearity under stacking then implies that the relation holds in general.

We recall the following definitions. When $c_- = 0= C$, $\Theta^{\pm}_{\OO}$ can be expressed in terms of the parameters $\{n_{\OO},m_{\OO}\}$ of a Wannier limit of the given state (Eq.~\eqref{eq:Theta_def_c=0}). In the Landau level limit, they are defined by Eq.~\eqref{eq:Theta_def_LL}.

\subsection{Relation between $\Theta^{\pm}_{\OO}$ and $\SO,\PO$}
Below we show that
$$\SO = \begin{cases}
    2 (\Theta^-_{\OO} - \Theta^+_{\OO}) \mod \MO & \MO \text{ even} \\
    2 (\Theta^-_{\OO} - \Theta^+_{\OO}) - \frac{C}{2} \mod \MO & \MO \text{ odd}.
\end{cases}$$
When $c_- = 0$, $\mathscr{S}_{\OO,\AI} = n_{\OO} \mod \MO$, and this equation can be verified immediately using Eq.~\eqref{eq:Theta_def_c=0}. In the LLL limit, $\mathscr{S}_{\OO,\text{LLL}} = \frac{1}{2} \mod \MO$, and the relation follows from Eq.~\eqref{eq:Theta_def_LL}. But there is also a derivation based on a topological action. We start with 
\begin{equation}
    \mathcal{L} = \frac{C}{4\pi} A \wedge dA +  \frac{\SO}{2\pi} \omega \wedge dA + \frac{\ell_{\OO}}{4\pi} \omega \wedge d\omega + \dots 
\end{equation}
and relabel $A \rightarrow A + \omega/2$. This gives
\begin{equation}
    \mathcal{L} \rightarrow \frac{C}{4\pi} A \wedge dA +  \frac{\SO + C/2}{2\pi} \omega \wedge dA + \frac{\ell_{\OO} + \SO + C/4}{4\pi} \omega \wedge d\omega + \dots 
\end{equation}
But this implies that
\begin{equation}
    \ell^-_{\OO} = \ell_{\OO} + \SO + \frac{C}{4} \mod \MO.
\end{equation}
Using Eq.~\eqref{eq:ells_Theta}, we see that 
\begin{equation}
    \SO = 2(\Theta^-_{\OO} - \Theta^+_{\OO}) + (t_- - t_+ - \frac{1}{4}) C\mod \MO.
\end{equation}
Expanding this out using Table~\ref{tab:Theta_vs_ells}, we obtain the claimed relation.
The relation between $\Theta^{\pm}_{\OO}$ and $\PO$ can then be deduced from Table~\ref{tab:PvsS}, which expresses $\PO$ in terms of linear combinations of $\SO$ for different $\OO$.

\subsection{Complete classification in terms of $\Theta^{\pm}_{\OO}$}\label{app:Gens-RSI-FT}

In this appendix we establish the relation between the field theoretic and real-space classifications by reexpressing the generators of the classification discussed in Sec.~\ref{sec:FT} using $\Theta^{\pm}_{\OO}$, $c_-, C$ and $\kappa$. These results follow from the relations stated in Sec.\ref{sec:rel-Theta-resp}.

\subsubsection{$M=4$}

In this case the generators in Eq.~\eqref{eq:FTresult_C4} satisfy the relations
\begin{align}\label{eq:RSdef_C4}
    I_1 &:= \mathscr{S}_{\alpha}-\ell_{\alpha}-\frac{c_-}{4} = 2 \Theta^-_{\alpha} - 4 \Theta^+_{\alpha} - \frac{3 c_-}{2}  \mod 8\nonumber \\ 
    I_2 &:= \nu_s = \Theta_\alpha^- + \Theta_\beta^- + 2\Theta_\gamma^- - \kappa/2 \mod 4\nonumber \\
    I_3 &:= \frac{1}{2}\left(\ell_{\OO}-\frac{c_-}{4}\right) = \Theta^+_{\alpha} + \frac{c_-}{2} \mod 2\nonumber \\
    I_4 &:= 2(\vec{\mathscr{P}}_{\alpha}-2 \vec{\mathscr{P}}_{s,\alpha})\cdot (1,0) =   2\Theta_\beta^- + 2\Theta_\gamma^-   \mod 4.
\end{align}
From this, we can conclude that
$\{c_-, C, \kappa, \Theta^{\pm}_{\OO}\}$ fully characterize invertible fermionic states with symmetry group $G_f$, thus proving a claim made in Ref.~\cite{zhang2023complete}. Note that the generators used there were different as they were chosen to make their quantization more apparent. Here we choose the generators that appear naturally in the topological action.

\subsubsection{$M=2$}

The generators of the cyclic groups in Eq.~\eqref{eq:FTresult_C2} can be expressed in terms of $\Theta^{-}_{\OO}$ as follows:
\begin{align}\label{eq:RSdef_C2}
    I_1 &:= \mathscr{S}_{\alpha} - \ell_{\alpha} - \frac{c_-}{4} = 2\Theta^-_{\alpha} - \frac{c_-}{2} \mod 4\nonumber \\
    I_2 &:= \nu_s =  -\frac{\kappa}{2} -c_- + \sum_{\OO}\Theta^-_{\OO} \mod 2 \nonumber \\
    (I_3, I_4) &:= 2 (\vec{\mathscr{P}}_{\alpha} - 2\vec{\mathscr{P}}_{s,\alpha}) = (2(\Theta^-_{\beta} + \Theta^-_{\gamma}) - c_- \mod 4, \quad 2(\Theta^-_{\beta} + \Theta^-_{\delta}) - c_- \mod 4) .
\end{align}

\subsubsection{$M=3$}
The generators of the cyclic groups in Eq.~\eqref{eq:FTresult_C3} can be expressed in terms of $\Theta^{\pm}_{\OO}$ as follows:

\begin{align}\label{eq:RSdef_C3}
    I_1 &:= \mathscr{S}_{\alpha} - \frac{c_-}{2} = 2 (\Theta^-_{\alpha} - \Theta^+_{\alpha}) - \frac{c_-}{2} \mod 3\nonumber \\
    I_2 &:= \nu_s = c_- + \sum_{\OO}\Theta^+_{\OO} \mod 3\nonumber \\
    I_3 &:= \ell_{\alpha} - \frac{c_-}{4} = 2 \Theta^+_{\alpha} + \frac{2c_-}{3} \mod 3\nonumber \\
    I_4 &:= 3 \vec{\mathscr{P}}_{\alpha} \cdot (1,0) = 2 (\Theta^-_{\alpha} - \Theta^-_{\beta} - \Theta^+_{\alpha} + \Theta^+_{\beta} - \kappa) \mod 3\nonumber \\
    I_5 &:= 3 \vec{\mathscr{P}}_{s,\alpha} \cdot (1,0) = 2(\Theta^+_{\beta} - \Theta^+_{\gamma}) \mod 3 .
\end{align}

\subsubsection{$M=6$}
The generators of the cyclic groups in Eq.~\eqref{eq:FTresult_C6} can be expressed in terms of $\Theta^{\pm}_{\OO}$ as follows:
\begin{align}\label{eq:RSdef_C6}
    I_1 &:= \mathscr{S}_{\alpha} - \ell_{\alpha} - \frac{c_-}{4} = 2 (\Theta^-_{\alpha} - \Theta^+_{\alpha}) - \frac{19 c_-}{6} \mod 12 \nonumber \\
    I_2 &:= \nu_s = \Theta^-_{\alpha} + 2\Theta^-_{\beta} + 3\Theta^-_{\gamma} -\frac{\kappa}{2} - 3 c_-\mod 6.\nonumber \\
    I_3 &:= \ell_{\alpha} - \frac{c_-}{4} = 2 \Theta^+_{\alpha} + \frac{8c_-}{3} \mod 3   
\end{align}

\section{Review of classification of band insulators in 2d}\label{app:BandInsulators}
\def\nRep{n_{\Rep}}
\subsection{Band structure combinatorics / K-theory}\label{app:BandCombinatorics}
Ref.~\cite{Kruthoff2017TIBandComb} showed that one can understand the classification of band insulators with space group symmetry (when the spatial symmetries commute with charge conservation, implying zero magnetic flux per unit cell, and there are no antiunitary symmetries involved) in terms of the representation theory of the valence (filled) bands. The classification in 2d for symmorphic wallpaper groups is obtained as follows.
\begin{enumerate}
    \item Fix a wallpaper-group $\Gwp$\ and an origin $\OO$ such that its stabilizer is isomorphic to the point group. By abuse of notation, we denote such subgroup of $\Gwp$ by $\Gpt$
    \item The classification is given by a K-theory of the BZ: $ 
    K_{\Gpt}^{0}(\TT^2) \cong \ZZ^{n_{\Ch}} \oplus \ZZ^{\nRep} $.  
    \item $n_{\Ch}=0$ if there are orientation reversing symmetries and $1$ otherwise. This invariant describes whether or not a non-zero Chern number $C$ is allowed. 
    \item Find the Brillouin zone ($\BZ$) and for each momentum $\kbs\in \BZ$, determine the little co-group $G_{\kbs} \subset \Gpt$ [the subgroup of the $\Gpt$ that maps $\kbs$ to itself].
    \item Find the high symmetry momenta: $\kbs$ such that for any $\kbs'$ close to $\kbs$, $G_{\kbs'}=G_{\kbs}$ implies $\kbs=\kbs'$. Denote the set of high symmetry momenta by $\HSM$. 
    \item Let $\HSM^{\star}=\HSM/\Gpt$ be the orbits of $\HSM$ under the $\Gpt$ action (also known as "stars"). Consider the abelian group $\Rmc=\bigoplus_{\kbs^{\star}\in \HSM^{\star}}\RU(G_{\kbs})$, where $\kbs$ is a representative of the orbit $\kbs^{\star}$ and $\RU(G_{\kbs})$ is the ring of complex representations of $G_{\kbs}$. \footnote{Note that $\RU(G_{\kbs})$ is isomorphic to $\ZZ^{m_{\kbs}}$ for some integer $m_{\kbs}$. Also note that $\RU(G)$ is the group completion of $\Rep(G)$.}
    \item For each pair of representatives $(\kbs_1,\kbs_2)$, denote by $l_{(\kbs_1,\kbs_2)}$ the line containing both points. Denote by $G_{(\kbs_1,\kbs_2)}\subset \Gpt$  the group that fixes every element of the line. Allowed elements of $\Rmc$ satify the condition 
    \begin{equation}
\Res^{G_{\kbs_1}}_{G_{(\kbs_1,\kbs_2)}}r_{\kbs_1} = \Res^{G_{\kbs_2}}_{G_{(\kbs_1,\kbs_2)}}r_{\kbs_2}
    \end{equation}
    where $r\in \Rmc$ and we denoted by $r_{\kbs}$ the component of $r$ in $\RU(G_{\kbs})$. $\Res^{G}_{H}: \RU(G)\to\RU(H)$ is the restriction map ($H$ is a subgroup of $G$). 
    \item $n_{\Rep}$ is found from counting the number of allowed elements of $\Rmc$. 
\end{enumerate}
When there are no reflections, $G_{(\kbs_1,\kbs_2)}$ is the trivial group so the only condition the restriction map imposes is that the dimension of the representation is independent of $\kbs$. Furthermore, $\RU(C_m)=\ZZ^{m}$, (where $C_m \simeq \mathbb{Z}_m$ denotes $m$-fold rotational symmetry) because the irreps of $C_m$ correspond to angular momenta $\ell =0,1,\dots, m-1$.

\subsection{Defining Invariants I: general}\label{app:BandComb:DefI}
\def\Umc{\mathcal{U}}

In the previous section, we described a way to find the classification using a K-theoretical argument. However, it is often useful to have an explicit construction of the invariants, i.e. state explicitly the eigenvalue of which matrix we are calculating. Our conventions for Fourier transforms, and other things, can be found in App.~\ref{app:ConvetionsFang}.

For every $g\in G$, let $\hat{U}_g$ be the unitary operator implementing $g$ in the many-body Hilbert space. Then, for every $\kbs \in \BZ$, there is a matrix $\tilde{U}_{g}(\kbs)$ such that 
\begin{equation}
    \hat{U}_{g} \alpha_{b}^{\dagger}(\kbs) \hat{U}_{g}^{\dagger} = \sum_{b} \alpha_{a}^{\dagger}(R_{g}\kbs) [\tilde{U}_{g}(\kbs)]_{ab}
\end{equation}
where $\alpha_{b}^{\dagger}(\kbs)$ is a creation operator of fermions at momentum $\kbs$ and band label $b$. $R_g$ is the matrix representation of $g$ acting on the spatial coordinates. See Tab.~\ref{tab:RgDef} for explicit definitions of $R_g$.
\begin{table}[t]
    \centering
    \begin{tabular}{|c|c|c|c|c|c}\hline
    $M$ & $2$ & $4$ & $3$ & $6$\\\hline
    $R_{\hbf_{M}}$&
        $\begin{pmatrix}
            -1 & 0 \\
            0 & -1
        \end{pmatrix}$ &
        $\begin{pmatrix}
            0 & -1 \\
            1 & 0
        \end{pmatrix}$ &
           $\begin{pmatrix}
            -\frac{1}{2} & -\frac{\sqrt{3}}{2} \\
            \frac{\sqrt{3}}{2}  & -\frac{1}{2}
        \end{pmatrix}$  &
           $\begin{pmatrix}
            \frac{1}{2} & -\frac{\sqrt{3}}{2} \\
            \frac{\sqrt{3}}{2}  & \frac{1}{2}
        \end{pmatrix}$ \\\hline
    \end{tabular}
    \caption{ Explicit matrix representation of $\hbf_M$. $\hbf_M$ is a anti-clockwise rotation by $2\pi/M$ radians around the origin. }
    \label{tab:RgDef}
\end{table}
At every $\kbs\in \BZ$, there is a matrix that projects to the filled bands $P_{\kbs}$\footnote{It satisfies \begin{enumerate}
    \item $P_{\kbs}^{\dagger}P_{\kbs}=\Id$;
    \item $-P_{\kbs}^{\dagger}h(\kbs)P_{\kbs}$ is a positive definite matrix.
\end{enumerate}}. The sewing matrix is defined by
\begin{equation}
    \Umc_{g}(\kbs) = P_{R_{g}\kbs}^{\dagger}\tilde{U}_{g}(\kbs)P_{\kbs}^{}.
\end{equation}
 For fixed $\kbs$, these matrices form a representation of $G_{\kbs}$
 \footnote{$\Umc_{g}(\kbs)\Umc_{h}(\kbs)= P_{\kbs}^{\dagger}\tilde{U}_{g}(\kbs) P_{\kbs}^{} P_{\kbs}^{\dagger}\tilde{U}_{h}(\kbs)P_{\kbs}^{} $. Then note that $P_{\kbs}P_{\kbs}^{\dagger}$ commutes with $\tilde{U}_{g}(\kbs)$ and $\tilde{U}_{g}(\kbs)\tilde{U}_{h}(\kbs)=\tilde{U}_{gh}(\kbs)$.}. The $r\in \mathcal{R}$ from Sec.~\ref{app:BandCombinatorics} is  constructed by setting $r_{\kbs}= \tilde{U}_{\cdot}(\kbs) \in \Rep(G_{\kbs})$.

\subsection{Defining Invariants II: orientation preserving}\label{app:DefInv:OrPreserving}
Recall that we fixed an origin $\OO$ such that $G_{\OO}\cong \Gpt$. For ease of notation, we will drop the $\OO$ in the name of the invariants below but recover it in the main text. For a momenta $\kbs\in \BZ$ and a positive integer $m$ that is a divisor of $M_{\kbs}$, we calculate the eigenvalues of $\Umc_{\hbf_{\OO}^m}(\kbs)$ where $\hbf_{\OO}$ is the generator of rotations around $\OO$. Let $\# \kbs_{j}^{(m)}$ be the number of times the $\ee^{\frac{2\pi \ii (j-1)}{m}}$ eigenvalue appears. 
\\
There are relations between the eigenvalue numbers, the filling ($\kappa$), and the Chern number ($\Ch$) \footnote{  App.~\ref{app:EigenvalsConstraintsDerivation} gives a derivation of these equations using results from Ref.~\cite{Fang2012PGS}.}:
\begin{align}
        \kappa &=\sum_{j=1}^{M_{\kbs}} \# \kbs_{j}^{(M_{\kbs})}; \quad \forall \kbs \in \HSM; \label{eq:App:FillingConstraintRot}\\
    \frac{\Ch}{M}&+\sum_{\kbs^{\star}\in\HSM^{\star}}  \sum_{j=1}^{M_{\kbs}} \frac{(j-1)}{M_{\kbs}}\#\kbs_{j}^{(M_{\kbs})} =  0 \mod 1 \label{eq:App:ChernConstraintRot}.
\end{align}
For later convenience, we define ``rotation invariants":
\begin{equation}\label{eq:RotInv}
    \left[\kbs^{(m)}_{j}\right] = \# \kbs_{j}^{(m)}- \# \Gamma_{j}^{(m)} 
\end{equation}
that satisfy \footnote{
By explicit calculation one can show that  $\sum_{\kbs^{\star}\in \HSM^{\star}}M_{\kbs}^{-1}\in \ZZ$ and $\frac{n(j-1)}{M}\# \Gamma^{(M/n)}_{j} = \frac{n(j-1)}{M}\sum_{i=1}^{n}\# \Gamma^{(M)}_{j+(i-1)M/n} \mod 1  $ when $n|M$. Using this, we can go from  Eq.~\eqref{eq:App:ChernConstraintRot} to Eq.~\eqref{eq:App:ChernConstraintRotInv}}
\begin{align}
        0 &=\sum_{j=1}^{M_{\kbs}} \left[\kbs_{j}^{(M_{\kbs})}\right]; \forall \kbs \in \HSM; \label{eq:App:FillingConstraintRotInv}\\
    \frac{\Ch}{M} &+\sum_{\kbs^{\star}\in \HSMsp} \frac{(j-1)}{M_{\kbs}}\left[\kbs_{j}^{(M_{\kbs})}\right] =0 \mod 1; \label{eq:App:ChernConstraintRotInv}
\end{align}
where $\HSMsp = \HSM\setminus\{\Gamma^{\star}\}$ are the set of orbits excluding the orbit of $\Gamma$.

We will see in Sec.~\ref{app:BandComb:Relation} that a complete set of invariants for a non-interacting band insulator are 
\begin{enumerate}
    \item The Chern number $\Ch$; 
    \item The filling factor $\kappa$;
    \item For each $\kbs^{\star}\in \HSM^{\star}$, take $\{\# \kbs_{j}^{(M_{\kbs})}| j=2,\dots, M_{\kbs}\}$.
\end{enumerate}
or equivalently 
\begin{enumerate}
    \item The Chern number $\Ch$;
    \item The eigenvalues at $\Gamma$: $\{\# \Gamma_{j}^{(M)}| j=2,\dots, M\}$ (whose sum is $\kappa$);
    \item For each $\kbs^{\star} \in \HSMsp$, take the rotation invariants $\{\left[\kbs_{j}^{(M_{\kbs})}\right]| j=2,\dots, M_{\kbs}\}$.
\end{enumerate}
An explicit set of invariants for each $M$ is summarized in Tab.~\ref{tab:basisNonInteracting} in the main text. These invariants are denoted by $J = (J_1,J_2,\dots, J_{\nRep+1} )$. Saying that this set of invariants is complete means that if $\psi,\psi'$ are two nIBI and $J_{j}[\psi] = J_{j}[\psi']$ for $j=1,2,\dots, 1+\nRep$, then $\psi \sim_S \psi'$ according to the nIBI classification (i.e. the K-theory). Where $\sim_{S}$ means stable equivalence as described in the main text. 

Note that the relation in Eq.~\eqref{eq:App:ChernConstraintRotInv} implies that not all combinations of invariants are allowed for invertible states. Another way to say this is that $J$ is not a surjective map.

\subsection{Relation between band structure combinatorics and eigenvalues}\label{app:BandComb:Relation}

For a cyclic group $C_M$, a representation is fully determined by the eigenvalues of the generator. Furthermore, the abelian monoid of representations of $C_M$ admits a product coming from tensor product of representations. We have
\begin{equation}
    \Rep(C_M) \cong \frac{\NN[z]}{(1-z^M)}
\end{equation}
where $\NN[z]$ is the ring of polynomials in $z$ with coeffients in $\NN=\ZZ^{\geq 0}$. $z$ corresponds to the representation with angular momentum 1. The quotient is because for $M$-fold rotations, angular momentum $M$ is trivial.

The set of eigenvalues with degeneracies $\{\#\kbs_{j}^{(M_{\kbs})}|j=1,\dots,M_{\kbs}\}$ defines a representation
\begin{equation}\label{eq:DefRepFromEigvals}
    p_{\kbs}(z) = \sum_{j=1}^{M_{\kbs}}  (\# \kbs_{j}^{M_{\kbs}}) z^{j-1} \in \Rep(C_{M_{\kbs}}).
\end{equation}
From Sec.~\ref{app:BandCombinatorics}, the classification is determined by the Chern number and the representation of the filled bands at high symmetry momenta subject to relations. For orientation preserving wallpaper groups, the only constraint between representations for a pair of momenta $(\kbs,\kbs')$ such that  $\kbs^{\star}\neq \kbs'^{\star}$ is that the total dimension is the same. In terms of eigenvalues (Eq.~\eqref{eq:DefRepFromEigvals}) this translates to Eq.~\eqref{eq:App:FillingConstraintRot}. Thus, we arrive at the first characterization at the end of Sec.~\ref{app:DefInv:OrPreserving}. The second characterization is a simple change of basis. 

One may wonder how Eq.~\eqref{eq:App:ChernConstraintRot}  affects the classification. In Ref.~\cite{Kruthoff2017TIBandComb}, it was mentioned that eigenvalues can be used to partially determine the Chern number but this does not modify the abelian group that determines the classification. 

\subsection{Conventions for non-interacting insulators}\label{app:ConvetionsFang}
\def\sbs{\boldsymbol{s}}

Our convention for Fourier transforms is to define a unit cell such that it contains one atom for each orbital. Then, each unit cell is identified with a vector $\rbs = n_1 \abss_1 +n_2\abss_2$ where $n_1,n_2\in \ZZ$ and $\abss_{1},\abss_{2}$ are lattice vectors that define the unit cell. Then, the Fourier transform of an orbital with creation operator $f^{\dagger}_{\rbs}$ is $\alpha^{\dagger}(\kbs)\propto \sum_{\rbs }\ee^{- \ii \rbs\cdot \kbs} f_{\rbs}^{\dagger}$. Therefore, under a translation that maps $f_{\rbs}\to f_{\rbs+\Rbs}$, we have $\alpha^{\dagger}(\kbs) \to \alpha^{\dagger}(\kbs) \ee^{\ii \kbs \cdot\Rbs}$, where $\Rbs$ is again of the form $n_1\abss_1 + n_2\abss_2$. With this convention, the Fourier transform of the single particle Hamiltonian is periodic with respect to the reciprocal lattice vectors ($h_{\kbs} = h_{\kbs + \boldsymbol{G}}$ with $\boldsymbol{G}$ a reciprocal lattice vector). However, lattice symmetries get momentum dependence because these symmetries will in general move orbitals between unit cells. 

We proceed to state our conventions for how unitary symmetries act on the first and second quantized picture. Let $G$ be a group of symmetries that are realized unitarily. Let $g \in G$ and $\hat{U}_g$ be the operator acting on the many-body Hilbert space and $\alpha_{a}^{\dagger}(\kbs)$ be a creation operator of a fermion with momentum $\kbs$ and $a$ is an orbital index that is the Fourier transform of $f_{\rbs,a}^{\dagger}$. We consider symmetries that act as  
\begin{equation}
    \hat{U}_{g}^{} f_{\rbs,a}^{\dagger} \hat{U}_{g}^{\dagger} = \sum_{b}f_{^{g}\rbs + \sbs_{g,a}, b} [U_g(\rbs)]_{ba}
\end{equation}
where $^{g}\rbs$ is the natural action of $g$ on the spatial coordinate. The extra vector $\sbs_{g,a}$ that depends on the orbital and group element is necessarily because $\rbs$ is a label for a unit cell and not the true position of the orbital. 

Upon taking the Fourier transform, the transformation reads
\begin{equation}
    \hat{U}_{g}^{}  \alpha_{a}^{\dagger}(\kbs) \hat{U}_{g}^{\dagger}= \sum_{b}\alpha_{b}^{\dagger}(R_g\kbs) [\tilde U_g(\kbs) ]_{ba}
\end{equation}
where $R_g$ is the matrix representation of $g$ on the spatial coordinates, that satisfies $R_g^{T}R_g=\text{Id}_2$. $\tilde{U}_g(\kbs)$ is a matrix in the orbital space which is the action of $g$ in the single-particle Hilbert space. The projection of $\tilde{U}(\kbs)$ to the occupied orbitals are the Sewing matrices in Ref.~\cite{Fang2012PGS}.

\subsection{Derivation of eigenvalue constraints}\label{app:EigenvalsConstraintsDerivation}

To derive Eq.~\eqref{eq:App:ChernConstraintRot}, we use Eqs.~24, 31,32, and 33 of Ref.~\cite{Fang2012PGS}:
\begin{equation}
    \begin{split}
        (-1)^{\Ch} &= \prod_{j\in \text{Occ.}} 
        \lambda^{(2)}_{j}(\Gamma)
        \lambda^{(2)}_{j}(X)
        \lambda^{(2)}_{j}(Y)
        \lambda^{(2)}_{j}(M) \quad [M=2]\\
        (i)^{-\Ch} &= \prod_{j\in \text{Occ.}} 
        (-1)^{F}\lambda^{(4)}_{j}(\Gamma)
        \lambda^{(4)}_{j}(M){\lambda^{(2)}_{j}(Y)} \quad [M=4]\\
        e^{-2\pi i\Ch/3} &= \prod_{j\in \text{Occ.}} 
        (-1)^{F}\lambda^{(3)}_{j}(\Gamma)
        \lambda^{(3)}_{j}(K)\lambda^{(3)}_{j}(K') \quad [M=3]\\
        e^{-\pi i\Ch/3} &= \prod_{j\in \text{Occ.}} 
        (-1)^{F}\lambda^{(6)}_{j}(\Gamma)
        \lambda^{(3)}_{j}(K)\lambda^{(2)}_{j}(M) \quad [M=6]
    \end{split}
\end{equation}
where $F=0$ for spinless fermions, and $F=1$ for spinful fermions. $\lambda_j^{m}(\kbs)$ is the $C_m$ eigenvalue of band $j$ at momenta $\kbs$. Note that $\Ch$ as defined in Eq.~\eqref{eq:ChernNumberDef} is negative the Chern number defined in Ref.~\cite{Fang2012PGS}.

We can write all the above equations compactly as 
\begin{equation}
    e^{-\frac{2\pi i}{M }\Ch} = \prod_{j\in \text{Occ.}} \left[\left[(-1)^{F}\right]^{1-\delta_{M,2}}\prod_{\kbs_{\star}\in \HSM_{\star}} \lambda_{j}^{(M_{\kbs})}(\kbs)\right]
\end{equation}
by noting that $[\lambda_{j}^{(m)}(\kbs)]^m = (-1)^F$.

Using the relation 
\begin{equation}
    \prod_{j\in \text{Occ.}} \lambda_{j}^{(m)}(\kbs) = \exp\left(\frac{2\pi i}{m}\sum_{l=1}^{m} (l-1)\# \kbs_{l}^{(m)}\right)
\end{equation}
and taking the logarithm, dividing by $2\pi i$, and setting $F=0$ because we are dealing with spinless fermions, we obtain Eq.~\eqref{eq:App:ChernConstraintRot}.

\section{Atomic insulators }\label{app:AI}

Atomic insulators are uniform copies of 0-dimensional systems.  
It is important to see how they fit in the general picture. This appendix is organized as follows: in Sec.~\ref{app:AI:Def} we explicitly construct the atomic insulator states for symmorphic wallpaper groups. In Sec.~\ref{app:AI:Ori}, we derive abstract formulas for band invariants in terms of the data to construct an AI for the orientation preserving wallpaper groups. In Sec.~\ref{app:AI:Summary}, we summarize, for each $M$, the various AIs with their band invariants.  

\subsection{Definition}\label{app:AI:Def}

Consider a site $\text{s}$ with stabilizer $\Gs \subset \Gwp$ and a representation $\rho \in \Rep(\Gs)$. We construct a $\Gwp$ invariant state using this data. At every position $\rbs$ of the Wyckoff position of $\text{s}$, put fermions with creation operators $f_{j}^{\dagger}(\rbs); j=1,\dots,\dim(\rho)$, with Hamiltonian
\begin{equation}
    H = -\sum_{\rbs\in \Lambda_{\text{s}}}\sum_{j=1}^{\abs{\dim(\rho)}} f^{\dagger}_{j}(\rbs)f^{}_{j}(\rbs)
\end{equation}
where $\Lambda_{\text{s}}$ are all points of the Wyckoff position of $\text{s}$. The action of $g\in \Gwp$ is 
\begin{equation}\label{eq:Actionf}
    \hat{U}_{g}f_{j}^{\dagger}(\rbs)\hat{U}_{g}^{\dagger} = \sum_{i} f_{i}^{\dagger}({^g}\rbs)(\rho(g'))_{ij}.
\end{equation}
where ${^g}\rbs$ is the action of $g$ on $\rbs$, which is equal to $g\cdot \rbs \cdot \overline g$ as elements of the 2d Euclidean group $\EE^2=\RR^2\rtimes \mathrm{O}(2)$. Here $g'$ is an element that satisfies $g = g'' g'$ where $g'\in G_{\pt}$ and $g''\not\in G_{\pt}$, where $g''$ is a translation not in $G_{\pt}$ while $g'\in G_{\pt}$.

We can take the Fourier transform of Eq.~\eqref{eq:Actionf} to determine $\tilde{U}_g(\kbs)$. In the current case $\Umc_g(\kbs)=\tilde{U}_g(\kbs)$.We can thus determine the invariants of App.~\ref{app:BandInsulators}. We will do this explicitly for the orientation preserving case in the next section.

\subsection{Orientation preserving case}\label{app:AI:Ori}
This sections aims to give an analytical derivation for the induced representation at every $\kbs$ starting from the data of an atomic insulator. Through out we fix an origin which we denote by $\OO$ that satisfy $M_{\OO} = M$. Explicit results for the relevant $\Gwp$ are presented in App.~\ref{app:AI:Summary}.

Take a position $\pt$ and a representation of the stabilizer $\rho \in \Rep(C_{M_{\pt}})$. Let $\hbf_{\pt}$ be a generator of $C_{M_{\pt}}$ that corresponds to the smallest clock-wise rotation. Any rotation can be decomposed as a rotation around $\OO$ and a translation. In other words, there exist a lattice vector $\tbs_{\pt}$ such that $T_{\tbs_{\pt}}\hbf_{\OO}^n = \hbf_{\pt} $ with $n= M/M_{\pt}$ and $T_{\tbs}$ is translation by $\tbs$ (See App.~\ref{app:AI:Summary} for explicit values of $\tbs_{\pt}$). The AI has $n$ fermions per unit cell, so there will be $n$ bands. Recall that $\alpha^{\dagger}(\kbs)$ are the creation operators obtained by taking the Fourier transform of $f^{\dagger}$. Instead of taking a Fouerier transform directly to Eq.~\ref{eq:Actionf}, we define $\alpha_{1}^{\dagger}$ as the Fourier transform of $f_{1}^{\dagger}$, and $\alpha^{\dagger}_{j+1}(\kbs) = \hat{U}_{\hbf_\OO}^{j}\alpha^{\dagger}_{j+1}(R_{\hbf_{\OO}}^{-j}\kbs) \hat{U}_{\hbf_\OO}^{-j} $ be the creation operator of the orbital at $\hbf_{\OO}^{j-1}\pt$, {i.e.} $\alpha_{j+1}^{\dagger}$ is a linear combination of $f_{j+1}^{\dagger}$.

The action of $\hat{U}_g$ is given by
\begin{equation}\label{eq:transformationAlphaCyclic}
    \hat{U}_{\hbf_{\OO}} \alpha_{j}^{\dagger}(\kbs)  \hat{U}_{\hbf_{\OO}}^{\dagger} = \begin{cases}
        \alpha_{j+1}^{\dagger}(R_{\hbf_{\OO}}\kbs) \quad &, j<n; \\
      \hat{U}_{\hbf_{\OO}^n}{}\alpha_1^{\dagger}(R_{\hbf_{\OO}^n}^{-1}R_{\hbf_{\OO}}\kbs)  \hat{U}_{\hbf_{\OO}^n}^{\dagger}&,j=n
    \end{cases}
\end{equation}
We now want to determine the action of $\hbf_{\OO}^n$ on $\alpha_{j}^{\dagger}(\kbs)$ from $\rho$. To do so, we start by noting that $T_{\tbs_{\pt}}\hbf_{\OO}^n = \hbf_{\pt}$ allows us to write 
\begin{equation}\label{eq:transformationAlpha1}
    \begin{split}
        \hat{U}_{T_{\tbs_{\pt}}}\hat{U}_{\hbf_{\OO}^n}{}\alpha_1^{\dagger}(R_{\hbf_{\OO}^n}^{-1}R_{\hbf_{\OO}}\kbs)  \hat{U}_{\hbf_{\OO}^n}^{\dagger}\hat{U}_{T_{\tbs_{\pt}}}^{\dagger}&= \hat{U}_{\hbf_\pt} 
    \alpha_{1}^{\dagger}(R^{-1}_{\hbf_{\pt}}R_{\hbf_\OO}\kbs)\hat{U}_{\hbf_\pt}^{\dagger} \\
    &= \frac{1}{\sqrt{N_{\text{UC}}}}\sum_{\rbs \in  \Lambda_{\pt}}
    \hat{U}_{\hbf_\pt} 
    f_{1}^{\dagger}(\rbs)\ee^{-\ii (R_{\hbf_{\pt}}\rbs)\cdot (R_{\hbf_\OO}\kbs)}\hat{U}_{\hbf_\pt}^{\dagger} \\
    &= \frac{1}{\sqrt{N_{\text{UC}}}}\sum_{\rbs \in \Lambda_{\pt}}
    f_{1}^{\dagger}(^{\hbf_{\pt}}\rbs) \ee^{-\ii (R_{\hbf_{\pt}}\rbs)\cdot (R_{\hbf_\OO}\kbs)} \rho(\hbf_{\pt})\\
    &= \alpha_1^{\dagger}(R_{\hbf_\OO}\kbs)\rho(\hbf_{\pt}) \\
\end{split}
\end{equation}
where $N_{\text{UC}}$ is the number of unit cells. This implies 
\begin{equation}
    \hat{U}_{\hbf_{\OO}^n}{}\alpha_1^{\dagger}(R_{\hbf_{\OO}^n}^{-1}R_{\hbf_{\OO}}\kbs)  \hat{U}_{\hbf_{\OO}^n}^{\dagger} = \alpha_1^{\dagger}(R_{\hbf_\OO}\kbs) \left[\rho(\hbf_{\pt}) \ee^{-\ii \tbs_{\pt}\cdot R_{\hbf_\OO}\kbs}\right].
\end{equation}

We now determine the eigenvalue spectrum analytically. We start with the case $^{\hbf_\OO}\kbs = \kbs$, so that the sewing matrix can be read off from Eq.~\eqref{eq:transformationAlphaCyclic} directly:
\begin{equation}
    \tilde{U}_{\hbf_{\OO}}(\kbs)_{a,b} =\begin{cases}
        1 & ;a=b+1\leq n \\
       \upsilon_{\pt}(R_{\hbf_{\OO}}\kbs)\equiv  \rho(\hbf_{\pt})\ee^{-\ii  \tbs_{\pt}\cdot R_{\hbf_\OO}\kbs} &; a=1,b=n \\
        0 &; \text{otherwise}.
    \end{cases}
\end{equation}
The eigenvalues of $\tilde{U}_{\hbf_{\OO}}(\kbs)$ are the different solutions for $\zeta$ in the equation $\zeta^n = \upsilon_{\pt}(R_{\hbf_\OO}\kbs)$. 

We can similarly determine the eigenvalues when the sewing matrix is $\tilde{U}_{\hbf_{\OO}^m}(\kbs)$. First, note $\tilde{U}_{\hbf_\OO}(\kbs) = \tau_{n}\Upsilon_{\pt}^{(n)}(R_{\hbf_{\OO}}\kbs),$ where $\tau_n$ is the $n$-by-$n$ shift matrix and $\Upsilon_{\pt}^{(k)}(R_{\hbf_{\OO}}\kbs)$ is the identity matrix with the $(k,k)$ element replaced by $\upsilon_{\pt}(R_{\hbf_{\OO}}\kbs)$. Then 
\begin{equation}\label{eq:foo}
    \tilde{U}_{\hbf_\OO^m}(\kbs)  = \prod^{\leftarrow}_{j=1,\dots,m} \tilde{U}_{\hbf_\OO}(R_{\hbf_{\OO}}^{j-1}\kbs) = \tau_n^m \prod^{\leftarrow}_{j=1,\dots,m} \Upsilon_{\pt}^{(n-j+1)_{n}}(R_{\hbf_{\OO}}^{j}\kbs)
\end{equation}
where $(a)_n \in \{1,\dots,n\}$ such that $(a)_n=a\mod n$. The $\leftarrow$ on top of $\Pi$ means that we multiply the terms in descending order $P_m \dots P_2 P_1$. The matrix $\tau_{n}^m$ has $k=\gcd(n,m)$ invariant subspaces of dimension $d=n/k$, with projectors $P_{a}: \CC^n \to \CC^{d}$ given by $(P_a)_{ij}= \delta_{i,jk+1-a}; a=1\dots, k$. The eigenvalues of $P_a^{\dagger}\tilde{U}_{\hbf_{\OO}^m}(\kbs)P_a$ are the different solutions for $\zeta_{(a)}$ in the equation 
\begin{equation}
    \zeta_{(a)}^{d}  = \prod_{j=1}^m \left[\upsilon_{\pt}(R_{\hbf_{\OO}}^{j}\kbs)\right]^{\delta_{(j)_{k},a}}= \rho(\hbf_{\pt})^{m/k} \exp(-\ii \tbs_{\pt}\cdot(\sum_{r=1}^{m/k} R_{\hbf_{\OO}^{k(r-1)+a}})\kbs).
\end{equation}
The eigenvalues of $\tilde{U}_{\hbf_{\OO}^m}(\kbs)$ are the eigenvalues of all $P_a^{\dagger}\tilde{U}_{\hbf_{\OO}^m}(\kbs)P_a; a=1,\dots, k$ together.

The expression simplifies in two special cases:
\begin{enumerate}
    \item $m|n$ so that $k=m$, then 
    \begin{equation}
        \zeta_{(a)}^{n/m} = \rho(\hbf_{\pt})\exp(-\ii \tbs_{\pt}\cdot R_{\hbf_{\OO}}^a\kbs); a=1,\dots,m.
    \end{equation}
    \item $\gcd(m,n)=1$ then there is only one block and 
    \begin{equation}
        \zeta^n_{(1)} = \rho(\hbf_{\pt})^{m} \exp(-\ii \tbs_{\pt}\cdot(\sum_{r=1}^{m} R_{\hbf_{\OO}}^{r})\kbs).
    \end{equation}
\end{enumerate}

\subsection{Summary of AIs for orientation preserving wallpaper groups }\label{app:AI:Summary}
\renewcommand{\arraystretch}{1.7}

In this section, we summarize the AIs for the 4 wallpaper groups labeled by $M=2,3,4,6$. We fix a coordinate system with origin $\zero$ in Wyckoff position $\alpha$. 

Let us start with some notation. We define the following lattice vectors: $\abss_{x}= (1,0)$, $\abss_{y}= (0,1)$, and $\abss_{w}= (1/2,\sqrt{3}/2)$. 
Below we identify the one-dimensional representations $\rho \in \Rep(C_M)$ with $\lambda= \rho(h_M)$ where $h_M$ is the smallest non-trivial clockwise rotation in $C_M$.

For $M=2,4$, we take the BZ to be the square $[-\pi,\pi] \times[-\pi,\pi]$ and denote the high symmetry momenta as $\Gm=(0,0)$, $\Xm=(\pi,0), \Ym=(0,\pi), \Mm=(\pi,\pi)$. For $M=3,6$, we take the BZ to be a hexagon with corners at the $C_6$ orbit of the point $(4\pi/3,0)$. We take the HSM to be $\Km = (4\pi/3,0)$, $\bKm = (-4\pi/3,0)$, $\Mm=(\pi,-\pi/\sqrt{3})$.

Below we present our results in the form of tables. We first summarize the information used to construct AIs: $\pt$ (site at which we specify the angular momentum), Wyckoff position (position to which $\pt$ belongs), $\tbs_{\pt}$ (vector such that $\hbf_{\pt} = T_{\tbs_{\pt}}\hbf_{\zero}^n$ where $\hbf_{\pt}$ is the smallest anti-clockwise rotation around $\pt$) and $n$ (the degeneracy of the Wyckoff position). Next to this table, we present another table with the eigenvalues at the high symmetry momenta for the atomic insulator induced from the respective Wyckoff position (WP) with rotation eigenvalue $\lambda$. Note that $\lambda$ is restricted to a $M_{\text{WP}}$-th root of unity. 

\paragraph{M=2:} there are four maximal Wyckoff positions:

\begin{tabular}{cc}
    \begin{minipage}{.5\linewidth}
\begin{equation*}
    \begin{array}{|c||c|c|c|c|}\hline
        \pt & \text{Wyckoff position} & \tbs_{\pt} & M_{\pt} & n \\ \hline
        \zero & \alpha & \zero & 2 & 1\\ 
        \half \abss_x +\half \abss_y  & \beta & \abss_x+\abss_y & 2 & 1\\
        \half \abss_y & \gamma & \abss_y & 2 & 1\\
        \half \abss_x & \delta & \abss_x & 2 & 1\\\hline
    \end{array}
\end{equation*}
        \label{tab:wkf_C2}
    \end{minipage} &

    \begin{minipage}{.5\linewidth}
\begin{equation*}
    \begin{array}{|c|c||c|c|c|c|}\hline
        \text{WP}& \rho & \Gm & \Mm & \Xm & \Ym   \\\hline 
       \alpha & \lambda & \lambda& \lambda& \lambda& \lambda\\
       \beta & \lambda & \lambda& \lambda& -\lambda& -\lambda\\
       \gamma & \lambda & \lambda& -\lambda& \lambda& -\lambda \\
       \delta & \lambda & \lambda& -\lambda& -\lambda& \lambda\\
       \hline
    \end{array}
\end{equation*}
    \end{minipage} 
\end{tabular}

\paragraph{M=4:} there are three maximal Wyckoff positions:

\begin{tabular}{cc}
    \begin{minipage}{.5\linewidth}
\begin{equation*}
    \begin{array}{|c||c|c|c|c|}\hline
        \pt & \text{Wyckoff position} & \tbs_{\pt} & M_{\pt} & n \\ \hline
       \zero & \alpha & \zero & 4 & 1\\ 
        \half \abss_x +\half \abss_y & \beta & \abss_x & 4 & 1\\
        \half \abss_x & \gamma_1 & \abss_x & 2 & 2\\
        \hline
    \end{array}
\end{equation*}
        \label{tab:wkf_C4}
    \end{minipage} &

    \begin{minipage}{.5\linewidth}
\begin{equation*}
     \begin{array}{|c|c||c|c|c|}\hline
        \text{WP}& \rho & \Gamma & \Mm & \Xm \\\hline 
        \alpha & \lambda & \lambda & \lambda & \lambda^2 \\
        \beta & \lambda & \lambda & -\lambda & -\lambda^2 \\
        \gamma_1 & \lambda & +\sqrt{\lambda},-\sqrt{\lambda} & +\sqrt{-\lambda},- \sqrt{-\lambda} & +1,-1  \\\hline
    \end{array}
\end{equation*}
    \end{minipage} 
\end{tabular}

\paragraph{M=3:} there are three maximal Wyckoff positions:

\begin{tabular}{cc}
    \begin{minipage}{.5\linewidth}
\begin{equation*}
    \begin{array}{|c||c|c|c|c|}\hline
        \pt & \text{Wyckoff position} & \tbs_{\pt} & M_{\pt} & n \\ \hline
        \zero & \alpha & \zero & 3 & 1\\ 
        \frac{\abss_x+\abss_w}{3} & \beta & \abss_x & 3 & 1\\
    \frac{-\abss_x +2 \abss_w}{3} & \gamma & \abss_w & 3 & 1\\\hline
    \end{array}
\end{equation*}
        \label{tab:wkf_C3}
    \end{minipage} &

    \begin{minipage}{.5\linewidth}
\begin{equation*}
    \begin{array}{|c|c||c|c|c|}\hline
        \text{WP}& \rho & \Gm & \Km & \bKm  \\\hline 
        \alpha & \lambda & \lambda & \lambda & \lambda \\
        \beta & \lambda & \lambda& \lambda \ee^{2\pi \ii /3} & \lambda\ee^{4\pi \ii /3} \\
        \gamma & \lambda & \lambda& \lambda\ee^{4\pi \ii /3} & \lambda\ee^{2\pi \ii /3}  \\\hline
    \end{array}
\end{equation*}
    \end{minipage} 
\end{tabular}

\paragraph{M=6:} there are three maximal Wyckoff positions:

\begin{tabular}{cc}
    \begin{minipage}{.4\linewidth}
    \begin{equation*}
    \begin{array}{|c||c|c|c|c|}\hline
        \pt & \text{Wyckoff position} & \tbs_{\pt} & M_{\pt} & n \\ \hline
        \zero & \alpha & \zero & 6 & 1\\ 
        \frac{\abss_x+\abss_w}{3} & \beta_1 & \abss_x & 3 & 2\\
        \half\abss_x & \gamma_1 & \abss_x & 2 & 3\\\hline
    \end{array}
\end{equation*}
        \label{tab:wkf_C6}
    \end{minipage} &
    \begin{minipage}{.6\linewidth}
\begin{equation*}
    \begin{array}{|c|c||c|c|c|}\hline
        \text{WP}& \rho & \Gm & \Km & \Mm \\\hline 
        \alpha & \lambda & \lambda & \lambda^2 & \lambda^3 \\
        \beta_1 & \lambda & \sqrt{\lambda},-\sqrt\lambda& \lambda\ee^{\frac{2\pi\ii}{3}},\lambda\ee^{\frac{4\pi\ii}{3}} & +1,-1 \\
        \gamma_1 & \lambda & \lambda^{1/3},\lambda^{1/3}\ee^{\frac{2\pi\ii}{3}},\lambda^{1/3}\ee^{\frac{4\pi\ii}{3}}& 1,\ee^{\frac{2\pi\ii}{3}},\ee^{\frac{4\pi\ii}{3}} & +\lambda,-\lambda,-\lambda  \\\hline
    \end{array}
\end{equation*}
    \end{minipage} 
\end{tabular}

\section{Generating set of non-interacting insulators}\label{app:GeneratingSet}

This appendix aims to formally show that within the non-interacting (``band structure combinatorics" classification), any state can be understood as the formal sum of a Chern insulator with $\Ch=1$ and atomic insulators.

\subsection{Generating set of atomic insulators}\label{app:GeneratingAI}
\def\MWP{\mathsf{MWP}}
We denote by $w[\ell]$ the atomic insulator induced from the Wyckoff position $w$ transforming with angular momentum $\ell$ ( $\lambda = e^{2\pi \ii \ell / M_{w}}$ in the tables of Sec.~\ref{app:AI:Summary}). We will omit the subscript of $w$ for ease of notation. For example, for $M=6$ the AI induced from $\beta_1$ with $\lambda = \ee^{\frac{2\pi \ell \ii}{3}}$ is denoted by $\beta[\ell]$, where $\ell =0,1,2$.

We start by pointing out some facts:
\begin{enumerate}
    \item The $J$ invariants of AIs induced from regular representations of the site group are independent of the Wyckoff position. In other words, $J[\sum_{\ell=0}^{M_w-1} w[\ell]]$ is independent of the Wyckoff position $w$. 
    \item Allowed representations in momentum space are subject to constraints relating to the total dimension at each orbit of HSM (Eq.~\eqref{eq:App:FillingConstraintRot}).
    \item For a given $M$, the number of HSM orbits and maximal Wyckoff positions are equal ($\abs{\HSM^{\star}} = \abs{\MWP}$) and the following holds:
    \begin{equation}
         \sum_{\kbs_\star \in\HSM_{\star }} M_{\kbs} = \sum_{\OO \in \MWP} M_{\OO}
    \end{equation}
    where $\MWP$ is the set of maximal Wyckoff positions. 
\end{enumerate}

The set of atomic insulators is given by 
\begin{equation}
    \AI  = \left\{\sum_{w\in \WP} \sum_{\ell=0}^{M_w-1} n_{w,\ell}w[\ell]| n_{w,\ell} \in \ZZ \right\}.
\end{equation}
where $\WP$ is the set of Wyckoff positions. Fact 1 above tells us that $J:\AI \to \ZZ^{1+\nRep} $ is not injective. As we are interested in the non-interacting classification, it is convenient to find a subset of $\AI$ where $J$ is injective and has the same image. Such a subset can be taken to be 
\begin{equation}
    \AI'  = \left\{n_{\alpha,0}\alpha[0]+\sum_{w\in \MWP} \sum_{\ell=1}^{M_w-1} n_{w,\ell}w[\ell]| n_{w,\ell} \in \ZZ \right\}.
\end{equation}
where we have used fact 1 above where $\MWP$ are the maximal Wyckoff positions. To see that $J[\AI']=J[\AI]$, note that we are excluding the AIs generated from $w[0]$ except for $\alpha[0]$. This works because using fact 1, we get $J[w[0]] =J[\sum_{\ell=0}^{M-1}\alpha[\ell] - \sum_{\ell'=1}^{M_{w}-1}w[\ell] ]$. Thus $w[0]$ can be expressed in terms of $\AI'$.

This shows that AIs in Tab.~\ref{tab:basisAI} are complete in the sense that any state in $\AI'$ can be written as a $\ZZ$-linear combination of them and $J[\AI] = J[\AI']$. We denote the set of AIs in  Tab.~\ref{tab:basisAI} by $\underline{\AI}$ and the elements as $\ai_{1},\ai_{2},\dots,\ai_{\nRep}$. We will omit the $M$ dependence as we did for $J$. Recall that $\nRep$ is the number of independent representation indices and depends on the wallpaper group (see App.~\ref{app:BandCombinatorics}) \footnote{The equality between the size of $\underline{\AI}$ and $\nRep$ can be deduced from the three facts at the start of the section of by direct computation. }. 
\begin{table}[t]
        \centering
       \begin{equation*}
    \begin{array}{|c|l|}\hline
       M  & \text{Basis} \\\hline
        2 & \alpha[0],\alpha[1], \beta[1],\gamma[1],\delta[1]\\
        3 &\alpha[0],\alpha[1],\alpha[2],\beta[1],\beta[2],\gamma[1],\gamma[2]\\
        4 &\alpha[0],\alpha[1],\alpha[2],\alpha[3],\beta[1],\beta[2],\beta[3],\gamma[1]\\
        6& \alpha[0],\alpha[1],\alpha[2],\alpha[3],\alpha[4],\alpha[5],\beta[1],\beta[2],\gamma[1]\\
        \hline
    \end{array}
\end{equation*}
        \caption{
        Basis of atomic insulators that are independent under the non-interacting classification. 
        }
        \label{tab:basisAI}
\end{table}

We next want to argue that we can achieve any of the non-interacting invariants compatible with $\Ch=0$ using atomic insulators. We construct a square matrix $T$ by evaluating $J_2,J_3,\dots, J_{\nRep}$ on the basis $\underline{\AI}$:
\begin{equation}
    (T)_{x,a} = J_{a+1}[\ai_{x}].
\end{equation}
The result can be read from Sec.~\ref{app:AI:Summary}. By explicit computation, we find $\abs{\det(T)} =M$. This means that $T$ is not invertible over $\ZZ$. However, this is expected because $T$ is not surjective due to the constraint in Eq.~\eqref{eq:App:ChernConstraintRotInv}.

To fix this, we can use the non-interacting invariants in Tab.~\ref{tab:basisNonInteractingFree}, that we denote by $\hat{J}$, where 
\begin{equation}
    \begin{split}
        \Nmc_{2} &= \frac{[\Mm_2^{(2)}] + [\Xm_2^{(2)}] + [\Ym_2^{(2)}] + \Ch}{2} \\
        \Nmc_{4} &= \frac{[\Mm_2^{(4)}]+2[\Mm_3^{(4)}]+3[\Mm_4^{(4)}] +2 [\Xm_2^{(2)}]  + \Ch}{4}\\
        \Nmc_{3} &= \frac{ [\Km_2^{(3)}] + 2[\Km_3^{(3)}] + [\bKm_2^{(3)}] + 2[\bKm_3^{(3)}] + \Ch}{3}\\
        \Nmc_{6,3} &= \frac{ 2[\Km_2^{(3)}] + 4[\Km_3^{(3)}] + \Ch}{3}\\
        \Nmc_{6,2} &= \frac{3[\Mm_2^{(2)}] + \Ch}{2}
    \end{split}.
\end{equation}
So that if we define $\hat{T}$ via 
\begin{equation}
    (\hat{T})_{x,a}  = \hat{J}_{a+1}[\ai_x].
\end{equation}
then $\abs{\det(\hat{T})}=1$, which means that $\hat{T}$ is invertible over $\ZZ$. 

Therefore, there is a one-to-one correspondence between the set $\AI'$ and the non-interacting invariants compatible with $\Ch=0$. In other words, any state with zero Chern number is equivalent to an atomic insulator under the non-interacting classification. 

\subsection{Complete generating set}\label{app:GeneratingComplete}

Let $\Psi_1$ be a Chern insulator with $\Ch=1$, see App.~\ref{app:ChernInsulatorsExamples} for the explicit models we use. Then, the set $\{\Psi_1, \ai_1,\ai_2,\dots,\ai_{\nRep}\}$ is a generating set of the non-interacting classification. To see this, construct the matrix $(1+\nRep)\times (1+\nRep)$ matrix $\tilde{T}$ defined as 
\begin{equation}
    \begin{split}
        \tilde{T}_{1,a} &= \Ch[\Psi_1]  \\
        \tilde{T}_{x,a} &= \hat{J}_{a}[\ai_{x-1}], \quad x =2,\dots,1+\nRep.
    \end{split}
\end{equation}
As all the atomic insulators have Chern number zero, the first row of $\tilde{T}$ is $(1,0,0,\dots,0)$, so that $\abs{\det(\tilde{T})} =\abs{\det(\hat{T})}$. Therefore, $\tilde{T}$ is invertible, which means that $\hat{J}$ is a bijection restricted to states of the form 
\begin{equation}\label{eq:nonInteractingDecomposition}
    \psi = \psi_1 \Psi_1 + \sum_{j=1}^{\nRep} \psi_{j+1}\ai_{j}, 
\end{equation}
where $\psi_j\in \ZZ$. Thus showing that any state in the non-interacting classification can be written as in Eq.~\eqref{eq:nonInteractingDecomposition}. 

To match the main text, we define $\Psi_{k} =  \ai_{k-1}$ as $k>1$ and denote the generating set as $\{\Psi_1,\Psi_2,\dots, \Psi_{\nRep+1}\}$. We then define the matrix $S$ by $J_{j} = \sum_k S_{jk}\hat{J}_{k}$. Then, we can write the matrix $O$ from the main text as 
\begin{equation}
    O_{jk} = J_j[\Psi_k] = \sum_{i}S_{ji}\hat{J}_{i}[\Psi_{k}] \Rightarrow O = S \tilde{T}.
\end{equation}
It is clear that $\det(S)\neq 0$, which together with $\det(\tilde{T}) = \pm 1$, imply that $\det(O)\neq 0$ and thus $O$ is invertible. For the sake of completeness, we report the matrix $O$ for each $M$:
\begin{itemize}
    \item $M=2:$
    \begin{equation}\label{eq:Omatrix_M2}
    O = \begin{pmatrix}
        1 & 0 & 0 & 0 & 0 & 0 \\
 0 & 1 & 0 & 0 & 0 & 0 \\
 1 & 0 & 1 & 1 & 1 & 1 \\
 -1 & 0 & 0 & 0 & -1 & -1 \\
 0 & 0 & 0 & -1 & 0 & -1 \\
 0 & 0 & 0 & -1 & -1 & 0 \\
    \end{pmatrix};
\end{equation}
\item $M=4:$
\begin{equation}\label{eq:Omatrix_M4}
    O = \begin{pmatrix}
        1 & 0 & 0 & 0 & 0 & 0 & 0 & 0 & 0 \\
 0 & 1 & 0 & 0 & 0 & 0 & 0 & 0 & 0 \\
 1 & 0 & 1 & 0 & 0 & 1 & 0 & 0 & 1 \\
 0 & 0 & 0 & 1 & 0 & 0 & 1 & 0 & 0 \\
 0 & 0 & 0 & 0 & 1 & 0 & 0 & 1 & 1 \\
 -1 & 0 & 0 & 0 & 0 & -1 & 0 & 1 & -1 \\
 0 & 0 & 0 & 0 & 0 & 0 & -1 & 0 & 1 \\
 0 & 0 & 0 & 0 & 0 & 1 & 0 & -1 & -1 \\
 0 & 0 & 0 & 0 & 0 & -1 & 1 & -1 & -1 \\
    \end{pmatrix};
\end{equation}
\item  $M=3:$
\begin{equation}\label{eq:Omatrix_M3}
    O = \begin{pmatrix}
        1 & 0 & 0 & 0 & 0 & 0 & 0 & 0 \\
 1 & 1 & 0 & 0 & 0 & 0 & 0 & 0 \\
 0 & 0 & 1 & 0 & 1 & 0 & 1 & 0 \\
 0 & 0 & 0 & 1 & 0 & 1 & 0 & 1 \\
 1 & 0 & 0 & 0 & -1 & 0 & -1 & 1 \\
 0 & 0 & 0 & 0 & 1 & -1 & 0 & -1 \\
 1 & 0 & 0 & 0 & -1 & 1 & -1 & 0 \\
 0 & 0 & 0 & 0 & 0 & -1 & 1 & -1 \\
    \end{pmatrix}
\end{equation}
\item $M=6:$
\begin{equation}\label{eq:Omatrix_M6}
    O = \begin{pmatrix}
       1 & 0 & 0 & 0 & 0 & 0 & 0 & 0 & 0 & 0 \\
 1 & 1 & 0 & 0 & 0 & 0 & 0 & 0 & 0 & 0 \\
 0 & 0 & 1 & 0 & 0 & 0 & 0 & 1 & 0 & 1 \\
 0 & 0 & 0 & 1 & 0 & 0 & 0 & 0 & 1 & 0 \\
 0 & 0 & 0 & 0 & 1 & 0 & 0 & 0 & 0 & 1 \\
 0 & 0 & 0 & 0 & 0 & 1 & 0 & 1 & 0 & 0 \\
 0 & 0 & 0 & 0 & 0 & 0 & 1 & 0 & 1 & 1 \\
 1 & 0 & 0 & 0 & 0 & 0 & 0 & -2 & 1 & 0 \\
 0 & 0 & 0 & 0 & 0 & 0 & 0 & 1 & -2 & 0 \\
 1 & 0 & 0 & 0 & 0 & 0 & 0 & 0 & 0 & -2 \\
    \end{pmatrix}
\end{equation}
\end{itemize}

\begin{table}[t]
        \centering
       \begin{equation*}
    \begin{array}{|c|l|}\hline
       M  & \text{Basis} \\\hline
        2 & \Ch, \#\Gamma_1^{(2)}, \#\Gamma_2^{(2)}, \Nmc_{2}, [X_2^{(2)}], [Y_2^{(2)}]\\
        3 & \Ch, \#\Gamma_1^{(3)}, \#\Gamma_2^{(3)}, \#\Gamma_3^{(3)}, \Nmc_{3}, [K_{3}^{(3)}], [\bar K_{2}^{(3)}], [\bar K_{3}^{(3)}]\\
        4 & \Ch, \#\Gamma_1^{(4)}, \#\Gamma_2^{(4)},\#\Gamma_3^{(4)},\#\Gamma_4^{(4)}, \Nmc_{4}, [M_3^{4}], [M_4^{4}], [X_2^{(2)}]\\
        6& \Ch, \#\Gamma_1^{(6)}, \#\Gamma_2^{(6)}, \#\Gamma_3^{(6)}, \#\Gamma_4^{(6)}, \#\Gamma_5^{(6)}, \#\Gamma_6^{(6)}, \Nmc_{6,3}, 
        [K_{3}^{(3)}], \Nmc_{6,2}\\
        \hline
    \end{array}
\end{equation*}
        \caption{
        Basis choice for linearly independent invariants in the non-interacting band insulator classification.
        }
        \label{tab:basisNonInteractingFree}
\end{table}

\begin{table}[t]
        \centering
       \begin{equation*}
    \begin{array}{|c|l|}\hline
       M  & \text{Basis} \\\hline
        2 & \text{QWZ},\alpha[0],\alpha[1], \beta[1],\gamma[1],\delta[1]\\
        3 & \text{Haldane},\alpha[0],\alpha[1],\alpha[2],\beta[1],\beta[2],\gamma[1],\gamma[2]\\
        4 & \text{QWZ},\alpha[0],\alpha[1],\alpha[2],\alpha[3],\beta[1],\beta[2],\beta[3],\gamma[1]\\
        6& \text{Haldane},\alpha[0],\alpha[1],\alpha[2],\alpha[3],\alpha[4],\alpha[5],\beta[1],\beta[2],\gamma[1]\\
        \hline
    \end{array}
\end{equation*}
        \caption{
        Complete basis for non-interacting states under the non-interacting classifications. 
        }
        \label{tab:basisComplete}
\end{table}

\subsection{Numerical results for Chern insulators with C=1}\label{app:ChernInsulatorsExamples}
\renewcommand{\arraystretch}{1.5}

Before we delve into technical details of the two examples, we would like to point here that we fix the overall phase of the rotation operators $\Cmop$ and their restrictions to $\Cmop\eval_D$ by requiring that the vacuum $\ket{\text{Vac}}$ is an eigenstate with eigenvalue, {i.e} $\Cmop\eval_D\ket{\text{Vac}} = +1\ket{\text{Vac}}$. The vacuum is the unique state $\ket{\text{Vac}}$ that satisfies $c_{\rbs}\ket{\text{Vac}} = 0$ for any annihilation operator $c_{\rbs}$.

\subsubsection{Chern insulators on the square lattice (QWZ)}\label{app:ChernInsulatorQiWuZhang}
We start with the Qi-Wu-Zhang model which was introduced as a model for the anomalous spin hall effect \cite{Qi2006QSHE_SquareLattice}. At every site of the square lattice, there are two fermions $c_{\ua}^{\dagger},c_{\da}^{\dagger}$. Letting $\tau^a$ be Pauli matrices acting on the $\sigma=\ua,\da$ indices, the Hamiltonian in momentum space is
\begin{equation}\label{eq:HamiltonianQWZMomentum}
    H_{\mathrm{QWZ}}=\sum_{\kbs }c_{\kbs}^{\dagger}\left[\vec{\tau} \cdot \vec{d}(\kbs) \right] c_{\kbs}
\end{equation}
with 
\begin{equation}
    \begin{split}
        d_x(\kbs) &=  t_x \sin(k_x)\\
        d_y(\kbs) &=  t_y \sin(k_y)\\
        d_z(\kbs) &= m + [t'_y \cos(k_x) + t'_y\cos(k_y)]
    \end{split}
\end{equation}
or in real space
\begin{equation}\label{eq:HamiltonianQWZ}
    H_{\mathrm{QWZ}} = \frac{1}{2}\sum_{j_x,j_y} m c_{j_x,j_y}^{\dagger}\tau^zc_{j_x,j_y}^{\,} +c_{j_x+1,j_y}^{\dagger}\left[t'_{x}\tau^z + t_x\ii \tau^x\right] c_{j_x,j_y}    +c_{j_x,j_y+1}^{\dagger}\left[t'_y\tau^z +t_y \ii \tau^y\right] c_{j_x,j_y} + \text{H.c.}.
\end{equation}
When $t_x=t_y = t$ and $t'_x = t'_y  = t'$, the Hamiltonian is invariant under $\hat{U}_{4}$, an order four rotation around the lattice sites. This rotation acts as 
\begin{equation}
    \begin{split}
    \hat{U}_{4}c^{\dagger}_{j_x,j_y}\hat{U}_{4}^{\dagger} = c^{\dagger}_{-j_y,j_x} \tilde{U}_{4};\quad 
        \tilde{U}_{4} 
    = \exp( \ii \pi\left(\frac{1-\tau^z}{4}\right)).
    \end{split}
\end{equation}
The model is also invariant under the translation operators $\hat{T}_{x}$ and $\hat{T}_y$ which act as 
\begin{equation}
    \begin{split}
        \hat{T}_{x}c^{\dagger}_{j_x,j_y} \hat{T}_{x}^{\dagger}= c^{\dagger}_{j_x+1,j_y}  \\
        \hat{T}_{y}c^{\dagger}_{j_x,j_y} \hat{T}_{y}^{\dagger}= c^{\dagger}_{j_x,j_y+1} . \\
    \end{split}
\end{equation}

\def\QWZ{{\mathrm{QWZ}}}
\def\QWZunder{\underline{\mathrm{QWZ}}}
We denote by $\QWZunder[t,t',m]$ the model with Hamiltonian given in Eq.~\eqref{eq:HamiltonianQWZ} and parameters $t_x=t_y=t$ and $t'_x=t'_y=t'$. We focus on the model $\QWZ = \QWZunder[1,1,1]$ which has Chern number $\Ch=+1$.

We next proceed to evaluate the real space invariants. Recall that we fix the overall phase of $\Cmop$ and their restrictions to subregions by requiring that the vacuum is an eigenstate with eigenvalue $+1$ (see start of App.~\ref{app:ChernInsulatorsExamples}). 

\paragraph{$M=4$ invariants:}
We set the sites to be at Wyckoff position $\alpha$. We calculate the $M=4$ RSIs with the $\Cmop$ operators: $\tilde{C}^{+}_{M_{\alpha}} = \hat{U}_4$,  $\tilde{C}^{+}_{M_{\beta}} = \hat{T}_x \hat{U}_4$ and  $\tilde{C}^{+}_{M_{\gamma}} = \hat{T}_x [\hat{U}_4]^2$.

We put the model on an $L$ by $L$ torus with periodic boundary conditions and diagonalize the Hamiltonian. We next numerically evaluate the $\Theta^{\pm}_{\OO}$'s using rectangular disks centered around $\OO$ with linear sizes of about $L/2$ to $3L/4$. We found that the invariants already converged for $L=24$ so we only report results for this system size in Tab.~\ref{tab:RealSpaceInvariantsQiWuZhang M = 4}. The eigenvalues at the HSM are $(\lambda_{\Gm}, \lambda_{\Mm}, \lambda_{\Xm}) = (\ii,1,-1)$ so the non-interacting invariants are 
\begin{equation}
    J[\QWZ] = (1, 0, 1, 0, 0, -1, 0, 0, 0) \quad ,  [\text{ for } M=4].
\end{equation}

\paragraph{$M=2$ invariants:} We next evaluate the $M=2$ case for the $\QWZ$ model.  We set the $\Cmop$ operators to be $\tilde{C}^{+}_{M_{\alpha}} = [\hat{U}_4]^2$,  $\tilde{C}^{+}_{M_{\beta}} = \hat{T}_x\hat{T}_y [\hat{U}_4]^2$,  $\tilde{C}^{+}_{M_{\gamma}} = \hat{T}_x [\hat{U}_4]^2$ and $\tilde{C}^{+}_{M_{\delta}} = \hat{T}_y [\hat{U}_4]^2$. We use the same system sizes as for the $M=4$ real space invariants. We report the results in Tab.~\ref{tab:RealSpaceInvariantsQiWuZhang M = 2}. The eigenvalues at the HSM are $(\lambda_{\Gm},\lambda_{\Mm},\lambda_{\Xm},\lambda_{\Ym})=(-1,1,-1,-1)$ so that the non-interacting invariants are
\begin{equation}
    J[\QWZ] = ({1,0,1,-1,0,0,}) \quad ,  [\text{ for } M=2].
\end{equation}

\begin{table}[t]
\begin{minipage}{.4\linewidth}
    \centering
    \begin{tabular}{|c|c|c|c|c|}\hline
         \diagbox{}{$\,\,\,\chi$} &  0 & 1&2  & 3\\ \hline
         $\Theta_{\alpha,\chi}^{+}$ & 0.5000 & 0.5000 & 1.5000 & 1.5000 \\
         $\Theta_{\beta,\chi}^{+}$ & 0.5000 & 1.5000 & 1.5000 & 0.5000 \\
         $\Theta_{\alpha,\chi}^{-}$ & 1.2500 & 3.7492 & 3.2500 & 3.7508\\
         $\Theta_{\beta,\chi}^{-}$ & 3.7500 & 3.2512 & 3.7500 &  1.2488\\
         $\Theta_{\gamma,\chi}^{-}$ & 1.7499 & 0.2501 & & \\ \hline
    \end{tabular}
    \caption{$M= 4 $ real-space invariants for the $\QWZ$ model of Sec.~\ref{app:ChernInsulatorQiWuZhang} rounded to four decimal places for a $24$ by $24$ system with disks with rectangular shape with $17^2$, $16^2$ and $16\times 17$ sites for the rotations around $\alpha$, $\beta$ and $\gamma$, respectively. }
    \label{tab:RealSpaceInvariantsQiWuZhang M = 4}
\end{minipage}
\quad\quad
\begin{minipage}{.4\linewidth}
    \centering
     \begin{tabular}{|c|c|c|}\hline
         \diagbox{}{$\,\,\,\chi$} &  0 & 1 \\ \hline
         $\Theta_{\alpha,\chi}^{-}$ & 1.7497 & 0.2503 \\
         $\Theta_{\beta,\chi}^{-}$ & 0.2501 & 1.7499 \\
         $\Theta_{\gamma,\chi}^{-}$ & 1.7499 & 0.2501 \\
         $\Theta_{\delta,\chi}^{-}$ & 1.7499 & 0.2501 \\\hline
    \end{tabular}
    \caption{$M= 2 $ real-space invariants for the $\QWZ$ model of Sec.~\ref{app:ChernInsulatorQiWuZhang} rounded to four decimal places for a $24$ by $24$ system with disks with rectangular shape with $17^2$, $16^2$, $17\times 16$ and $16\times 17$  sites for the rotations around $\alpha$, $\beta$, $\gamma$ and $\delta$ respectively. }
    \label{tab:RealSpaceInvariantsQiWuZhang M = 2}
\end{minipage}
\end{table}

We extract $\Theta^{\pm}_{M_\OO}$ for the QWZ model from Tab.~\ref{tab:RealSpaceInvariantsQiWuZhang M = 4} and Tab.~\ref{tab:RealSpaceInvariantsQiWuZhang M = 2} by assigning the closest value allowed by the quantization conditions in the main text. We report these values in  Tab.~\ref{tab:RealSpaceInvariantsQiWuZhangExact}.

\begin{table}[t]
    \centering
    \begin{minipage}{0.4\textwidth}
        \begin{tabular}{|c|c|c|c|c|}\hline
        \multicolumn{5}{|c|}{$M=4$}\\\hline
      \diagbox{}{$\,\,\,\chi$} &  0 & 1&2  & 3\\ \hline 
  $\Theta_{\alpha,\chi}^{+}$ & $\frac{1}{2}$ & $\frac{1}{2}$ & $\frac{3}{2}$ & $\frac{3}{2}$ \\
    $\Theta_{\beta,\chi}^{+}$ & $\frac{1}{2}$ & $\frac{3}{2}$  & $\frac{3}{2}$  & $\frac{1}{2}$  \\
    $\Theta_{\alpha,\chi}^{-}$ & $\frac{5}{4}$  & $\frac{15}{4}$  & $\frac{13}{4}$ & $\frac{15}{4}$\\
    $\Theta_{\beta,\chi}^{-}$ & $\frac{15}{4}$  & $\frac{13}{4}$  & $\frac{15}{4}$  &  $\frac{5}{4}$ \\
    $\Theta_{\gamma,\chi}^{-}$ & $\frac{7}{4}$  & $\frac{1}{4}$ & &  \\ \hline
    \end{tabular}
    \end{minipage}
    \begin{minipage}{0.4\textwidth}
     \begin{tabular}{|c|c||c|}\hline
     \multicolumn{3}{|c|}{$M=2$}\\\hline
    \diagbox{}{$\,\,\,\chi$} &  0 & 1\\ \hline 
    $\Theta_{\alpha,\chi}^{-}$ &  $\frac{7}{4}$ &  $\frac{1}{4}$ \\
         $\Theta_{\beta,\chi}^{-}$ &  $\frac{1}{4}$ &  $\frac{7}{4}$ \\
         $\Theta_{\gamma,\chi}^{-}$ &  $\frac{7}{4}$ &  $\frac{1}{4}$\\
         $\Theta_{\delta,\chi}^{-}$ &  $\frac{7}{4}$ &  $\frac{1}{4}$\\ \hline
    \end{tabular}
    \end{minipage}
    \caption{Real space invariants for the QWZ model obtained using the numerical evaluation and quantization condition from the main text.  }
    \label{tab:RealSpaceInvariantsQiWuZhangExact}
\end{table}

\subsubsection{Chern insulator $C_6$: Haldane model }\label{app:ChernInsulator:HaldaneModel}
\def\ebs{\boldsymbol{e}}
\def\Hald{\mathrm{Haldane}}
Consider the Haldane models with $t,\lambda\in \RR$ defined on the honeycomb lattice as
\begin{equation}\label{eq:HamiltonianHaldane}
  \begin{split}
        H_{\text{Haldane}}[t,\lambda] &= - t\sum_{\Rbs} \left[ c_{\Rbs A}^{\dagger}(c_{\Rbs B}^{}+c_{\Rbs+\ebs_2 B}^{}+c_{\Rbs+\ebs_3 B}^{})  + \text{h.c.}\right]+ \\
      &\quad  +\lambda \sum_{\Rbs} \left[ \ii c_{\Rbs A}^{\dagger}(c_{\Rbs+\ebs_1 A}^{} - c_{\Rbs+\ebs_2 A}^{} + c_{\Rbs+\ebs_3 A}^{}) -  \ii c_{\Rbs B}^{\dagger}(c_{\Rbs+\ebs_1 B}^{} - c_{\Rbs+\ebs_2 B}^{} + c_{\Rbs+\ebs_3 B}^{}) +\text{h.c.}\right] 
  \end{split}
\end{equation}
where $\ebs_1=[1,0]$, $\ebs_2=[1/2,\sqrt{3}/2]$, $\ebs_3 = \ebs_2-\ebs_1$. The system is invariant under the unitary operators $\hat{U}_{6}$ that act as 
\begin{equation}
    \hat{U}_{6} c_{\rbs}^{\dagger} \hat{U}_{6}^{\dagger} = c_{C_6\rbs}^{\dagger}.
\end{equation}
The model is also invariant under the translation acting as $T_{\Rbs} c^{\dagger}_{\rbs} T_{\Rbs}^{\dagger} = c^{\dagger}_{\rbs+\Rbs}$. Here $c_{\Rbs A/B} = c_{\Rbs +\tbs_{A/B}}$ where $\tbs_{A/B}= [ 1/2,\pm \sqrt{{3}}/6] $ \footnote{It seems like the model should only depend on $\ell$ modulo 3 because the site group of atom positions is $C_3$. However, the parity of $\ell$ in some sense determines the relative phase between the A and B sublattice. This sign is important because of $t$. }

Going to momentum space $c_{\kbs}= \frac{1}{\sqrt{N}}\sum_{\Rbs }\ee^{\ii\kbs\cdot\Rbs}[c_{\Rbs,A}, c_{\Rbs,B}]^\top$, the Hamiltonian becomes 
\begin{equation}
    \begin{split}
        H_{\text{Haldane}}[t,\lambda]&= \sum_{\kbs} c^{\dagger}_{\kbs} [\vec{\tau}\cdot \vec{d}(\kbs)]c_{\kbs}^{\,}. \\
        d_{x}(\kbs)-\ii d_y(\kbs) &= - t (1+\ee^{-\ii \kbs \cdot \ebs_2}+\ee^{-\ii \kbs \cdot \ebs_3}) \\
        d_{z}(\kbs) &= 2\lambda \left(\sin(\kbs\cdot\ebs_1) - \sin(\kbs\cdot\ebs_2) + \sin(\kbs\cdot\ebs_3)\right)\\
    \end{split}
\end{equation}

We denote by $\Haldane(t,\lambda)$ the model specified by Eq.~\eqref{eq:HamiltonianHaldane}. It turns out that as long as $t\neq 0\neq \lambda$, the system is gapped and has a unique ground state with periodic boundary conditions. We specialize to the model $\Hald \equiv \Haldane[1,\frac{1}{3\sqrt{3}}]$ which has $\Ch=+1$. 

We set the hexagon center to be at Wyckoff position $\alpha$ (so that sites belong to $\gamma$ for $M=6$ and to $\beta$ and $\gamma$ when $M=3$). 

We proceed to evaluate the real space invariants. Recall that we fix the overall phase of $\Cmop$ and their restrictions to subregions by requiring that the vacuum is an eigenstate with eigenvalue $+1$ (see start of App.~\ref{app:ChernInsulatorsExamples}). 

\paragraph{$M=6:$}We calculate the $M=6$ RSIs with the $\Cmop$ operators: $\tilde{C}^{+}_{M_{\alpha}} = \hat{U}_6$,  $\tilde{C}^{+}_{M_{\beta}} = \hat{T}_{\ebs_1} [\hat{U}_6]^2$ and  $\tilde{C}^{+}_{M_{\gamma}} = \hat{T}_{\ebs_1}[\hat{U}_6]^3$. We put the model on an $L$ by $L$ torus with periodic boundary conditions and diagonalize the Hamiltonian. We numerically evaluate the $\Theta^{\pm}_{\OO}$'s using hexagonal disks centered around $\OO$ with linear sizes of about $L/2$ to $3L/4$. Instead of calculating $\Theta^{\pm}_{\alpha}$ directly, we evaluate the invariants $\Theta^{\pm}_{(\alpha,s),\chi}$ with $s=2,3$ by treating the system as if $M_{\alpha}=s$ with the $s$-fold rotation operator $[\tilde{C}^{+}_{M_{\alpha}}]^{s}$. These invariants capture the same information as $\Theta_{\alpha}^{\pm}$ (see App.~\ref{app:RelationC6-RSI}). We use this alternative set of invariants because they converge faster as a function of system size compared to $\Theta^{\pm}_{\alpha}$.

We report results for $L=24$ in     Tab.~\ref{tab:RealSpaceInvariantsHaldaneN M = 6}. The eigenvalues at the HSM are $(\lambda_{\Gm}, \lambda_{\Km}, \lambda_{\Mm}) = (1,\ee^{2\pi \ii/3},-1)$ so the non-interacting invariants are 
\begin{equation}
    J[\Hald] = (1, 1, 0, 0, 0, 0, 0, 1, 0, 1) \quad ,  [\text{ for } M=6].
\end{equation}
Using the values in Tab.~\ref{tab:RealSpaceInvariantsHaldaneN M = 6} and Eq.~\eqref{eq:Relations M = 6}, we predict 
\begin{equation}\label{eq:PredictionTheta}
    \begin{split}
        (\Theta_{\alpha,0}^{+},\Theta_{\alpha,1}^{+},\Theta_{\alpha,2}^{+},\Theta_{\alpha,3}^{+},\Theta_{\alpha,4}^{+},\Theta_{\alpha,5}^{+}) &= (\frac{8}{3},\frac{5}{3},\frac{5}{3},\frac{8}{3},\frac{5}{3},\frac{5}{3})\\
    (\Theta_{\alpha,0}^{-},\Theta_{\alpha,1}^{-},\Theta_{\alpha,2}^{-},\Theta_{\alpha,3}^{-},\Theta_{\alpha,4}^{-},\Theta_{\alpha,5}^{-}) &= 
    (\frac{5}{12},\frac{35}{12},\frac{5}{12},\frac{59}{12},\frac{53}{12},\frac{59}{12}).
    \end{split}
\end{equation}
We calculated $\Theta^{\pm}_{\alpha}- \Theta^{\pm}_{\alpha,\text{LLL}}$ for a 36 by 36 system and disks with 384 and 600 sites inside the disk. We report the values in Tab.~\ref{tab:PredictionThetaAlpha} together with the expected value as predicted by Eq.~\eqref{eq:PredictionTheta}.

\begin{table}[b]
    \centering
    \begin{tabular}{|c||c|c|c|c|c|c|}\hline
    & \multicolumn{6}{c|}{$\chi$}\\\hline
        $n_{\text{sites}}$ &  0 & 1 & 2 &3 &4 &5\\ \hline
         384 &1.4601 &  3.9063 & 1.4601 &  0.0462 & 5.5811 &0.0462 \\
         600 & 1.4784 & 3.9532 & 1.4784  & 0.0233 & 5.5433 &  0.0233\\ 
         $\text{Exact}$ & $\frac{3}{2}$ & $4$ & $\frac{3}{2}$ & $0$ & $\frac{11}{2}$ & $ 0$\\\hline
    \end{tabular}
    \caption{Values $\Theta^{-}_{\alpha,\chi}- \Theta^{-}_{\alpha,\text{LLL}}$ evaluated on the $\Hald$ model on a 36 by 36 torus and a disc with $n_{\text{sites}}$ number of sites on the disk. "Exact" corresponds to using the prediction in Eq.~\eqref{eq:PredictionTheta}.}
    \label{tab:PredictionThetaAlpha}
\end{table}

\paragraph{$M=3:$} We evaluate the $M=3$ invariants with $\tilde{C}^{+}_{M_{\alpha}} = [\hat{U}_6]^2$, $\tilde{C}^{+}_{M_{\beta}} = T_{\ebs_1}[\hat{U}_6]^2$ and $\tilde{C}^{+}_{M_{\gamma}} = T_{\ebs_2}[\hat{U}_6]^2$. We use the same system sizes as for $M=3$. We report the results in Tab.~\ref{tab:RealSpaceInvariantsHaldaneN M = 3}. The band eigenvalues at the HSM are $(\lambda_{\Gm},\lambda_{\Km},\lambda_{\bKm}) = (1,\ee^{2\pi \ii/3},\ee^{2\pi \ii/3})$. We extract $\Theta^{\pm}_{M_\OO}$ for the Haldane model from Tab.~\ref{tab:RealSpaceInvariantsHaldaneN M = 3} and Tab.~\ref{tab:RealSpaceInvariantsHaldaneN M = 6} by assigning to closest value allowed by the quantization conditions in the main text. We report the values in  Tab.~\ref{tab:RealSpaceInvariantsHaldane}. 

\begin{table}[t]
\begin{minipage}{.4\linewidth}
    \centering
    \begin{tabular}{|c||c|c|c|} \hline
     \diagbox{}{$\,\,\,\chi$}     &  0 & 1 & 2\\ \hline
        $\Theta_{\alpha,3,\chi}^{+}$  & 0.6664 &2.6668 &2.6668 \\
        $\Theta_{\beta,\chi }^{+}$  & 2.6667 & 0.6667 &  2.6667\\
        $\Theta_{\alpha,2,\chi}^{-}$ & 1.7500 & 0.2500 &\\
        $\Theta_{\alpha,3,\chi}^{-}$ & 0.1667 & 1.1667 & 0.1667 \\
        $\Theta_{\beta,\chi }^{-}$ & 0.1667 & 0.1667 & 1.1667\\
        $\Theta_{\gamma,2}^{-}$ & 0.2499 & 1.7501 &\\ \hline
    \end{tabular}
    \caption{ $M=6$ real-space invariants for a the model $\Hald$ for a $24$ by $24$ system with hexagonal disks with 294, 166 and 150 sites around $\alpha$, $\beta$ and $\gamma$, respectively. }
    \label{tab:RealSpaceInvariantsHaldaneN M = 6}
\end{minipage}
\quad\quad
\begin{minipage}{.4\linewidth}
    \centering
    \begin{tabular}{|c||c|c|c|} \hline
         \diagbox{}{$\,\,\,\chi$} &  0 & 1 & 2\\ \hline
        $\Theta_{\alpha,\chi}^{+}$  & 0.6664 &2.6668 &2.6668 \\
        $\Theta_{\beta,\chi }^{+}$  & 2.6667 & 0.6667 &  2.6667\\
        $\Theta_{\gamma,\chi}^{-}$ & 2.6667 & 0.6667 &  2.6667\\
        $\Theta_{\alpha,\chi}^{-}$ & 0.1667 & 1.1667 & 0.1667 \\
        $\Theta_{\beta,\chi }^{-}$ & 0.1667 & 0.1667 & 1.1667\\
        $\Theta_{\gamma,\chi}^{-}$ & 0.1667 & 0.1667 & 1.1667\\ \hline 
    \end{tabular}
    \caption{ $M=3$ real-space invariants for a the model $\Hald$ for a $24$ by $24$ system with hexagonal disks with 294, 166 and 150 sites around $\alpha$, $\beta$ and $\gamma$, respectively. }
    \label{tab:RealSpaceInvariantsHaldaneN M = 3}
\end{minipage}
\end{table}

\begin{table}[t]
    \centering
       \begin{minipage}{.4\linewidth}
            \begin{tabular}{|c|c|c|c|}\hline
            \multicolumn{4}{|c|}{ $M=6$}\\\hline
        \diagbox{}{$\,\,\,\chi$} &  0 & 1&2  \\\hline
        $\Theta_{\alpha,3,\chi}^{+}$ & $\frac{2}{3}$ & $\frac{8}{3}$ & $\frac{8}{3}$\\
        $\Theta_{\beta,\chi }^{+}$ & $\frac{8}{3}$ & $\frac{2}{3}$ & $\frac{8}{3}$\\
        $\Theta_{\alpha,2,\chi}^{-}$ & $\frac{7}{4}$& $\frac{1}{4}$ & \\
        $\Theta_{\alpha,3,\chi}^{-}$ & $\frac{1}{6}$ & $\frac{7}{6}$ & $\frac{1}{6}$\\
        $\Theta_{\beta,\chi }^{-}$ & $\frac{1}{6}$ & $\frac{1}{6}$ & $\frac{7}{6}$\\
        $\Theta_{\gamma,2}^{-}$  &$\frac{1}{4}$ & $\frac{7}{4}$ &  \\ \hline  
    \end{tabular}
       \end{minipage}
       \begin{minipage}{.4\linewidth}
            \begin{tabular}{|c|c||c|c|c|}\hline
            \multicolumn{4}{|c|}{ $M=3$}\\\hline
        \diagbox{}{$\,\,\,\chi$} &  0 & 1&2  \\\hline
     $\Theta_{\alpha,\chi}^{+}$ & $\frac{2}{3}$ & $\frac{8}{3}$ & $\frac{8}{3}$\\
        $\Theta_{\beta,\chi }^{+}$ & $\frac{8}{3}$ & $\frac{2}{3}$ & $\frac{8}{3}$\\
        $\Theta_{\gamma,\chi }^{+}$ & $\frac{8}{3}$ & $\frac{2}{3}$ & $\frac{8}{3}$\\
        $\Theta_{\alpha,\chi}^{-}$ & $\frac{1}{6}$ & $\frac{7}{6}$ & $\frac{1}{6}$\\
        $\Theta_{\beta,\chi }^{-}$ & $\frac{1}{6}$ & $\frac{1}{6}$ & $\frac{7}{6}$\\
        $\Theta_{\gamma,\chi }^{-}$ & $\frac{1}{6}$ & $\frac{1}{6}$ & $\frac{7}{6}$\\\hline 
    \end{tabular}
       \end{minipage}
    \caption{ Real space invariants for the Haldane model obtained using the numerical evaluation and quantization condition from the main text.  }
    \label{tab:RealSpaceInvariantsHaldane}
\end{table}

\section{Relations between real-space invariants for $C_6$ symmetric systems}\label{app:RelationC6-RSI}

Consider a system that is p6 symmetric. We want to argue that we can determine $\Theta_{\alpha,\chi}^{\pm}$ from $\Theta_{(\alpha,3),\chi}^{\pm},\Theta_{(\alpha,2),\chi}^{-}$ and $c_-$. Here $\Theta_{(\alpha,s),\chi}^{\pm}$ are the real space invariants using the rotation operator $[\Cmop]^{6/s}$, {i.e.} we calculate the real-space invariant using an s-fold rotation instead of the six-fold rotation. We omit the $\chi$ dependence below.

We show this first in the $C=c_-=0$ case. We can write 
\begin{equation}
    \begin{split}
        \Theta_{\alpha,\AI}^{+} &= m_{\alpha} \mod 3 ,\\
        \Theta_{(\alpha,3),\AI}^{+} &= m_{\alpha} \mod 3,\\
        \Theta_{\alpha,\AI}^{-} &= \frac{n_{\alpha}}{2}+m_{\alpha} \mod 6,\\
        \Theta_{(\alpha,2),\AI}^{-} &= \frac{n_{\alpha}}{2}+m_{\alpha} \mod 2,\\
        \Theta_{(\alpha,3),\AI}^{-} &= \frac{n_{\alpha}}{2}+m_{\alpha} \mod 3/2.
    \end{split}
\end{equation}
where $n_{\alpha}$ and $m_{\alpha}$ is the charge and $C_6$ rotation charge localized at $\alpha$. It is clear that we can write 
\begin{equation}
    \begin{split}
        \Theta_{\alpha,\AI}^{+} & = \Theta_{(\alpha,3),\AI}^{+} \mod 3 \\
        -3[\Theta_{(\alpha,2),\AI}^{-}]_2+ 4[\Theta_{(\alpha,3),\AI}^{-}]_{3/2} & = -3(\frac{n_{\alpha}}{2}+m_{\alpha} + 2 k_1) +4(\frac{n_{\alpha}}{2}+m_{\alpha} + \frac{3}{2} k_2) \\
        &=\frac{n_{\alpha}}{2}+m_{\alpha} + (-6k_1 +6 k_2)
    \end{split}
\end{equation}
for some integers $k_1,k_2 $. Then, taking modulo 6 reduction and identifying the RHS with $\Theta_{\alpha,\AI}^{-}$, we find
\begin{equation}\label{eq:RelationThetaM6alpha}
  \Theta_{\alpha,\AI}^{-} = -3\Theta_{(\alpha,2),\AI}^{-}+ 4\Theta_{(\alpha,3),\AI}^{-} \mod 6.
\end{equation}

Finally, for a system with $c_-\neq 0$, we use the relation in Eq.~\eqref{eq:Theta_def_gen} to obtain
\begin{equation}\label{eq:Relations M = 6}
    \begin{split}
        \Theta_{\alpha}^{+} &= c_-\Theta_{\alpha,\text{LLL}}^{+} + (\Theta_{\alpha,3}^{+}-c_-\Theta_{(\alpha,3),\text{LLL}}^{+}) \\ 
        &= \Theta_{\alpha,3}^{+} -c_-  \\ 
        \Theta_{\alpha}^{-} &= c_-\Theta_{\alpha,\text{LLL}}^{-} -3\left(\Theta_{\alpha,2}^{-}-c_-\Theta_{(\alpha,2),\text{LLL}}^{-}\right) + 4 \left(\Theta_{\alpha,3}^{-}-c_-\Theta_{(\alpha,3),\text{LLL}}^{-}\right)\\ 
       &= -3 \Theta_{\alpha,2}^{-} + 4 \Theta_{\alpha,3}^{-} - c_-
    \end{split}
\end{equation}

\newpage

\end{widetext}

\clearpage
\bibliography{refs}

\end{document}